\documentclass[journal]{IEEEtran}
\IEEEoverridecommandlockouts
\usepackage{lineno}
\usepackage{cite}
\usepackage{hyperref}
\usepackage{amsmath,amssymb,amsfonts}
\usepackage{amsthm}
\usepackage{mdwmath}
\usepackage{mdwtab}
\DeclareMathOperator*{\argmax}{argmax}

\usepackage{graphicx}
\usepackage{textcomp}
\usepackage{xcolor}
\usepackage{float}
\usepackage{subfigure}
\usepackage{mathrsfs}
\usepackage{mathtools}
\usepackage{multirow}
\usepackage{algpseudocode}
\usepackage[linesnumbered,ruled,vlined]{algorithm2e}
\usepackage[justification=centering]{caption}
\usepackage{setspace}
\usepackage{footmisc}
\usepackage{caption} 
\usepackage{dsfont}
\usepackage{bbm}
\usepackage{cases}
\usepackage{stfloats,enumerate}
\theoremstyle{definition}
\newtheorem{remark}{Remark}
\newtheorem{corollary}{Corollary}

\newtheorem{theorem}{Theorem}

\newtheorem{lemma}{Lemma}

\makeatletter
\newcommand{\biggg}{\bBigg@{3}}
\newcommand{\Biggg}{\bBigg@{3.5}}
\makeatother
\usepackage{bm}
\expandafter\def\expandafter\normalsize\expandafter{%
	\normalsize%
	\setlength\abovedisplayskip{5pt}%
	\setlength\belowdisplayskip{5pt}%
	\setlength\abovedisplayshortskip{4pt}%
	\setlength\belowdisplayshortskip{4pt}%
}
\begin{document}
	\title{Continuous-Aperture Array (CAPA)-Based Wireless Communications: Capacity Characterization}
	\author{Boqun Zhao, Chongjun Ouyang, Xingqi Zhang, and Yuanwei Liu\vspace{-10pt}
		\thanks{An earlier version of this paper was presented in part at the IEEE Global Communications Conference, Cape Town, South Africa, in Dec. 2024 \cite{ouyang2024performance}.}
		\thanks{B. Zhao and X. Zhang are with the Department of Electrical and Computer Engineering, University of Alberta, Edmonton AB, T6G 2R3, Canada (email: \{boqun1, xingqi.zhang\}@ualberta.ca).}
		\thanks{C. Ouyang was with the School of Electrical and Electronic Engineering, University College Dublin, Dublin, D04 V1W8, Ireland, and is now with the School of Electronic Engineering and Computer Science, Queen Mary University of London, London, E1 4NS, U.K. (e-mail: c.ouyang@qmul.ac.uk).}
		\thanks{Y. Liu is with the Department of Electrical and Electronic Engineering, The University of Hong Kong, Hong Kong (e-mail: yuanwei@hku.hk).}
	}
	\maketitle
	\begin{abstract}
		The capacity limits of continuous-aperture array (CAPA)-based wireless communications are characterized. To this end, an analytically tractable transmission framework is established for both uplink and downlink CAPA systems. Based on this framework, closed-form expressions for the single-user channel capacity are derived. The results are further extended to a multiuser case by characterizing the capacity limits of a two-user channel and proposing the associated capacity-achieving decoding and encoding schemes. In the uplink case, the capacity-achieving detectors and sum-rate capacity are derived, and the capacity region is characterized. In the downlink case, the uplink-downlink duality is established by deriving the uplink-to-downlink and downlink-to-uplink transformations under the same power constraint, based on which the optimal source current distributions and the achieved sum-rate capacity and capacity region are characterized. For comparison, the uplink and downlink sum-rates achieved by the linear zero-forcing scheme are also analyzed. To gain further insights, several case studies are presented by specializing the derived results into various array structures, including the planar CAPA, linear CAPA, and planar spatially discrete array (SPDA). { Numerical results are provided to reveal that the channel capacity achieved by CAPAs converges towards a finite upper bound as the aperture size increases; and CAPAs offer superior capacity over the conventional SPDAs.}
	\end{abstract} 
	
	\section{Introduction}
	Multiple-antenna technology is considered one of the fundamental building blocks of modern wireless communication systems. At its core lies the principle of utilizing an increased number of antenna elements and expanded array apertures to enhance spatial degrees of freedom (DoFs) and array gains, ultimately improving channel capacity \cite{tse2005fundamentals,heath2018foundations}. In this context, technologies such as multiple-input multiple-output (MIMO) and massive MIMO have played a crucial role in advancing cellular networks from the third generation (3G) to the fourth (4G), fifth (5G), and beyond \cite{bjornson2019massive,zhang2020prospective,liu2024near}.
	
	In general, the number of spatial DoFs and array gains scale with the number of antenna elements and the array aperture size. Motivated by this principle, various advanced MIMO variants and novel array architectures have been proposed. Examples include holographic MIMO \cite{pizzo2020spatially}, gigantic MIMO \cite{bjornson2024enabling}, extremely large aperture arrays \cite{de2020non}, reconfigurable intelligent surfaces \cite{basar2019wireless}, dynamic metasurface antennas \cite{shlezinger2021dynamic}, and fluid \cite{wong2020fluid}, movable \cite{zhu2023movable}, or pinching \cite{liu2025pinching} antennas. Extensive theoretical analyses and experimental studies have demonstrated the substantial performance gains achievable through these emerging multiple-antenna technologies. Despite their diverse architectures, these novel multiple-antenna systems share a common evolutionary trend: \emph{larger} aperture sizes, \emph{denser} antenna configurations, \emph{higher} operating frequencies, and \emph{more flexible} structures. {This progression paves the way for the development of an (approximately) continuous electromagnetic (EM) aperture \cite{hu2018beyond,dardari2020communicating,capa_single_4}, commonly referred to as a \emph{continuous-aperture array (CAPA)} \cite{sayeed2010continuous,zhang2023pattern,liu2024capa}, which is to be focused on in this paper.}
	
	A CAPA functions as a single, large-aperture antenna with a continuous current distribution, which comprises a (virtually) infinite number of radiating elements integrated with electronic circuits and driven by a limited number of radio-frequency (RF) chains. Unlike conventional spatially discrete arrays (SPDAs), a CAPA fully exploits the entire aperture surface and allows for precise control over the current distribution \cite{liu2024capa}. This capability significantly enhances spatial DoFs and array gains, as demonstrated in various research endeavors; see Section {\ref{Section: Theoretical Investigations of CAPAs}} for more relevant references. However, the continuous nature of EM field interactions in CAPAs necessitates a fundamental shift in system modeling. Instead of the \emph{matrix-based} models used for SPDAs, CAPA-based systems require an EM-theoretic approach, leading to \emph{a continuous Hilbert–Schmidt operator-based} integral framework \cite{poon2005degrees,noise,capa_single_0,migliore2008electromagnetics,migliore2018horse,capa_single_1,wan2023mutual}. This transition from discrete to continuous modeling is not just a mathematical refinement but a paradigm shift in the design and analysis of wireless transmission systems. To fully unlock the potential of CAPAs, a novel conceptual framework tailored to their unique characteristics is essential.
	\subsection{Prior Works}
	\subsubsection{Hardware Implementation of CAPAs}
	{The development of semi-CAPA and CAPA architectures has a long history in antenna design research, dating back to pioneering contributions in the 1960s \cite{wheeler1965simple,staiman1968new}. With recent advancements in materials science and array fabrication, researchers have successfully developed CAPA prototypes based on designs such as metasurface-based leaky-wave antennas \cite{smith2017analysis,araghi2021holographic}, optically driven tightly coupled arrays \cite{prather2017optically}, and interdigital transducer-based grating antennas \cite{yuan2024interdigital}; see \cite{gong2023holographic,liu2024capa} for further details. Some of these prototypes have even reached commercialization, with early test reports demonstrating their significant potential for enhancing coverage and throughput in practical wireless networks \cite{staff2019holographic,sazegar2022full}.
		
		It is worth noting that most CAPA prototypes available today are semi-continuous; however, they are effectively treated as continuous antennas, as emphasized by their inventors \cite{staff2019holographic,sazegar2022full}. Achieving a true CAPA would require modulating phase and amplitude at each aperture point in an analog manner. One possible realization involves using a metasurface composed of materials such as photographic emulsion, which, when illuminated by a tailored wave, could enable unique phase and amplitude modulation at each point. This approach would ultimately generate the desired continuous current distribution when a reference wave propagates over the surface \cite{bjornson2019massive,gong2023holographic}. Such advancements align with the objectives of leading enterprises, which are striving for an advanced level of spatial continuity in antenna design.
		
		Besides the aforementioned array architectures, several key challenges must be carefully addressed to ensure the \emph{practicality} and \emph{feasibility} of CAPA and semi-CAPA systems. Some of the most critical challenges include:
		\begin{itemize}
			\item \textbf{EM Coupling:} When antennas are densely or continuously packed, mutual coupling (MC) effects become significant. MC occurs where EM waves transmitted by one antenna are absorbed by adjacent antennas, which alters their circuitry and distorting the overall spatial channel \cite{ivrlavc2010toward}. In a CAPA or an ultra-dense semi-CAPA, EM coupling can lead to severe performance degradation, increased interference, and complex interactions that are difficult to predict and mitigate. 
			\item \textbf{Complex Feed Network:} A CAPA or an ultra-dense semi-CAPA requires a highly complex feed network to properly excite the aperture or individual elements. Designing such a network for a continuous surface or a large number of densely packed elements introduces substantial complexity, increases insertion losses, and raises overall power consumption \cite{mailloux2017phased,balanis2016antenna}.
			\item \textbf{Material Limitations:} The materials used to construct and sustain a CAPA or an ultra-dense semi-CAPA must meet stringent physical requirements. Challenges such as thermal management, mechanical durability, and environmental susceptibility (e.g., humidity, temperature variations, and material aging) impose additional constraints on long-term reliability and scalability \cite{mailloux2017phased,balanis2016antenna}.
		\end{itemize}
		Taken together, these challenges indicate that realizing a fully functional CAPA or a densely packed semi-CAPA remains elusive, even with recent advancements in materials and fabrication techniques. Addressing these issues requires collaborative efforts from both the communication and antenna communities.}
	\subsubsection{Theoretical Design and Analysis of CAPAs}\label{Section: Theoretical Investigations of CAPAs}
	Although research on the hardware implementation of an ideal CAPA is still in its infancy, increasing efforts have been devoted to the theoretical design and analysis of CAPA- or semi-CAPA-based communications. Many of these studies rely on simplified models, such as neglecting the effects of EM coupling and considering narrowband scenarios. One of the most widely studied topics is the number of EM DoFs \cite{liu2024near}, which has been investigated using various analytical tools, including diffraction theory \cite{miller2000communicating}, Nyquist-Shannon sampling theorem \cite{pizzo2022nyquist,poon2005degrees}, and Landau's theorem \cite{do2023parabolic}. Beyond DoFs, the array gain achieved by CAPAs has also been analyzed in \cite{ouyang2024reactive}. These initial studies collectively highlight the advantages of CAPAs over traditional SPDAs.
	
	{In addition, significant research attention has been given to characterizing the channel capacity of CAPA-based point-to-point communication systems. These studies leverage Green’s function-based models to describe the EM channel and apply Hilbert operator theory to derive capacity expressions under various propagation environments \cite{capa_single_0,migliore2008electromagnetics,noise,capa_single_1,migliore2018horse,wan2023mutual}. Recently, this research has been extended to scenarios where a CAPA-based reconfigurable surface assists communications between SPDAs \cite{xu2023exploiting}.} Another key research direction focuses on continuous beamforming design for CAPA-based communications. One prevalent approach relies on discretization-based methods, such as Fourier series expansion, to approximate continuous signals and simplify the associated optimization problems \cite{capa_single_4,zhang2023pattern}. Additionally, calculus of variations has been applied to design low-complexity beamforming strategies \cite{wang2024beamforming} and explore optimal beamforming structures \cite{wang2024optimal,per7}.
	
	{As mentioned earlier, these research developments have relied on simplified system models that primarily consider the spatially continuous nature of CAPAs while omitting other crucial factors, such as the impact of MC and the challenges posed by wideband scenarios. To address these limitations, the authors in \cite{castellanos2024electromagnetic} proposed an EM-based array manifold that accounts for complex physical interactions, including MC. This framework enables the modeling of arbitrary antenna configurations, including CAPAs, with greater physical accuracy.
		
		Regarding wideband transmission, the work in \cite{pizzo2022fourier} introduced a small-scale fading model for CAPA-based communications using $k$-space (wavenumber domain) analysis. This approach provides a wavenumber (spatial-frequency) representation of how the aperture is excited, which illustrates how energy is distributed across different propagation directions (angles) at each frequency. Such insights can be instrumental in tackling spatial-wideband effects in CAPA-based wideband communications. Another research direction involves circuit-based modeling. A recent example in \cite{akrout2023super} employs multi-port circuit theory combined with Chu's antenna model to investigate the impact of MC on the wideband behavior of ultra-dense arrays.}
	\subsection{Motivation and Contributions}
	Despite these advantages, the fundamental theoretical limits of CAPA-based communications remain largely unexplored. Building on existing research in CAPA-based single-user communications, this paper takes a step forward by investigating the capacity limits of CAPA-based multiuser systems and establishing an analytically tractable framework for multiuser CAPA communications. { Specifically, we consider a scenario where a \emph{CAPA} serves multiple users equipped with \emph{traditional single-antenna devices}. For simplicity, we focus on a narrowband model, while extending the analysis to wideband scenarios remains an important avenue for future research.} The main contributions are summarized as follows.
	\begin{itemize}
		\item We propose a transmission framework for uplink and downlink CAPA communications using EM field theories, where continuous operators model both signals and spatial responses. Based on this framework, we design optimal detectors and source current distributions to achieve the single-user channel capacity and derive closed-form expressions for the corresponding capacity.
		\item We extend our analysis to multiuser CAPA systems by studying a two-user case. For uplink communications, we propose using successive decoding to achieve capacity and derive closed-form expressions for the optimal detectors of each user. Based on this, we derive closed-form expressions for the per-user rate and the sum-rate capacity, and characterize the capacity region. For downlink communications, we establish the uplink-downlink duality by presenting the uplink-to-downlink and downlink-to-uplink transformations under the same sum power constraint and derive closed-form expressions for the capacity-achieving source current distributions. Additionally, we characterize the downlink capacity limits by deriving the per-user rate, sum-rate capacity, and capacity region. {For comparison, we also analyze the uplink and downlink sum-rates achieved using the linear zero-forcing (ZF) detector and precoder, respectively.}   
		\item To unveil further insights, we present several case studies exploring the line-of-sight (LoS) channel model. The derived results are specialized to various array structures, including the planar CAPA, linear CAPA, and planar SPDA. We derive closed-form expressions for the sum-rate capacity in each case and perform asymptotic analyses by setting the aperture size to infinity.
		\item We present numerical results illustrating that the capacity achieved by the CAPA increases with aperture size and converges to a finite upper limit. Our results demonstrate that, compared to CAPA, the capacity of the conventional SPDA is limited by its array occupation ratio. {Moreover, CAPA can theoretically achieve greater capacity than conventional SPDA, whose capacity converges to that of CAPA as the array occupation ratio approaches one.}
	\end{itemize}
	
	The remainder of this paper is organized as follows. Section \ref{Section: System Model} introduces the transmission framework for CAPA-based communications. Sections \ref{sec_uplink} and \ref{Section: downlink} delve into the capacity analysis and the capacity-achieving schemes for uplink and downlink transmission, respectively. Section \ref{Section: Special Cases} provides case studies for several specific array structures. Section \ref{section_numerical} presents numerical results to demonstrate the superior performance of CAPAs. Finally, Section \ref{conclusion} concludes the paper.
	
	\begin{figure}[!t]
		\centering
		\includegraphics[height=0.23\textwidth]{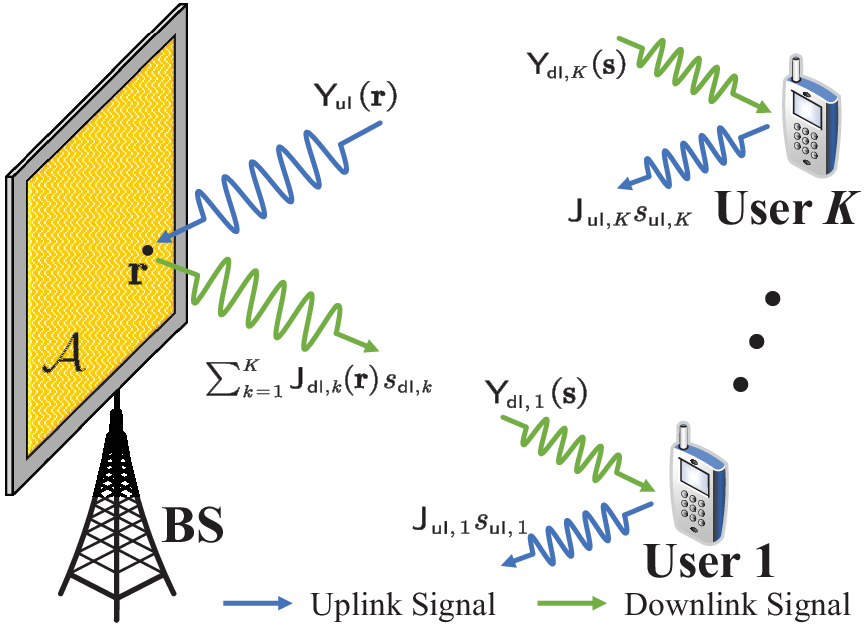}
		\caption{Illustration of CAPA-based communications.}
		\vspace{-5pt}
		\label{Figure: System_Model}
	\end{figure}
	
	\section{System Model}\label{Section: System Model}
	{Consider a multiuser multiple-input single-output (MISO) channel where a base station (BS) simultaneously communicates with $K$ users, as shown in {\figurename} {\ref{Figure: System_Model}}. The BS is equipped with a CAPA featuring an aperture $\mathcal{A}\subseteq{\mathbbmss{R}}^{3\times1}$ with size $A=\int_{\mathcal{A}}{\rm{d}}{\mathbf{s}}$. Each user $k\in\{1,\ldots,K\}$ is equipped with a single conventional antenna. Let $\mathbf{s}_k\in\mathbbmss{R}^{3\times1}$ denote the location of user $k$, and let ${\mathcal{A}}_{k}\subseteq{\mathbbmss{R}}^{3\times1}$ represent the effective aperture of the user's antenna. We consider that the aperture size of each user's antenna, i.e., $A_{{\mathsf{u}},k}=\int_{{\mathcal{A}}_k}{\rm{d}}{\mathbf{s}}$, is on the order of a sub-wavelength scale, which satisfies $A_{{\mathsf{u}},k}\ll A$. For instance, if an isotropic antenna is used, we have $A_{{\mathsf{u}},k}=\frac{\lambda^2}{4\pi}$, where $\lambda$ denotes the wavelength. In this context, the variations in signals and channel responses across the aperture are negligible, which means that each user can only resolve a single data stream.}
	\subsection{Signal Model}
	\subsubsection{Uplink Communications}
	We commence with the uplink scenario, where the users simultaneously transmit their respective messages to the BS. Let us define ${\mathsf{J}}_{{\mathsf{ul}},k}({\mathbf{s}})\in{\mathbbmss{C}}$ as the continuous distribution of source currents generated by user $k$ at point $\mathbf{s}\in{\mathcal{A}}_{k}$ to convey the normalized coded date symbol $s_{{\mathsf{ul}},k}\in{\mathbbmss{C}}$ with ${\mathbbmss{E}}\{\lvert s_{{\mathsf{ul}},k}\rvert^2\}=1$. The transmitted signal by user $k$ is thus given by $x_{{\mathsf{ul}},k}({\mathbf{s}})={\mathsf{J}}_{{\mathsf{ul}},k}({\mathbf{s}})s_{{\mathsf{ul}},k}$ for $\mathbf{s}\in{\mathcal{A}}_{k}$.
	
	The excited electric field ${\mathsf{E}}_{{\mathsf{ul}},k}(\mathbf{r})\in{\mathbbmss{C}}$ by user $k$ at point $\mathbf{r}\in{\mathcal{A}}$ can be expressed as follows \cite{balanis2016antenna}:
	\begin{align}
		{\mathsf{E}}_{{\mathsf{ul}},k}(\mathbf{r})&=\int_{{\mathcal{A}}_{k}}{\mathsf{H}}(\mathbf{r},{\mathbf{s}}){x}_{{\mathsf{ul}},k}({\mathbf{s}}){\rm{d}}{\mathbf{s}},\label{User_k_electric_radiation_field}
	\end{align}
	where ${\mathsf{H}}(\mathbf{r},{\mathbf{s}})\in{\mathbbmss{C}}$ denotes the spatial channel response from $\mathbf{s}$ to $\mathbf{r}$. Given that the aperture size of each user is of a sub-wavelength scale, the variations in currents and channel responses across the transmit aperture are negligible. Consequently, we have ${\mathsf{H}}(\mathbf{r},{\mathbf{s}})\approx{\mathsf{H}}(\mathbf{r},{\mathbf{s}}_{k})\triangleq\mathsf{H}_k(\mathbf{r})$ and ${\mathsf{J}}_{{\mathsf{ul}},k}({\mathbf{s}})\approx {\mathsf{J}}_{{\mathsf{ul}},k}({\mathbf{s}}_k) \triangleq {\mathsf{J}}_{{\mathsf{ul}},k}$ for $\mathbf{s}\in{\mathcal{A}}_{k}$. It follows that
	\begin{align}\label{Simplified_electric_radiation_field}
		{\mathsf{E}}_{{\mathsf{ul}},k}(\mathbf{r})\approx \mathsf{H}_k(\mathbf{r}){\mathsf{J}}_{{\mathsf{ul}},k}s_{{\mathsf{ul}},k}\int_{{\mathcal{A}}_k}{\rm{d}}{\mathbf{s}}
		=\mathsf{H}_k(\mathbf{r}){\mathsf{J}}_{{\mathsf{ul}},k} A_{{\mathsf{u}},k} s_{{\mathsf{ul}},k}.
	\end{align}
	The total observed electric field ${\mathsf{Y}}_\mathsf{ul}(\mathbf{r})$ at point $\mathbf{r}\in{\mathcal{A}}$ is the sum of the information-carrying electric fields $\{{\mathsf{E}}_{{\mathsf{ul}},k}(\mathbf{r})\}_{k=1}^{K}$, along with a random noise field ${\mathsf{N}}_\mathsf{ul}(\mathbf{r})\in{\mathbbmss{C}}$, i.e.,
	\begin{subequations}
		\begin{align}
			{\mathsf{Y}}_{\mathsf{ul}}(\mathbf{r})&=\sum\nolimits_{k=1}^{K}{\mathsf{E}}_{{\mathsf{ul}},k}(\mathbf{r})+{\mathsf{N}}_\mathsf{ul}(\mathbf{r})\\
			&\approx\sum\nolimits_{k=1}^{K}\mathsf{H}_k(\mathbf{r}){\mathsf{J}}_{{\mathsf{ul}},k} A_{{\mathsf{u}},k} s_{{\mathsf{ul}},k}
			+{\mathsf{N}}_\mathsf{ul}(\mathbf{r}),\label{Total_electric_radiation_field}
		\end{align}
	\end{subequations}
	where ${\mathsf{N}}_\mathsf{ul}(\mathbf{r})$ accounts for thermal noise \cite{noise}. The noise field is modeled as a zero-mean complex Gaussian random process satisfying ${\mathbbmss{E}}\{{\mathsf{N}}_\mathsf{ul}(\mathbf{r}){\mathsf{N}}_\mathsf{ul}^{*}({\mathbf{r}}')\}={\sigma}^2\delta(\mathbf{r}-{\mathbf{r}}')$ \cite{noise}, where $\delta(\cdot)$ represents the Dirac delta function, and $\sigma^2$ describes the noise intensity. ${\mathsf{N}}_{\mathsf{ul}}(\mathbf{r})$ and $\{s_{{\mathsf{ul}},k}\}_{k=1}^{K}$ are assumed to be uncorrelated. After observing ${\mathsf{Y}}_{\mathsf{ul}}(\mathbf{r})$, the BS should employ a properly designed detector along with maximum-likelihood (ML) decoding to recover the data information encoded in $\{s_{{\mathsf{ul}},k}\}_{k=1}^{K}$, which will be detailed in Section \ref{Section: Single-User Case: Uplink Capacity}.
	\subsubsection{Downlink Communications}
	At the BS, the downlink electrical signal is designed as follows:
	\begin{align}
		{x}_{{\mathsf{dl}}}({\mathbf{r}})=\sum\nolimits_{k=1}^{K}{\mathsf{J}}_{{\mathsf{dl}},k}({\mathbf{r}})s_{{\mathsf{dl}},k},
	\end{align}
	where $s_{{\mathsf{dl}},k}\in{\mathbbmss{C}}$ represents the coded date symbol dedicated to user $k$ with ${\mathbbmss{E}}\{\lvert s_{{\mathsf{dl}},k}\rvert^2\}=1$, and ${\mathsf{J}}_{{\mathsf{dl}},k}({\mathbf{r}})\in{\mathbbmss{C}}$ (${\mathbf{r}}\in{\mathcal{A}}$) is the associated current distribution to convey $s_{{\mathsf{dl}},k}$. As a result, the excited electric field at point $\mathbf{s}\in{\mathcal{A}}_{k}$ within user $k$'s aperture can be written as ${\mathsf{E}}_{{\mathsf{dl}},k}(\mathbf{s})=\int_{{\mathcal{A}}}{\mathsf{H}}(\mathbf{s},{\mathbf{r}}){x}_{{\mathsf{dl}}}({\mathbf{r}}){\rm{d}}{\mathbf{r}}$, viz.
	\begin{align}\label{do_User_k_electric_radiation_field}
		{\mathsf{E}}_{{\mathsf{dl}},k}(\mathbf{s})=\int_{{\mathcal{A}}}{\mathsf{H}}(\mathbf{s},{\mathbf{r}})\left(\sum\nolimits_{k'=1}^{K}{\mathsf{J}}_{{\mathsf{dl}},k'}({\mathbf{r}})
		s_{{\mathsf{dl}},k'}\right){\rm{d}}{\mathbf{r}}.
	\end{align}
	Since the variations of the electric field and channel response along the receive aperture are negligible, we have ${\mathsf{E}}_{{\mathsf{dl}},k}(\mathbf{s})\approx{\mathsf{E}}_{{\mathsf{dl}},k}(\mathbf{s}_k)\triangleq {\mathsf{E}}_{{\mathsf{dl}},k}$ and ${\mathsf{H}}(\mathbf{s},{\mathbf{r}})\approx{\mathsf{H}}(\mathbf{s}_k,{\mathbf{r}})$ for $\mathbf{s}\in{\mathcal{A}}_k$. By further assuming channel reciprocity between the uplink and downlink, we have ${\mathsf{H}}(\mathbf{s}_k,{\mathbf{r}})={\mathsf{H}}({\mathbf{r}},\mathbf{s}_k)=\mathsf{H}_k(\mathbf{r})$, which gives
	\begin{align}\label{do_Simplified_electric_radiation_field}
		{\mathsf{E}}_{{\mathsf{dl}},k}(\mathbf{s})\approx{\mathsf{E}}_{{\mathsf{dl}},k}=\!
		\int_{{\mathcal{A}}}{\mathsf{H}}_k({\mathbf{r}})\left(\sum\nolimits_{k'=1}^{K}{\mathsf{J}}_{{\mathsf{dl}},k'}({\mathbf{r}})
		s_{{\mathsf{dl}},k'}\!\right){\rm{d}}{\mathbf{r}}.
	\end{align}
	Accordingly, the observation of user $k$ at point $\mathbf{s}\in{\mathcal{A}}_{k}$ can be written as follows:
	\begin{subequations}
		\begin{align}
			{\mathsf{Y}}_{{\mathsf{dl}},k}(\mathbf{s})&={\mathsf{E}}_{{\mathsf{dl}},k}(\mathbf{s})+{\mathsf{N}}_{{\mathsf{dl}},k}(\mathbf{s})
			\approx{\mathsf{E}}_{{\mathsf{dl}},k}+{\mathsf{N}}_{{\mathsf{dl}},k}(\mathbf{s})\\
			&=s_{{\mathsf{dl}},k}\int_{{\mathcal{A}}}{\mathsf{H}}_k({\mathbf{r}}){\mathsf{J}}_{{\mathsf{dl}},k}({\mathbf{r}}){\rm{d}}{\mathbf{r}}+{\mathsf{N}}_{{\mathsf{dl}},k}(\mathbf{s})
			\nonumber\\
			&+\sum\nolimits_{k'\ne k}s_{{\mathsf{dl}},k'}\int_{{\mathcal{A}}}{\mathsf{H}}_k({\mathbf{r}}){\mathsf{J}}_{{\mathsf{dl}},k'}({\mathbf{r}}){\rm{d}}{\mathbf{r}},  \label{do_Total_electric_radiation_field}  
		\end{align}
	\end{subequations}
	where ${\mathsf{N}}_{{\mathsf{dl}},k}(\mathbf{s})$ denotes the thermal noise subject to ${\mathbbmss{E}}\{{\mathsf{N}}_{{\mathsf{dl}},k}(\mathbf{s}){\mathsf{N}}_{{\mathsf{dl}},k}^{*}({\mathbf{s}}')\}={\sigma}_k^2\delta(\mathbf{s}-{\mathbf{s}}')$. 
	\subsection{Channel Model}
	To facilitate theoretical investigations into fundamental performance limits, we focus our discussion on LoS channels. In this case, ${\mathsf{H}}(\mathbf{r},{\mathbf{s}})$ is modeled as follows \cite{ouyang2024reactive}:
		\begin{align}\label{CAPA_LoS_Channel_Model}
			{\mathsf{H}}(\mathbf{r},{\mathbf{s}})={\mathsf{H}}^{\mathsf{em}}(\mathbf{r},{\mathbf{s}}){\mathsf{H}}^{\mathsf{pa}}(\mathbf{r},{\mathbf{s}}),
		\end{align}
		where ${\mathsf{H}}^{\mathsf{pa}}(\mathbf{r},{\mathbf{s}})\triangleq\sqrt{\frac{\lvert{\mathbf{e}}^{\mathsf{T}}({\mathbf{s}}-{\mathbf{r}})\rvert}{\lVert{\mathbf{r}}-{\mathbf{s}}\rVert}}$ models the impact of the projected aperture of the BS array, and where ${\mathbf{e}}\in{\mathbbmss{R}}^{3\times1}$ is the normal vector of the CAPA at the BS. Note that the projected aperture is reflected by the projection of the normal vector onto the wave propagation direction. Furthermore, the function
		\begin{align}\label{Green_Function_Full_Version}
			{\mathsf{H}}^{\mathsf{em}}(\mathbf{r},{\mathbf{s}})\triangleq\frac{{\rm{j}}k_0\eta{\rm{e}}^{-{\rm{j}}k_0\lVert{\mathbf{r}}-{\mathbf{s}}\rVert}
			}{4\pi \lVert{\mathbf{r}}-{\mathbf{s}}\rVert}
		\end{align}
		models the influence of free-space EM propagation \cite{ouyang2024reactive}, where $\eta=120\pi\,\Omega$ is the impedance of free space, and $k_0=\frac{2\pi}{\lambda}$ with $\lambda$ being the wavelength denotes the wavenumber.
		
		{If we consider a rich-scattering environment, ${\mathsf{H}}(\mathbf{r},{\mathbf{s}})$ can be modeled as a stochastic process \cite{pizzo2020spatially,pizzo2022fourier}. Additionally, the channel model in \eqref{CAPA_LoS_Channel_Model} neglects the impact of MC. If MC is taken into account, we could model ${\mathsf{H}}(\mathbf{r},{\mathbf{s}})$ as follows:
		\begin{align}
			{\mathsf{H}}(\mathbf{r},{\mathbf{s}})=\int_{{\mathcal{A}}}\mathsf{Z}_{\mathsf{mc}}(\mathbf{r},\mathbf{r}'){\mathsf{H}}_{\mathsf{nmc}}(\mathbf{r}',{\mathbf{s}}){\rm{d}}\mathbf{r}',
		\end{align}
		which generalizes the commonly used linear MC model \cite{ivrlavc2010toward}. Here, the operator $\mathsf{Z}_{\mathsf{mc}}(\mathbf{r},\mathbf{r}')$ captures the MC effects within the CAPA, while ${\mathsf{H}}_{\mathsf{nmc}}(\mathbf{r}',{\mathbf{s}})$ represents the spatial channel response without considering MC. Modeling $\mathsf{Z}_{\mathsf{mc}}(\mathbf{r},\mathbf{r}')$ requires circuit-theoretic knowledge \cite{akrout2023super,ivrlavc2010toward}, which is beyond the scope of this paper. A detailed study of MC effects on CAPA-based communications is left for future work, although some initial simulation results are provided in Section \ref{section_numerical}.
		
		For the rest of this paper, we adopt the LoS channel model as an illustrative example. However, it is important to emphasize that our proposed approaches are not constrained to any specific channel type, nor do they rely on assumptions about MC. Since all the derived results are based on continuous operator formulations, they remain applicable in a wide range of channel environments.}
	\section{Uplink Capacity}\label{sec_uplink}
	This section characterizes the channel capacity for CAPA-based uplink communications. We commence with the single-user case, and then proceed to explore the two-user case, while extensions to scenarios with an arbitrary number of users will be addressed in future work.
	{ \subsection{Single-User Case}\label{Section: Single-User Case: Uplink Capacity}
		For $K=1$, we omit the user index $k$ in the subscripts and denote ${\mathcal{A}}_{\mathsf{u}}\subseteq{\mathbbmss{R}}^{3\times1}$, $\mathbf{s}_{\mathsf{u}}\in{\mathbbmss{R}}^{3\times1}$, and $A_{\mathsf{u}}$ as the user's array aperture, its center location, and aperture size, respectively. We use a detector ${\mathsf{V}}_{\mathsf{ul}}(\mathbf{r})$ to recover $s_{{\mathsf{ul}}}$ from ${\mathsf{Y}}_{\mathsf{ul}}(\mathbf{r})$, which yields
		\begin{equation}\label{Single_User_Detection}
			\begin{split}
				\int_{{\mathcal{A}}}{\mathsf{V}}_{\mathsf{ul}}^{*}(\mathbf{r}){\mathsf{Y}}_{\mathsf{ul}}(\mathbf{r}){\rm{d}}{\mathbf{r}}
				&=s_{{\mathsf{ul}}}{\mathsf{J}}_{{\mathsf{ul}}} A_{\mathsf{u}}\int_{{\mathcal{A}}}{\mathsf{V}}_{\mathsf{ul}}^{*}(\mathbf{r}){\mathsf{H}}(\mathbf{r}){\rm{d}}{\mathbf{r}}\\
				&+\int_{{\mathcal{A}}}{\mathsf{V}}_{\mathsf{ul}}^{*}(\mathbf{r}){\mathsf{N}}_\mathsf{ul}(\mathbf{r}){\rm{d}}{\mathbf{r}}.
			\end{split}
		\end{equation}
		Since ${\mathsf{N}}_\mathsf{ul}(\mathbf{r})$ is a zero-mean complex Gaussian random field, $\int_{{\mathcal{A}}}{\mathsf{V}}_{\mathsf{ul}}^{*}(\mathbf{r}){\mathsf{N}}_\mathsf{ul}(\mathbf{r}){\rm{d}}{\mathbf{r}}$ is complex Gaussian distributed. Its mean and variance are calculated as follows.
		\vspace{-5pt}
		\begin{lemma}\label{Lemma_Noise_Distribution}
			Under the assumption that ${\mathsf{N}}_\mathsf{ul}(\mathbf{r})$ is a zero-mean complex Gaussian process with ${\mathbbmss{E}}\{{\mathsf{N}}_\mathsf{ul}(\mathbf{r}){\mathsf{N}}_\mathsf{ul}^{*}({\mathbf{r}}')\}={\sigma}^2\delta(\mathbf{r}-{\mathbf{r}}')$, we have $\int_{{\mathcal{A}}}{\mathsf{V}}_{\mathsf{ul}}^{*}(\mathbf{r}){\mathsf{N}}_\mathsf{ul}(\mathbf{r}){\rm{d}}{\mathbf{r}}\sim{\mathcal{CN}}(0,\sigma^2
			\int_{{\mathcal{A}}}\lvert{\mathsf{V}}_{\mathsf{ul}}(\mathbf{r})\rvert^2{\rm{d}}{\mathbf{r}})$.
		\end{lemma}
		\vspace{-5pt}
		\begin{IEEEproof}
			Please refer to Appendix \ref{Proof_Lemma_Noise_Distribution} for more details.
		\end{IEEEproof}
		The signal-to-noise ratio (SNR) for decoding $s_{{\mathsf{ul}}}$ reads
		\begin{equation}\label{Single_User_SNR}
			\gamma_\mathsf{ul}=
			\frac{A_{\mathsf{u}}^2\lvert {\mathsf{J}}_{{\mathsf{ul}}}\rvert^2
				\lvert\int_{{\mathcal{A}}}{\mathsf{V}}_{\mathsf{ul}}^{*}(\mathbf{r}){\mathsf{H}}(\mathbf{r}){\rm{d}}{\mathbf{r}}\rvert^2}
			{\sigma^2
				\int_{{\mathcal{A}}}\lvert{\mathsf{V}}_{\mathsf{ul}}(\mathbf{r})\rvert^2{\rm{d}}{\mathbf{r}}}.
		\end{equation}
		It can be observed from \eqref{Single_User_SNR} that the optimal detector satisfies
		\begin{align}\label{Optimal_Detector_MISO_SU_Answer}
			{\mathsf{V}}_{\mathsf{ul}}^{\star}(\mathbf{r})&\propto\argmax_{\int_{{\mathcal{A}}}\lvert{\mathsf{V}}_{\mathsf{ul}}(\mathbf{r})\rvert^2{\rm{d}}{\mathbf{r}}=1}{\left\lvert
				\int_{{\mathcal{A}}}{\mathsf{V}}_{\mathsf{ul}}^{*}(\mathbf{r}){\mathsf{H}}(\mathbf{r}){\rm{d}}{\mathbf{r}}
				\right\rvert^2}=\frac{{\mathsf{H}}({\mathbf{r}})}{{\mathsf{h}}_{\mathsf{u}}^{{1}/{2}}},
		\end{align}
		where ${\mathsf{h}}_{\mathsf{u}}\triangleq \int_{\mathcal{A}}\lvert{\mathsf{H}}({\mathbf{r}})\rvert^2{\rm{d}}{\mathbf{r}}$.
		\vspace{-5pt}
		\begin{remark}
			The results in \eqref{Optimal_Detector_MISO_SU_Answer} suggest that the capacity-achieving detector for uplink single-user CAPA communications aligns with the spatial channel response ${\mathsf{H}}(\mathbf{r})$, which is the continuous version of \emph{maximal-ratio combining (MRC)}.
		\end{remark}
		\vspace{-5pt}
		Substituting \eqref{Optimal_Detector_MISO_SU_Answer} into \eqref{Single_User_SNR} gives
		\begin{equation}\label{Single_User_SNR_2}
			\gamma_\mathsf{ul}=\frac{A_{\mathsf{u}}^2\lvert {\mathsf{J}}_{{\mathsf{ul}}}\rvert^2}{\sigma^2}
			\int_{\mathcal{A}}\lvert{\mathsf{H}}({\mathbf{r}})\rvert^2{\rm{d}}{\mathbf{r}}
			=\overline{\gamma}_\mathsf{ul}
			\int_{{\mathcal{A}}}\lvert{\mathsf{G}}(\mathbf{r})\rvert^2{\rm{d}}{\mathbf{r}},
		\end{equation}
		where $\overline{\gamma}_\mathsf{ul}=\frac{A_{\mathsf{u}}^2\lvert {\mathsf{J}}_{{\mathsf{ul}}}\rvert^2k_0^2\eta^2}{4\pi\sigma^2}$ and ${\mathsf{G}}(\mathbf{r})=\frac{{\rm{e}}^{-{\rm{j}}k_0\lVert{\mathbf{r}}-{\mathbf{s}}_{\mathsf{u}}\rVert}}{\sqrt{4\pi} \lVert{\mathbf{r}}-{\mathbf{s}}_{\mathsf{u}}\rVert}
		\sqrt{\frac{\lvert{\mathbf{e}}^{\mathsf{T}}({\mathbf{s}}_{\mathsf{u}}-{\mathbf{r}})\rvert}{\lVert{\mathbf{r}}-{\mathbf{s}}_{\mathsf{u}}\rVert}}$. Note that $\frac{{\rm{e}}^{-{\rm{j}}k_0\lVert{\mathbf{r}}-{\mathbf{s}}_{\mathsf{u}}\rVert}}{\sqrt{4\pi} \lVert{\mathbf{r}}-{\mathbf{s}}_{\mathsf{u}}\rVert}$ and $\sqrt{\frac{\lvert{\mathbf{e}}^{\mathsf{T}}({\mathbf{s}}_{\mathsf{u}}-{\mathbf{r}})\rvert}{\lVert{\mathbf{r}}-{\mathbf{s}}_{\mathsf{u}}\rVert}}$ characterize the impact of spherical-wave propagation and projected aperture in arbitrary homogeneous mediums, respectively. Thus, we define $\int_{{\mathcal{A}}}\lvert{\mathsf{G}}(\mathbf{r})\rvert^2{\rm{d}}{\mathbf{r}}\triangleq {\mathsf{g}}_{\mathsf{u}}$ as the channel gain. The remaining term $\overline{\gamma}_\mathsf{ul}$ can be treated as the uplink transmit SNR. As a result, building upon \eqref{Single_User_SNR_2}, the single-user uplink channel capacity can be written as follows:
		\begin{align}\label{SU_Channel_Capacity}
			{\mathsf{C}}_{\mathsf{ul}}=\log_2(1+\gamma_{\mathsf{ul}})=\log_2(1+{\overline{\gamma}_{\mathsf{ul}}}{\mathsf{g}}_{\mathsf{u}}).
	\end{align}}

	{ \subsection{Two-User Case}}
	We now extend the single-user scenario to the two-user one.
	\subsubsection{Optimal Uplink Detection}
	When $K=2$, the observed electric field in \eqref{Total_electric_radiation_field} is given by
	\begin{equation}\label{Two_User_Total_electric_radiation_field}
		{\mathsf{Y}}_{\mathsf{ul}}(\mathbf{r})
		=\mathsf{H}_1(\mathbf{r}){\mathsf{J}}_{{\mathsf{ul}},1} A_{{\mathsf{u}},1} s_{{\mathsf{ul}},1}
		+\mathsf{H}_2(\mathbf{r}){\mathsf{J}}_{{\mathsf{ul}},2} A_{{\mathsf{u}},2} s_{{\mathsf{ul}},2}
		+{\mathsf{N}}_\mathsf{ul}(\mathbf{r}).
	\end{equation}
	The capacity of an uplink multiuser channel can be achieved using \emph{successive interference cancellation (SIC) decoding} \cite{tse2005fundamentals}. Specifically, the message sent by one user is first decoded by treating the message from the other user as interference. Once decoded, this interference is removed, and the other message is then decoded without inter-user interference (IUI).
	
	Given the uplink signal model in \eqref{Two_User_Total_electric_radiation_field}, there are two possible SIC orders: $1\rightarrow2$ and $2\rightarrow1$. Here, we discuss the channel capacity achieved by the decoding order $2\rightarrow1$. In this case, $s_{{\mathsf{ul}},2}$ is decoded from ${\mathsf{Y}}_{\mathsf{ul}}(\mathbf{r})$ by treating $\mathsf{H}_1(\mathbf{r}){\mathsf{J}}_{{\mathsf{ul}},1} A_{{\mathsf{u}},1} s_{{\mathsf{ul}},1}$ as interference. After $s_{{\mathsf{ul}},2}$ is decoded and subtracted from ${\mathsf{Y}}_{\mathsf{ul}}(\mathbf{r})$, $s_{{\mathsf{ul}},1}$ is decoded without IUI from the following signal:
	\begin{align}\label{Two_User_SIC_electric_radiation_field}
		{\mathsf{Z}}(\mathbf{r})\triangleq \mathsf{H}_1(\mathbf{r}){\mathsf{J}}_{{\mathsf{ul}},1} A_{{\mathsf{u}},1} s_{{\mathsf{ul}},1}
		+{\mathsf{N}}_\mathsf{ul}(\mathbf{r}),
	\end{align}
	Following the derivation steps to obtain \eqref{Optimal_Detector_MISO_SU_Answer}, the BS can use the detector ${\mathsf{V}}_{{\mathsf{ul}},1}(\mathbf{r})=\frac{{\mathsf{H}_1}({\mathbf{r}})}
	{\sqrt{\int_{{\mathcal{A}}}\lvert{\mathsf{H}}_1({\mathbf{r}})\rvert^2{\rm{d}}{\mathbf{r}}}}$ to recover $s_{{\mathsf{ul}},1}$ from \eqref{Two_User_SIC_electric_radiation_field}. Therefore, the SNR for decoding the message sent by user $1$ is given by
	\begin{align}
		\gamma_{\mathsf{ul},1}=\overline{\gamma}_{\mathsf{ul},1}{\mathsf{g}}_{1},
	\end{align}
	where $\overline{\gamma}_{\mathsf{ul},k}=\frac{A_{\mathsf{u},k}^2\lvert {\mathsf{J}}_{{\mathsf{ul}},k}\rvert^2k_0^2\eta^2}{4\pi\sigma^2}$ denotes the transmit SNR of user $k\in\{1,2\}$, and ${\mathsf{g}}_{k}=\int_{{\mathcal{A}}}\lvert{\mathsf{G}}_k(\mathbf{r})\rvert^2{\rm{d}}{\mathbf{r}}$ denotes the associated channel gain with ${\mathsf{G}}_k(\mathbf{r})=\frac{{\rm{e}}^{-{\rm{j}}k_0\lVert{\mathbf{r}}-{\mathbf{s}}_{k}\rVert}}{\sqrt{4\pi} \lVert{\mathbf{r}}-{\mathbf{s}}_{k}\rVert}
	\sqrt{\frac{\lvert{\mathbf{e}}^{\mathsf{T}}({\mathbf{s}}_{k}-{\mathbf{r}})\rvert}{\lVert{\mathbf{r}}-{\mathbf{s}}_{k}\rVert}}$. 
	
	Next, we investigate the decoding of $s_{{\mathsf{ul}},2}$ from ${\mathsf{Y}}_{\mathsf{ul}}(\mathbf{r})$, where the interference-plus-noise term is given by ${\mathsf{Z}}(\mathbf{r})$ as shown in \eqref{Two_User_SIC_electric_radiation_field}. Since ${\mathsf{N}}_\mathsf{ul}(\mathbf{r})$ and $s_{{\mathsf{ul}},1}$ are uncorrelated, the autocorrelation function of the random field ${\mathsf{Z}}(\mathbf{r})$ is given by
	\begin{align}\label{Autocorrelatioon_Interference_Noise}
		\begin{split}
			{\mathbbmss{E}}\{{\mathsf{Z}}(\mathbf{r}){\mathsf{Z}}^*(\mathbf{r}')\}&=
			{\mathsf{G}}_1(\mathbf{r}){\mathsf{G}}_1^*(\mathbf{r}')
			\overline{\gamma}_{\mathsf{ul},1}{\sigma}^2\\
			&+{\sigma}^2\delta(\mathbf{r}-{\mathbf{r}}')\triangleq {\mathsf{R}}_{{\mathsf{Z}}}(\mathbf{r},{\mathbf{r}}').
		\end{split}
	\end{align}
	To obtain the optimal detector that maximizes the achievable rate of user $2$, we first need to design an invertible linear transformation ${\mathsf{W}}_{{\mathsf{Z}}}(\mathbf{r}',\mathbf{r})$ to whiten ${\mathsf{Z}}(\mathbf{r})$. This transformation ${\mathsf{W}}_{{\mathsf{Z}}}(\mathbf{r}',\mathbf{r})$ must satisfy the following two conditions:
	\begin{enumerate}[i]
		\item \emph{Whitening:} The autocorrelation function of ${\mathsf{Z}}_{\mathsf{w}}(\mathbf{r}')\triangleq \int_{{\mathcal{A}}}{\mathsf{W}}_{{\mathsf{Z}}}(\mathbf{r}',\mathbf{r}){\mathsf{Z}}(\mathbf{r}){\rm{d}}\mathbf{r}$ aligns with the delta function, i.e.,
		\begin{align}
			{\mathbbmss{E}}\{{\mathsf{Z}}_{\mathsf{w}}(\mathbf{r}){\mathsf{Z}}_{\mathsf{w}}^*(\mathbf{r}')\}\propto\delta(\mathbf{r}'-\mathbf{r});
		\end{align}
		\item \emph{Invertibility:} Given an arbitrary function $f(\mathbf{r})$ defined on $\mathbf{r}\in{\mathcal{A}}$, there exists a transformation $\overline{\mathsf{W}}_{{\mathsf{Z}}}(\mathbf{x},\mathbf{r}')$ such that 
		\begin{align}
			\int_{{\mathcal{A}}}\overline{\mathsf{W}}_{{\mathsf{Z}}}(\mathbf{x},\mathbf{r}')\int_{{\mathcal{A}}}{\mathsf{W}}_{{\mathsf{Z}}}(\mathbf{r}',\mathbf{r})f(\mathbf{r})
			{\rm{d}}\mathbf{r}{\rm{d}}\mathbf{r}'=f(\mathbf{x}).
		\end{align}
	\end{enumerate}
	Applying an invertible transformation to ${\mathsf{Y}}_{\mathsf{ul}}(\mathbf{r})$ is information lossless and has no impact on the channel capacity \cite{tse2005fundamentals}. To proceed, we introduce the following lemmas.
	\vspace{-5pt}
	\begin{lemma}\label{Lemma_IN_Whiten}
		The invertible linear transformation used to whiten the interference-plus-noise term ${\mathsf{Z}}(\mathbf{r})$ is given by
		\begin{align}\label{whiten_transform}
			{\mathsf{W}}_{{\mathsf{Z}}}(\mathbf{r}',\mathbf{r})=\delta(\mathbf{r}'-\mathbf{r})+\mu_1
			{\mathsf{G}}_1(\mathbf{r}'){\mathsf{G}}_1^*(\mathbf{r}),    
		\end{align}
		where $\mu_1=-\frac{1}{{\mathsf{g}}_{1}}\pm\frac{1}{{\mathsf{g}}_{1}
			\sqrt{1+\overline{\gamma}_{\mathsf{u},1}{\mathsf{g}}_{1}}}$. The autocorrelation function of ${\mathsf{Z}}_{\mathsf{w}}(\cdot)$ satisfies ${\mathbbmss{E}}\{{\mathsf{Z}}_{\mathsf{w}}(\mathbf{r}){\mathsf{Z}}_{\mathsf{w}}^*(\mathbf{r}')\}=
		{\sigma}^2\delta(\mathbf{r}-\mathbf{r}')$. Additionally, the inversion of ${\mathsf{W}}_{{\mathsf{Z}}}(\mathbf{r}',\mathbf{r})$ is given by $\overline{\mathsf{W}}_{{\mathsf{Z}}}(\mathbf{r},\mathbf{r}')=\delta(\mathbf{r}-\mathbf{r}')-\frac{\mu_1}{1+\mu_1 {\mathsf{g}}_{1}}
		{\mathsf{G}}_1(\mathbf{r}){\mathsf{G}}_1^*(\mathbf{r}')$.
	\end{lemma}
	\vspace{-5pt}
	\begin{IEEEproof}
		Please refer to Appendix \ref{Proof_Lemma_IN_Whiten} for more details.
	\end{IEEEproof}
    Next, we exploit ${\mathsf{W}}_{{\mathsf{Z}}}(\mathbf{r}',\mathbf{r})$ to transform ${\mathsf{Y}}_{\mathsf{ul}}(\mathbf{r})$ as follows:
	\begin{equation}\label{MMSE_Filtering_Signal}
		\int_{{\mathcal{A}}}{\mathsf{W}}_{{\mathsf{Z}}}(\mathbf{r}',\mathbf{r}){\mathsf{Y}}_{\mathsf{ul}}(\mathbf{r}){\rm{d}}\mathbf{r}
		=\overline{\mathsf{H}}_2(\mathbf{r}'){\mathsf{J}}_{{\mathsf{ul}},2} A_{{\mathsf{u}},2} s_{{\mathsf{ul}},2}
		+{\mathsf{Z}}_{\mathsf{w}}(\mathbf{r}'),
	\end{equation}
	where $\overline{\mathsf{H}}_2(\mathbf{r}')\triangleq\int_{{\mathcal{A}}}{\mathsf{W}}_{{\mathsf{Z}}}(\mathbf{r}',\mathbf{r}){\mathsf{H}}_2(\mathbf{r}){\rm{d}}\mathbf{r}$. By noting that ${\mathbbmss{E}}\{{\mathsf{Z}}_{\mathsf{w}}(\mathbf{r}){\mathsf{Z}}_{\mathsf{w}}^*(\mathbf{r}')\}=
	{\sigma}^2\delta(\mathbf{r}-\mathbf{r}')$, the model presented in \eqref{MMSE_Filtering_Signal} is similar to that in the single-user case. Motivated by this, we design an MRC-based detector $\mathsf{V}_{\mathsf{ul},2}(\mathbf{r}')$ by treating $\overline{\mathsf{H}}_2(\mathbf{r}')$ as the equivalent channel response, which yields
	\begin{equation}
		{\mathsf{V}}_{\mathsf{ul},2}(\mathbf{r}')=\frac{{\overline{\mathsf{H}}_2}({\mathbf{r}'})}
		{\sqrt{\int_{{\mathcal{A}}}\lvert\overline{\mathsf{H}}_2({\mathbf{r}'})\rvert^2{\rm{d}}{\mathbf{r}}}}.
	\end{equation}
	Following the derivation steps to obtain \eqref{Single_User_SNR_2}, the SNR for decoding $s_{{\mathsf{ul}},2}$ can be written as follows:
	\begin{equation}\label{Two_User_MMSE_SNR}
		\begin{split}
			\gamma_{\mathsf{ul},2}=
			\frac{A_{{\mathsf{u}},2}^2\lvert {\mathsf{J}}_{{\mathsf{ul}},2}\rvert^2}{\sigma^2}
			\int_{{\mathcal{A}}}\lvert\overline{\mathsf{H}}_2(\mathbf{r}')\rvert^2{\rm{d}}{\mathbf{r}}'.
		\end{split}
	\end{equation}
	A closed-form expression for $\gamma_{\mathsf{ul},2}$ is given as follows.
	\vspace{-5pt}
	\begin{theorem}\label{Theorem_MMSE_SNR_First_User}
		By first whitening ${\mathsf{Z}}(\mathbf{r})$ with ${\mathsf{W}}_{{\mathsf{Z}}}(\mathbf{r}',\mathbf{r})$ and then employing the MRC detector ${\mathsf{V}}_{\mathsf{ul},2}(\mathbf{r}')$, the resultant SNR in decoding $x_{\mathsf{u},2}$ can be written as follows:
		\begin{equation}\label{Two_User_MMSE_SNR_Closed_Form}
			\begin{split}
				\gamma_{\mathsf{ul},2}=\overline{\gamma}_{\mathsf{ul},2}\mathsf{g}_2( 1-{\overline{\gamma}_{\mathsf{ul},1}\mathsf{g}_1\lvert \rho \rvert^2}({1+\overline{\gamma}_{\mathsf{ul},1}\mathsf{g}_1})^{-1}),
			\end{split}
		\end{equation}
		where $\rho\triangleq\frac{\int_{{\mathcal{A}}}{\mathsf{G}}_1^*({\mathbf{r}}){\mathsf{G}}_2({\mathbf{r}}){\rm{d}}{\mathbf{r}}}{\sqrt{{\mathsf{g}}_1{\mathsf{g}}_2}}$ is defined as the channel correlation factor with $\lvert\rho\rvert\in[0,1]$.  
	\end{theorem}
	\vspace{-5pt}
	\begin{IEEEproof}
		Please refer to Appendix \ref{Proof_Theorem_MMSE_SNR_First_User} for more details.
	\end{IEEEproof}
	After obtaining the capacity-achieving detectors under the SIC order $2\rightarrow1$, we summarize the entire decoding procedure in Table \ref{SIC_Decoding_Two_User} for ease of reference. The presented decoding procedure can be directly extended to the other SIC order $1\rightarrow2$ by exchanging the user indices.
	
	\begin{table}[!t]
		\centering
		\setlength{\abovecaptionskip}{3pt}
		\begin{tabular}{|c|}
			\hline
			\begin{minipage}{3in}
				\vspace{1pt}
				\begin{algorithmic}[1]
					\State {\footnotesize Whiten the interference-plus-noise term ${\mathsf{Z}}(\mathbf{r})$: 
						{\setlength\abovedisplayskip{2pt}
							\setlength\belowdisplayskip{2pt}
							\begin{equation}
								\int_{{\mathcal{A}}}{\mathsf{W}}_{{\mathsf{Z}}}(\mathbf{r}',\mathbf{r}){\mathsf{Y}}_{\mathsf{ul}}(\mathbf{r}){\rm{d}}\mathbf{r}
								=\overline{\mathsf{H}}_2(\mathbf{r}'){\mathsf{J}}_{{\mathsf{ul}},2} A_{{\mathsf{u}},2} s_{{\mathsf{ul}},2}
								+{\mathsf{Z}}_{\mathsf{w}}(\mathbf{r}');\nonumber
							\end{equation}
					}}
					\State {\footnotesize Use the MRC detector and ML decoder to recover $s_{{\mathsf{ul}},2}$:
						{\setlength\abovedisplayskip{2pt}
							\setlength\belowdisplayskip{2pt}
							\begin{equation}
								\int_{{\mathcal{A}}}{\mathsf{V}}_{\mathsf{ul},2}^{*}(\mathbf{r}')\int_{{\mathcal{A}}}{\mathsf{W}}_{{\mathsf{Z}}}(\mathbf{r}',\mathbf{r}){\mathsf{Y}}_{\mathsf{ul}}(\mathbf{r}){\rm{d}}\mathbf{r}{\rm{d}}\mathbf{r}'
								\rightarrow s_{{\mathsf{ul}},2}; \nonumber
							\end{equation}
					}}
					\State {\footnotesize Employ SIC to subtract $\mathsf{H}_2(\mathbf{r}){\mathsf{J}}_{{\mathsf{ul}},2} A_{{\mathsf{u}},2} s_{{\mathsf{ul}},2}$ from ${\mathsf{Y}}_{\mathsf{ul}}(\mathbf{r})$;}
					\State {\footnotesize Use the MRC detector and ML decoder to recover $s_{{\mathsf{ul}},1}$:
						{\setlength\abovedisplayskip{2pt}
							\setlength\belowdisplayskip{2pt}
							\begin{equation}
								\int_{{\mathcal{A}}}{\mathsf{V}}_{\mathsf{ul},1}^{*}(\mathbf{r})
								(\mathsf{H}_1(\mathbf{r}){\mathsf{J}}_{{\mathsf{ul}},1} A_{{\mathsf{u}},1} s_{{\mathsf{ul}},1}
								+{\mathsf{N}}_\mathsf{ul}(\mathbf{r})){\rm{d}}\mathbf{r}
								\rightarrow s_{{\mathsf{ul}},1}. \nonumber
							\end{equation}
					}}
				\end{algorithmic}
				\vspace{0pt}
			\end{minipage}
			\\ \hline
		\end{tabular}
		\caption{SIC decoding for uplink CAPA communications.}
		\label{SIC_Decoding_Two_User}
		\vspace{-7pt}
	\end{table}
	\subsubsection{Capacity Characterization}
	After obtaining the decoding SNR of each $s_{{\mathsf{ul}},k}$ for $k=1,2$, the uplink achievable rate of user $k$ can be calculated as ${\mathsf{R}}_{\mathsf{ul},k}^{2\rightarrow1}=\log_2(1+\gamma_{\mathsf{ul},k})$, i.e.,
	\begin{align}
		{\mathsf{R}}_{\mathsf{ul},1}^{2\rightarrow1}&=\log_2(1+\overline{\gamma}_{\mathsf{ul},1}{\mathsf{g}}_{1}),\label{User1_Rate}\\
		{\mathsf{R}}_{\mathsf{ul},2}^{2\rightarrow1}&=\log_2\bigg(1+\overline{\gamma}_{\mathsf{ul},2}\mathsf{g}_2\bigg( 1-\frac{\overline{\gamma}_{\mathsf{ul},1}\mathsf{g}_1\lvert \rho \rvert^2}{{1+\overline{\gamma}_{\mathsf{ul},1}\mathsf{g}_1}}\bigg)\bigg).\label{User2_Rate}
	\end{align}
	Accordingly, the sum-rate capacity is calculated as follows:
	\begin{equation}\label{Sum_Rate_Capacity}
		\begin{split}
			{\mathsf{C}}_{\mathsf{ul}}^{2\rightarrow1}&={\mathsf{R}}_{\mathsf{ul},1}^{2\rightarrow1}+{\mathsf{R}}_{\mathsf{ul},2}^{2\rightarrow1}\\
			&=\log_2\Big(1+\overline{\gamma}_{\mathsf{ul},1}\overline{\gamma}_{\mathsf{ul},2}{\mathsf{g}}_{1}
			{\mathsf{g}}_{2}\overline{\rho}+\sum\nolimits_{k=1}^{2}\overline{\gamma}_{\mathsf{ul},k}{\mathsf{g}}_{k}\Big),
		\end{split}
	\end{equation}
	where $\overline{\rho}=1-\lvert{\rho}\rvert^2\in[0,1]$. Following the same derivation steps used to obtain \eqref{Sum_Rate_Capacity}, we can calculate the sum-rate capacity achieved under the SIC order $1\rightarrow2$ as follows.
	\vspace{-5pt}
	\begin{corollary}\label{Corollary_Sum_Rate_Capacity}
		Under the SIC order $1\rightarrow2$, the achievable rates of user 1 and user 2, as well as the sum-rate capacity, can be expressed as follows:
		\begin{align}
			&{\mathsf{R}}_{\mathsf{ul},1}^{1\rightarrow2}=\log_2\bigg(1+\overline{\gamma}_{\mathsf{ul},1}\mathsf{g}_1\bigg( 1-\frac{\overline{\gamma}_{\mathsf{ul},2}\mathsf{g}_2\lvert \rho \rvert^2}{1+\overline{\gamma}_{\mathsf{ul},2}\mathsf{g}_2} \bigg)\bigg),\\
			&{\mathsf{R}}_{\mathsf{ul},2}^{1\rightarrow2}=\log_2(1+\overline{\gamma}_{\mathsf{ul},2}{\mathsf{g}}_{2}),\\
			&{\mathsf{C}}_{\mathsf{ul}}^{1\rightarrow2}=\log_2\Big(1+\overline{\gamma}_{\mathsf{ul},1}\overline{\gamma}_{\mathsf{ul},2}{\mathsf{g}}_{1}
			{\mathsf{g}}_{2}\overline{\rho}+\sum\nolimits_{k=1}^{2}\overline{\gamma}_{\mathsf{ul},k}{\mathsf{g}}_{k}\Big).\label{Sum_Rate_Capacity_1_2}
		\end{align}
	\end{corollary}
	\vspace{-5pt}
	Comparing \eqref{Sum_Rate_Capacity} with \eqref{Sum_Rate_Capacity_1_2} yields the following conclusion.
	\vspace{-5pt}
	\begin{theorem}
		The sum-rate capacity of the considered uplink CAPA-based channel is always the same, i.e.,
		\begin{align}
			{\mathsf{C}}_{\mathsf{ul}}=\log_2\Big(1+\overline{\gamma}_{\mathsf{ul},1}\overline{\gamma}_{\mathsf{ul},2}{\mathsf{g}}_{1}
			{\mathsf{g}}_{2}\overline{\rho}+\sum\nolimits_{k=1}^{2}\overline{\gamma}_{\mathsf{ul},k}{\mathsf{g}}_{k}\Big),\label{Sum_Rate_Capacity_Exact}
		\end{align}
		regardless of the decoding order.
	\end{theorem}
	\vspace{-5pt}
	\vspace{-5pt}
	\begin{remark}\label{remark_uplink sum-rate capacity}
		The sum-rate capacity expression given in \eqref{Sum_Rate_Capacity_Exact} is determined by the \emph{channel gain} of each user and the \emph{channel correlation factor}. This expression is applicable to an arbitrary aperture \emph{regardless of its location, shape, and size}. Furthermore, although only the LoS channel model is considered, the above derivations can be directly extended to other channel types, as they are \emph{not channel-specific}.
	\end{remark}
	\vspace{-5pt}
	Having obtained the sum-rate capacity, we now study the capacity region. The capacity region comprises all achievable rate pairs $({\mathsf{R}}_{1},{\mathsf{R}}_{2})$, where ${\mathsf{R}}_{k}$ denotes the achievable rate of user $k\in\{1,2\}$. The capacity region is defined by the rate pairs such that \cite{tse2005fundamentals}
	\begin{align}
		&{\mathsf{R}}_{1}\leq\log_2(1+\overline{\gamma}_{\mathsf{ul},1}{\mathsf{g}}_{1}),{\mathsf{R}}_{2}\leq\log_2(1+\overline{\gamma}_{\mathsf{ul},2}{\mathsf{g}}_{2}),\\
		&{\mathsf{R}}_{1}+{\mathsf{R}}_{2}\leq \log_2\Big(1\!+\!\overline{\gamma}_{\mathsf{ul},1}\overline{\gamma}_{\mathsf{ul},2}{\mathsf{g}}_{1}
		{\mathsf{g}}_{2}\overline{\rho}\!+\!\sum\nolimits_{k=1}^{2}\!\overline{\gamma}_{\mathsf{ul},k}{\mathsf{g}}_{k}\Big).
	\end{align}
	This region can be achieved through SIC decoding along with time sharing, which forms a pentagon.
	{\subsubsection{Sum-Rate Achieved by Zero-Forcing Detector}
		To attain the uplink sum-rate capacity in \eqref{Sum_Rate_Capacity_Exact}, a non-linear detector based on SIC is necessary. However, in practical systems, the linear ZF detector is frequently favored to lower decoding complexity. In the subsequent analysis, we aim to evaluate the sum-rate performance of the ZF detector and contrast it with the sum-rate capacity achieved by the non-linear detector.
		
		Let $\mathsf{V}_{\mathsf{zf},k}(\mathbf{r})$ denote the ZF detector for user $k$ with $k\in\{1,2\}$. In order to nullify the IUI, i.e., $\int_{{\mathcal{A}}}\mathsf{V}_{\mathsf{zf},k}^*(\mathbf{r})\mathsf{H}_{k'}(\mathbf{r}){\rm{d}}{\mathbf{r}}=0$ for $k'\ne k$, and maximize the channel gain of user $k$, $\mathsf{V}_{\mathsf{zf},k}(\mathbf{r})$ is designed as the projection of $\mathsf{H}_{k}(\mathbf{r})$ onto the space orthogonal to $\mathsf{H}_{k'}(\mathbf{r})$, which is given by
		\begin{equation}\label{v_zf}
			\mathsf{V}_{\mathsf{zf},k}(\mathbf{r})=\mathsf{H}_k(\mathbf{r})-\frac{\int_{\mathcal{A}}{\mathsf{H}_k}(\mathbf{r})\mathsf{H}_{k^{\prime}}^{*}(\mathbf{r})\mathrm{d}\mathbf{r}}{\int_{\mathcal{A}}{\left| \mathsf{H}_{k^{\prime}}(\mathbf{r}) \right|^2}\mathrm{d}\mathbf{r}}\mathsf{H}_{k^{\prime}}(\mathbf{r}).
		\end{equation}
		Substituting \eqref{v_zf} into $\mathsf{R}_{\mathsf{ul},k}^{\mathsf{zf}}=\log _2( 1+\gamma _{\mathsf{ul},k}^{\mathsf{zf}} ) $ with
		\begin{align}
			\resizebox{1\hsize}{!}{$\gamma _{\mathsf{ul},k}^{\mathsf{zf}}\!=\!\frac{A_{\mathsf{u},k}^{2}| \mathsf{J}_{\mathsf{ul},k} |^2| \int_{\mathcal{A}}{\mathsf{V}_{\mathsf{zf},k}^{*}}(\mathbf{r})\mathsf{H}_k(\mathbf{r})\mathrm{d}\mathbf{r} |^2}{\sigma ^2\!\int_{\mathcal{A}}\!{| \mathsf{V}_{\mathsf{zf},k}(\mathbf{r}) |^2}\mathrm{d}\mathbf{r}+A_{\mathsf{u},k^{\prime}}^{2}\!| \mathsf{J}_{\mathsf{ul},k^{\prime}} |^2| \int_{\mathcal{A}}\!{\mathsf{V}_{\mathsf{zf},k}^{*}}(\mathbf{r})\mathsf{H}_{k^{\prime}}(\mathbf{r})\mathrm{d}\mathbf{r}|^2}$} \nonumber
		\end{align}
		and performing some simple manipulations, we can obtain the sum-rate achieved by the ZF detector as follows:
		\begin{equation}
			\mathsf{R}_{\mathsf{ul}}^{\mathsf{zf}}=\sum\nolimits_{k=1}^2{\log _2\Big( 1+\overline{\gamma }_{\mathsf{ul},k}\mathsf{g}_k\big( 1-\left| \rho \right|^2 \big) \Big)}.	
		\end{equation}
		From the results, it can be observed that $\mathsf{R}_{\mathsf{ul}}^{\mathsf{zf}}\leq\mathsf{C}_{\mathsf{ul}}$, and the gap between them is further illustrated in the numerical results presented in Section~\ref{section_numerical}.}
	\section{Downlink Capacity}\label{Section: downlink}
	In this section, we characterize the channel capacity for CAPA-based downlink communications.
		{\subsection{Single-User Case}
		For $K=1$, \eqref{do_Total_electric_radiation_field} simplifies to
		\begin{align}\label{do_Single_User_electric_radiation_field}
			{\mathsf{Y}}_{{\mathsf{dl}}}(\mathbf{s})=s_{{\mathsf{dl}}}\int_{{\mathcal{A}}}{\mathsf{H}}({\mathbf{r}}){\mathsf{J}}_{{\mathsf{dl}}}({\mathbf{r}}){\rm{d}}{\mathbf{r}}
			+{\mathsf{N}}_{{\mathsf{dl}}}(\mathbf{s}),
		\end{align}
		where ${\mathbbmss{E}}\{{\mathsf{N}}_{{\mathsf{dl}}}(\mathbf{s}){\mathsf{N}}_{{\mathsf{dl}}}^{*}({\mathbf{s}}')\}={\sigma}_{\mathsf{u}}^2\delta(\mathbf{s}-{\mathbf{s}}')$. Applying a detector ${\mathsf{V}}_{{\mathsf{dl}}}(\mathbf{s})$ to ${\mathsf{Y}}_{{\mathsf{dl}}}(\mathbf{s})$ yields
		\begin{equation}\label{do_Single_User_Detection}
			\begin{split}
				\int_{{\mathcal{A}}_{\mathsf{u}}}{\mathsf{V}}_{{\mathsf{dl}}}^*(\mathbf{s}){\mathsf{Y}}_{{\mathsf{dl}}}(\mathbf{s}){\rm{d}}{\mathbf{s}}
				&=s_{{\mathsf{dl}}}\int_{{\mathcal{A}}}{\mathsf{H}}({\mathbf{r}}){\mathsf{J}}_{{\mathsf{dl}}}({\mathbf{r}}){\rm{d}}{\mathbf{r}}
				\int_{{\mathcal{A}}_{\mathsf{u}}}{\mathsf{V}}_{{\mathsf{dl}}}^*(\mathbf{s}){\rm{d}}{\mathbf{s}}\\
				&+\int_{{\mathcal{A}}_{\mathsf{u}}}{\mathsf{V}}_{{\mathsf{dl}}}^*(\mathbf{s}){\mathsf{N}}_{{\mathsf{dl}}}(\mathbf{s}){\rm{d}}{\mathbf{s}}.
			\end{split}
		\end{equation}
		Similar to \textbf{Lemma \ref{Lemma_Noise_Distribution}}, we have $\int_{{\mathcal{A}}_{\mathsf{u}}}{\mathsf{V}}_{{\mathsf{dl}}}^*(\mathbf{s}){\mathsf{N}}_{{\mathsf{dl}}}(\mathbf{s}){\rm{d}}{\mathbf{s}}\sim{\mathcal{CN}}(0,{\sigma}_{\mathsf{u}}^2
		\int_{{\mathcal{A}}_{\mathsf{u}}}\lvert{\mathsf{V}}_{{\mathsf{dl}}}(\mathbf{s})\rvert^2{\rm{d}}{\mathbf{s}})$. As a result, the SNR for decoding $s_{{\mathsf{dl}}}$ is given by
		\begin{align}\label{do_Single_User_SNR_Ini1}
			\gamma_{\mathsf{dl}}=
			\frac{\lvert\int_{{\mathcal{A}}_{\mathsf{u}}}{\mathsf{V}}_{{\mathsf{dl}}}^*(\mathbf{s}){\rm{d}}{\mathbf{s}}\rvert^2
				\lvert\int_{{\mathcal{A}}}{\mathsf{H}}({\mathbf{r}}){\mathsf{J}}_{{\mathsf{dl}}}({\mathbf{r}}){\rm{d}}{\mathbf{r}}\rvert^2}
			{{\sigma}_{\mathsf{u}}^2
				\int_{{\mathcal{A}}_{\mathsf{u}}}\lvert{\mathsf{V}}_{{\mathsf{dl}}}(\mathbf{s})\rvert^2{\rm{d}}{\mathbf{s}}}.  
		\end{align}
		Recalling that the variations of the signal along ${{\mathcal{A}}_{\mathsf{u}}}$ are negligible, we have $\int_{{\mathcal{A}}_{\mathsf{u}}}{\mathsf{V}}_{{\mathsf{dl}}}^*(\mathbf{s}){\rm{d}}{\mathbf{s}}\approx {\mathsf{V}}_{{\mathsf{dl}}}^*(\mathbf{s}_{\mathsf{u}})A_{\mathsf{u}}$ and $\int_{{\mathcal{A}}_{\mathsf{u}}}\lvert{\mathsf{V}}_{{\mathsf{dl}}}(\mathbf{s})\rvert^2{\rm{d}}{\mathbf{s}}\approx 
		\lvert{\mathsf{V}}_{{\mathsf{dl}}}(\mathbf{s}_{\mathsf{u}})\rvert^2A_{\mathsf{u}}$, which simplifies \eqref{do_Single_User_SNR_Ini1} as follows:
		\begin{align}
			\gamma_{\mathsf{dl}}\approx
			\frac{A_{\mathsf{u}}
				\lvert\int_{{\mathcal{A}}}{\mathsf{H}}({\mathbf{r}}){\mathsf{J}}_{{\mathsf{dl}}}({\mathbf{r}}){\rm{d}}{\mathbf{r}}\rvert^2}
			{{\sigma}_{\mathsf{u}}^2}. \label{do_Single_User_SNR}
		\end{align}
		We note that $\gamma_{\mathsf{dl}}$ does not rely on ${\mathsf{V}}_{{\mathsf{dl}}}(\mathbf{s})$. Moreover, it is observed from \eqref{do_Single_User_SNR} that the SNR is maximized when the current distribution satisfies ${\mathsf{J}}_{{\mathsf{dl}}}({\mathbf{r}})\propto{\mathsf{H}}^*({\mathbf{r}})$, i.e.,
		\begin{equation}\label{Optimal_Precoder_MISO_SU_Answer}
			{\mathsf{J}}_{{\mathsf{dl}}}({\mathbf{r}})=\frac{{\mathsf{H}}^*({\mathbf{r}})}{{\mathsf{h}}_{\mathsf{u}}^{1/2}}
			\sqrt{\int_{{\mathcal{A}}}\lvert{\mathsf{J}}_{{\mathsf{dl}}}({\mathbf{r}})\rvert^2{\rm{d}}{\mathbf{r}}},
		\end{equation}
		where $\int_{{\mathcal{A}}}\lvert{\mathsf{J}}_{{\mathsf{dl}}}({\mathbf{r}})\rvert^2{\rm{d}}{\mathbf{r}}$ is proportional to the transmit power.
		\vspace{-5pt}
		\begin{remark}
			The results in \eqref{Optimal_Precoder_MISO_SU_Answer} suggest that the capacity-achieving source current for downlink single-user CAPA communications aligns with ${\mathsf{H}}({\mathbf{r}})$, which can be regarded as the continuous version of \textit{maximal-ratio transmission (MRT)}.  
		\end{remark}
		\vspace{-5pt}
		Inserting \eqref{Optimal_Precoder_MISO_SU_Answer} into \eqref{do_Single_User_SNR} gives
		\begin{equation}\label{do_Single_User_Resulting_SNR}
			\gamma_\mathsf{dl}=\frac{A_{\mathsf{u}}\int_{{\mathcal{A}}}\lvert{\mathsf{J}}_{{\mathsf{dl}}}({\mathbf{r}})\rvert^2{\rm{d}}{\mathbf{r}}}
			{{\sigma}_{\mathsf{u}}^2}
			\int_{\mathcal{A}}\lvert{\mathsf{H}}({\mathbf{r}})\rvert^2{\rm{d}}{\mathbf{r}}={\overline{\gamma}_{\mathsf{dl}}}{\mathsf{g}}_{\mathsf{u}},
		\end{equation}
		where ${\overline{\gamma}_{\mathsf{dl}}}=\frac{A_{\mathsf{u}}\int_{{\mathcal{A}}}\lvert{\mathsf{J}}_{{\mathsf{dl}}}({\mathbf{r}})\rvert^2{\rm{d}}{\mathbf{r}}k_0^2\eta^2}
		{4\pi{\sigma}_{\mathsf{u}}^2}$ can be interpreted as the downlink transmit SNR. Based on \eqref{do_Single_User_Resulting_SNR}, the single-user downlink capacity is given by
		\begin{align}\label{do_SU_Channel_Capacity}
			{\mathsf{C}}_\mathsf{dl}=\log_2(1+\gamma_\mathsf{dl})=\log_2(1+{\overline{\gamma}_{\mathsf{dl}}}{\mathsf{g}}_{\mathsf{u}}).
	\end{align}}
	{\subsection{Two-User Case}}
		In this part, we focus on the two-user scenario.
		\subsubsection{Optimal Multiuser Precoding}
		The capacity of a downlink multiuser Gaussian channel can be achieved using dirty-paper coding (DPC) \cite{goldsmith}. We consider the encoding order $\varepsilon_K \rightarrow\varepsilon_{K-1}\rightarrow\ldots\rightarrow\varepsilon_1$ with $\{\varepsilon_k\}_{k=1}^{K}=\{1,\ldots,K\}$. For user $\varepsilon_k$, the dirty-paper encoder treats the interference from user $\varepsilon_{k'}$ for $k'>k$ as non-causally known, while its decoder regards the interference from user $\varepsilon_{k'}$ for $k'<k$ as additional noise. Let ${\mathsf{V}}_{{\mathsf{dl}},k}(\mathbf{s})$ denote the detector used by user $k$ to recover $s_{{\mathsf{dl}},k}$ from \eqref{do_Total_electric_radiation_field}. By applying DPC and using ML decoding at each user, the signal-to-noise-plus-interference ratio (SINR) for user $\varepsilon_{k}$ is given by
		\begin{align}
			\gamma _{\mathsf{dl},\varepsilon_{k}}
			=\frac{\overline{\mathsf{v}}_{\varepsilon_{k}}\lvert \int_{{\mathcal{A}}}{\mathsf{H}}_{\varepsilon_{k}}({\mathbf{r}}){\mathsf{J}}_{{\mathsf{dl}},\varepsilon_{k}}({\mathbf{r}}){\rm{d}}{\mathbf{r}} \rvert^2}{\sigma _{\varepsilon_{k}}^{2}{\mathsf{v}}_{\varepsilon_{k}}+\sum_{k'<k}\overline{\mathsf{v}}_{\varepsilon_{k}}\lvert \int_{{\mathcal{A}}}{\mathsf{H}}_{\varepsilon_{k}}({\mathbf{r}}){\mathsf{J}}_{{\mathsf{dl}},\varepsilon_{k'}}({\mathbf{r}}){\rm{d}}{\mathbf{r}} \rvert^2},
		\end{align}
		where ${\mathsf{v}}_{\varepsilon_{k}}=\int_{\mathcal{A}_{\varepsilon_{k}}}{\lvert \mathsf{V}_{{\mathsf{dl}},\varepsilon_{k}}(\mathbf{s}) \rvert^2}\mathrm{d}\mathbf{s}$ and $\overline{\mathsf{v}}_{\varepsilon_{k}}=\lvert \int_{\mathcal{A}_{\varepsilon_{k}}}{\mathsf{V}}_{{\mathsf{dl}},\varepsilon_{k}}^{*}(\mathbf{s})\mathrm{d}\mathbf{s} \rvert^2$. Since the variations of the signal along each user's aperture is negligible, we have
		\begin{align}
			\overline{\mathsf{v}}_{\varepsilon_{k}}\approx \lvert\mathsf{V}_{{\mathsf{dl}},\varepsilon_{k}}({\mathbf{s}}_{\varepsilon_{k}})\rvert^2A_{{\mathsf{u}},\varepsilon_{k}}^2,
			{\mathsf{v}}_{\varepsilon_{k}}\approx 
			\lvert\mathsf{V}_{{\mathsf{dl}},\varepsilon_{k}}({\mathbf{s}}_{\varepsilon_{k}})\rvert^2A_{{\mathsf{u}},\varepsilon_{k}},
		\end{align}
		which simplifies $\gamma _{\mathsf{dl},\varepsilon_{k}}$ as follows:
		\begin{align}
			\gamma_{\mathsf{dl},\varepsilon_{k}}\approx 
			\frac{A_{{\mathsf{u}},\varepsilon_{k}}\lvert \int_{{\mathcal{A}}}{\mathsf{H}}_{\varepsilon_{k}}({\mathbf{r}}){\mathsf{J}}_{{\mathsf{dl}},\varepsilon_{k}}({\mathbf{r}}){\rm{d}}{\mathbf{r}} \rvert^2}
			{\sigma_{\varepsilon_{k}}^{2}+A_{{\mathsf{u}},\varepsilon_{k}}\sum_{k'<k}{\lvert \int_{{\mathcal{A}}}{\mathsf{H}}_{\varepsilon_{k}}({\mathbf{r}}){\mathsf{J}}_{{\mathsf{dl}},\varepsilon_{k'}}({\mathbf{r}}){\rm{d}}{\mathbf{r}} \rvert^2}}.
		\end{align}
		The sum-rate is given by $\sum_{k=1}^{K}\log_2(1+\gamma_{\mathsf{dl},\varepsilon_{k}})$. 
		
		Under the two-user case, without loss of generality, we consider the DPC order $2\rightarrow 1$. The achieved rates of user $1$ and user $2$ are given by
		\begin{align}
			\mathsf{R}_{\mathsf{dl},1}^{2\rightarrow 1}&=\log _2\left( 1+\frac{A_{{\mathsf{u}},1}\lvert \int_{{\mathcal{A}}}{\mathsf{H}}_{1}({\mathbf{r}}){\mathsf{J}}_{{\mathsf{dl}},1}({\mathbf{r}}){\rm{d}}{\mathbf{r}} \rvert^2}{\sigma_{1}^{2}} \right) ,\label{do_R1}\\
			\mathsf{R}_{\mathsf{dl},2}^{2\rightarrow 1}&=\log _2\left( 1+\frac{A_{{\mathsf{u}},2}\lvert \int_{{\mathcal{A}}}{\mathsf{H}}_{2}({\mathbf{r}}){\mathsf{J}}_{{\mathsf{dl}},2}({\mathbf{r}}){\rm{d}}{\mathbf{r}} \rvert^2}{\sigma _{2}^{2}+A_{{\mathsf{u}},2}\lvert \int_{{\mathcal{A}}}{\mathsf{H}}_{2}({\mathbf{r}}){\mathsf{J}}_{{\mathsf{dl}},1}({\mathbf{r}}){\rm{d}}{\mathbf{r}} \rvert^2} \right) ,\label{do_R2}
		\end{align}
		respectively. The problem of downlink sum-rate maximization is formulated as follows:
		\begin{align}\label{do_nonconvex}
			\mathsf{C}_{\mathsf{dl}}^{2\rightarrow 1}=\max\nolimits_{\sum_{k=1}^{2}{\int_{\mathcal{A}}{\lvert {\mathsf{J}}_{{\mathsf{dl}},k}({\mathbf{r}}) \rvert^2}\mathrm{d}\mathbf{r}}\le {\mathsf{P}}} \mathsf{R}_{\mathsf{dl},1}^{2\rightarrow 1}+\mathsf{R}_{\mathsf{dl},2}^{2\rightarrow 1},
		\end{align} 
		where $\mathsf{P}$ reflects the power budget. Following the same derivation approach, the sum-rate achieved by the DPC order $1\rightarrow2$ can also be derived. For downlink Gaussian channels, the maximal DPC sum-rate is independent of the DPC encoding order, and thus $\mathsf{C}_{\mathsf{dl}}^{2\rightarrow 1}$ is essentially the sum-rate capacity \cite{el2011network}.
		
		Problem \eqref{do_nonconvex} is non-convex, making the numerical identification of the maximum a nontrivial task. However, as demonstrated in \cite{vishwanath2003duality}, there exists a duality between the uplink and downlink scenarios. This duality establishes the equivalence of the DPC capacity region of a downlink channel with the capacity region of its dual uplink channel (as described in \eqref{Two_User_Total_electric_radiation_field}), where users are subject to a sum power constraint. By defining $\overline{\gamma}_{\mathsf{du},k} \triangleq \frac{A_{{\mathsf{u}},k}k_{0}^{2}\eta ^2}{4\pi \sigma _{k}^{2}}{\mathsf{P}}_k$ as the transmit SNRs of the dual uplink channel for $k=1,2$, the downlink sum-rate capacity is equal to the following dual uplink capacity \cite{goldsmith,el2011network,vishwanath2003duality}:
		\begin{align}\label{problem_dual}
			\begin{split}
				{\mathsf{C}}_{\mathsf{du}}=\max_{{\mathsf{P}}_1,{\mathsf{P}}_2\ge 0,{\mathsf{P}}_1+{\mathsf{P}}_2\le {\mathsf{P}}} \log _2(1&+\overline{\gamma}_{\mathsf{du},1}\mathsf{g}_1 +
				\overline{\gamma}_{\mathsf{du},2}\mathsf{g}_2\\
				&+\overline{\gamma}_{\mathsf{du},1}\overline{\gamma}_{\mathsf{du},2}\mathsf{g}_1\mathsf{g}_2\overline{\rho}),    
			\end{split}
		\end{align}
		which is a convex problem and can be solved as follows. 
		\vspace{-5pt}
		\begin{lemma}
			The optimal power allocation for the dual uplink is given by
			\begin{align}\label{optimal_p}
				( {\mathsf{P}}_{1}^{\star},{\mathsf{P}}_{2}^{\star} ) =\begin{cases}
					( {\mathsf{P}},0 )&		\xi \ge {\mathsf{P}}\\
					( 0,{\mathsf{P}} )&		\xi \le -{\mathsf{P}}\\
					( ({{\mathsf{P}}-\xi})/{2},({{\mathsf{P}}+\xi})/{2} )&		\mathsf{else}\\
				\end{cases},
			\end{align}
			where $\xi =\frac{{A_{{\mathsf{u}},1}}\mathsf{g}_1/{\sigma_{1}^{2}}-{A_{{\mathsf{u}},2}}\mathsf{g}_2/{\sigma_{2}^{2}}}{{A_{{\mathsf{u}},1}A_{{\mathsf{u}},2}k_{0}^{2}\eta ^2}\mathsf{g}_1\mathsf{g}_2\overline{\rho}/({4\pi \sigma_{1}^{2}\sigma_{2}^{2}})}$.    
		\end{lemma}
		\vspace{-5pt}
		\begin{IEEEproof}
			The results can be readily obtained by utilizing the Karush–Kuhn–Tucker (KKT) conditions.
		\end{IEEEproof}
		Upon obtaining the optimal power allocation policy \eqref{optimal_p} for the dual uplink channel,  we recover the corresponding downlink capacity-achieving source currents as follows.
		\vspace{-5pt}
		\begin{theorem}\label{theorem_optimal_current}
			Given the DPC order $2\rightarrow1$ and the power constraint $\sum_{k=1}^{2}{\int_{\mathcal{A}}{\lvert {\mathsf{J}}_{{\mathsf{dl}},k}({\mathbf{r}}) \rvert^2}\mathrm{d}\mathbf{r}}\le {\mathsf{P}}$, the downlink capacity-achieving source currents for user $1$ and user $2$ are given by
			\begin{align}
				{\mathsf{J}}_{{\mathsf{dl}},1}({\mathbf{r}})&=\sqrt{{\mathsf{P}}_{1}}\frac{\hat{\mathsf{H}}_{1}^{*}(\mathbf{r})-\frac{{\mathsf{P}}_{2}\int_{\mathcal{A}}
						\hat{\mathsf{H}}_{1}^{*}(\mathbf{r})\hat{\mathsf{H}}_{2}(\mathbf{r}){\rm{d}}{\mathbf{r}}}{1+{\mathsf{P}}_{2}\int_{\mathcal{A}}
						\lvert\hat{\mathsf{H}}_{2}(\mathbf{r})\rvert^2{\rm{d}}{\mathbf{r}}}\hat{\mathsf{H}}_{2}^{*}(\mathbf{r})}{\sqrt{\int_{\mathcal{A}}
						\lvert\hat{\mathsf{H}}_{1}(\mathbf{r})\rvert^2{\rm{d}}{\mathbf{r}}-\frac{{\mathsf{P}}_{2}\lvert\int_{\mathcal{A}}
							\hat{\mathsf{H}}_{1}^{*}(\mathbf{r})\hat{\mathsf{H}}_{2}(\mathbf{r}){\rm{d}}{\mathbf{r}}\rvert^2}{1+{\mathsf{P}}_{2}\int_{\mathcal{A}}
							\lvert\hat{\mathsf{H}}_{2}(\mathbf{r})\rvert^2{\rm{d}}{\mathbf{r}}}}},\label{J1}\\
				{\mathsf{J}}_{{\mathsf{dl}},2}({\mathbf{r}})&=\sqrt{{\mathsf{P}}_{2}}\frac{\hat{\mathsf{H}}_{2}^{*}(\mathbf{r})\sqrt{1+\lvert\int_{\mathcal{A}}
						\hat{\mathsf{\mathsf{H}}}_{2}(\mathbf{r}){\mathsf{J}}_{{\mathsf{dl}},1}({\mathbf{r}}){\rm{d}}{\mathbf{r}}\rvert^2}}{\int_{\mathcal{A}}
					\lvert\hat{\mathsf{H}}_{2}(\mathbf{r})\rvert^2{\rm{d}}{\mathbf{r}}},\label{J2}
			\end{align}
			respectively, where $\hat{\mathsf{H}}_{k}(\mathbf{r})=\frac{\sqrt{A_{{\mathsf{u}},k}}}{\sigma_k}\mathsf{H}_{k}(\mathbf{r})$ for $k=1,2$. Additionally, the optimal source currents satisfy 
			$\sum_{k=1}^{2}{\int_{\mathcal{A}}{\lvert {\mathsf{J}}_{{\mathsf{dl}},k}({\mathbf{r}}) \rvert^2}\mathrm{d}\mathbf{r}}={\mathsf{P}}_{1}+{\mathsf{P}}_{2}= {\mathsf{P}}$.
		\end{theorem}
		\vspace{-5pt}
		\begin{IEEEproof}
			Please refer to Appendix \ref{proof_theorem_optimal_current} for more details.    
		\end{IEEEproof}
		\vspace{-5pt}
		\begin{remark}\label{uplink-to-downlink transformation}
			\textbf{Theorem \ref{theorem_optimal_current}} establishes the \emph{uplink-to-downlink transformation} that takes as inputs an \emph{uplink transmit power allocation policy} and an \emph{SIC decoding order}, and outputs a set of \emph{downlink source currents} with the same sum power as the uplink dual channel. As shown in Appendix \ref{proof_theorem_optimal_current} and Section \ref{Downlink Capacity: Capacity Characterization}, the output source currents achieve the \emph{same rates} as the dual uplink channel using the input power allocation policy and SIC decoding with the specified decoding order.
		\end{remark}
		\vspace{-5pt}
		Having obtained the capacity-achieving source currents\footnote{{ We note that the derivation of the optimal source currents is heuristic and inspired by their SPDA counterpart. Specifically, based on \cite[Section IV]{vishwanath2003duality}, for a dual uplink SPDA-based multiuser MISO channel, the optimal DPC-based precoders can be expressed as a weighted sum of the channel vectors, where the weighting factors are determined by the channel gains and channel correlation coefficients. Motivated by this, it is reasonable to infer that the optimal current distributions in a CAPA system can also be represented as a linear combination of the continuous spatial responses, i.e., ${\mathsf{J}}_{{\mathsf{dl}},1}({\mathbf{r}})=a\hat{\mathsf{H}}_{1}^{*}(\mathbf{r})+b\hat{\mathsf{H}}_{2}^{*}(\mathbf{r})$ and ${\mathsf{J}}_{{\mathsf{dl}},2}({\mathbf{r}})=c\hat{\mathsf{H}}_{1}^{*}(\mathbf{r})+d\hat{\mathsf{H}}_{2}^{*}(\mathbf{r})$ for $a,b,c,d\in{\mathbbmss{C}}$. This intuition follows from the fact that the optimal current should reside in the signal space spanned by $\hat{\mathsf{H}}_{1}^{*}(\mathbf{r})$ and $\hat{\mathsf{H}}_{2}^{*}(\mathbf{r})$. To determine the weighting coefficients $\{a,b,c,d\}$, we compare the achieved rates with those in the dual uplink system, leading to the results presented in \textbf{Theorem} \ref{theorem_optimal_current} and the corresponding per-user rate. However, we acknowledge that this derivation remains largely intuitive and heuristic. For more complex scenarios---such as multiuser systems with an arbitrary number of users or configurations where each user is equipped with multiple antennas or even a CAPA---it may be challenging to directly apply this method to obtain the optimal currents and the corresponding dual transformation. A deeper investigation of these scenarios is left for future work.}} under the DPC order $2\rightarrow1$, we summarize the entire precoding procedure in {Table \ref{DPC_encoding_Two_User}} for ease of reference. The presented DPC procedure can be directly extended to the other order $1\rightarrow2$ by exchanging the user indices.
	
	\begin{table}[!t]
		\centering
		\setlength{\abovecaptionskip}{3pt}
		\begin{tabular}{|c|}
			\hline
			\begin{minipage}{3in}
				\vspace{1pt}
				\begin{algorithmic}[1]
					\State{\footnotesize Encode the data information into the symbols $\{s_{{\mathsf{dl}},k}\}_{k=1}^2$ using DPC with the order $2\rightarrow1$;}
					\State{\footnotesize Determine the dual power allocation policy $\{{\mathsf{P}}_k\}_{k=1}^2$ according to the specific target quality of service requirements;}
					\State{\footnotesize Generate the source current ${\mathsf{J}}_{{\mathsf{dl}},1}({\mathbf{r}})$:
						{\setlength\abovedisplayskip{2pt}
							\setlength\belowdisplayskip{2pt}
							\begin{equation}
								{\mathsf{J}}_{{\mathsf{dl}},1}({\mathbf{r}})=\sqrt{{\mathsf{P}}_{1}}\frac{\hat{\mathsf{H}}_{1}^{*}(\mathbf{r})-\frac{{\mathsf{P}}_{2}\int_{\mathcal{A}}
										\hat{\mathsf{H}}_{1}^{*}(\mathbf{r})\hat{\mathsf{H}}_{2}(\mathbf{r}){\rm{d}}{\mathbf{r}}}{1+{\mathsf{P}}_{2}\int_{\mathcal{A}}
										\lvert\hat{\mathsf{H}}_{2}(\mathbf{r})\rvert^2{\rm{d}}{\mathbf{r}}}\hat{\mathsf{H}}_{2}^{*}(\mathbf{r})}{\sqrt{\int_{\mathcal{A}}
										\lvert\hat{\mathsf{H}}_{1}(\mathbf{r})\rvert^2{\rm{d}}{\mathbf{r}}-\frac{{\mathsf{P}}_{2}\lvert\int_{\mathcal{A}}
											\hat{\mathsf{H}}_{1}^{*}(\mathbf{r})\hat{\mathsf{H}}_{2}(\mathbf{r}){\rm{d}}{\mathbf{r}}\rvert^2}{1+{\mathsf{P}}_{2}\int_{\mathcal{A}}
											\lvert\hat{\mathsf{H}}_{2}(\mathbf{r})\rvert^2{\rm{d}}{\mathbf{r}}}}}; \nonumber
							\end{equation}
					}}
					\State{\footnotesize Generate the source current ${\mathsf{J}}_{{\mathsf{dl}},2}({\mathbf{r}})$:
						{\setlength\abovedisplayskip{2pt}
							\setlength\belowdisplayskip{2pt}
							\begin{equation}
								{\mathsf{J}}_{{\mathsf{dl}},2}({\mathbf{r}})=\sqrt{{\mathsf{P}}_{2}}\frac{\hat{\mathsf{H}}_{2}^{*}(\mathbf{r})\sqrt{1+\lvert\int_{\mathcal{A}}
										\hat{\mathsf{\mathsf{H}}}_{2}(\mathbf{r}){\mathsf{J}}_{{\mathsf{dl}},1}({\mathbf{r}}){\rm{d}}{\mathbf{r}}\rvert^2}}{\int_{\mathcal{A}}
									\lvert\hat{\mathsf{H}}_{2}(\mathbf{r})\rvert^2{\rm{d}}{\mathbf{r}}}; \nonumber
							\end{equation}
					}}
					\State{\footnotesize Formulate the transmit signal ${x}_{{\mathsf{dl}}}({\mathbf{r}})=\sum_{k=1}^{K}{\mathsf{J}}_{{\mathsf{dl}},k}({\mathbf{r}})s_{{\mathsf{dl}},k}$.}
				\end{algorithmic}
				\vspace{0pt}
			\end{minipage}
			\\ \hline
		\end{tabular}
		\caption{DPC for downlink CAPA communications.}
		\label{DPC_encoding_Two_User}
		\vspace{-10pt}
	\end{table}
	
	\subsubsection{Capacity Characterization}\label{Downlink Capacity: Capacity Characterization}
	As proved in Appendix \ref{proof_theorem_optimal_current}, given the power allocation policy $({\mathsf{P}}_1,{\mathsf{P}}_2)$, the achievable rates of user $1$ and user $2$ under the DPC order $2\rightarrow1$ are given by
		\begin{align}
			\mathsf{R}_{\mathsf{dl},1}^{2\rightarrow 1}&=\log _2\left( 1+\overline{\gamma }_{\mathsf{dl},1}\mathsf{g}_1\left( 1-\frac{\overline{\gamma }_{\mathsf{dl},2}\mathsf{g}_2|\rho |^2}{1+\overline{\gamma }_{\mathsf{dl},2}\mathsf{g}_2} \right) \right),\label{Dual_Uplink_R1}\\
			\mathsf{R}_{\mathsf{dl},2}^{2\rightarrow 1}&=\log _2\left( 1+\overline{\gamma }_{\mathsf{dl},2}\mathsf{g}_2 \right),\label{Dual_Uplink_R2}    
		\end{align}
		respectively, where $\overline{\gamma}_{\mathsf{dl},k} = \frac{A_{{\mathsf{u}},k}k_{0}^{2}\eta ^2}{4\pi \sigma _{k}^{2}}{\mathsf{P}}_k^{\star}$ for $k\in\{1,2\}$. By comparing the above two expressions with the results in {\textbf{Corollary \ref{Corollary_Sum_Rate_Capacity}}}, the following observations are made.
		\vspace{-5pt}
		\begin{remark}
			The downlink user rates achieved by the DPC order $2\rightarrow 1$ equal those achieved in the dual uplink channel under the SIC decoding order $1\rightarrow 2$.
		\end{remark}
		\vspace{-5pt}
		The rates achieved by the DPC order $1\rightarrow 2$ can be derived with similar steps, which are omitted for brevity. In the sequel, we derive the downlink-to-uplink transformation.
		\vspace{-5pt}
		\begin{theorem}\label{theorem_downlink-to-uplink transformation}
			Given the source currents $\{{\mathsf{J}}_{{\mathsf{dl}},k}({\mathbf{r}})\}_{k=1}^{2}$ such that $\sum_{k=1}^{2}{\int_{\mathcal{A}}{\lvert {\mathsf{J}}_{{\mathsf{dl}},k}({\mathbf{r}}) \rvert^2}\mathrm{d}\mathbf{r}}\le {\mathsf{P}}$, the power allocation policy of the dual uplink channel, which achieves the same sum-rate as the downlink channel under the DPC order $2\rightarrow1$, is given by
			\begin{align}
				{\mathsf{P}}_{2}&=\frac{\lvert\int_{\mathcal{A}}
					\hat{\mathsf{\mathsf{H}}}_{2}(\mathbf{r}){\mathsf{J}}_{{\mathsf{dl}},2}({\mathbf{r}}){\rm{d}}{\mathbf{r}}\rvert^2}{\int_{\mathcal{A}}\lvert\hat{\mathsf{\mathsf{H}}}_{2}(\mathbf{r})\rvert^2
					{\rm{d}}{\mathbf{r}}(1+\lvert\int_{\mathcal{A}}
					\hat{\mathsf{\mathsf{H}}}_{2}(\mathbf{r}){\mathsf{J}}_{{\mathsf{dl}},1}({\mathbf{r}}){\rm{d}}{\mathbf{r}}\rvert^2)},\label{P2}\\
				{\mathsf{P}}_{1}&=\frac{\lvert\int_{\mathcal{A}}
					\hat{\mathsf{\mathsf{H}}}_{1}(\mathbf{r}){\mathsf{J}}_{{\mathsf{dl}},1}({\mathbf{r}}){\rm{d}}{\mathbf{r}}\rvert^2}
				{\int_{\mathcal{A}}\lvert\hat{\mathsf{\mathsf{H}}}_{1}(\mathbf{r})\rvert^2{\rm{d}}{\mathbf{r}}-\frac{{\mathsf{P}}_{2}\lvert\int_{\mathcal{A}}
						\hat{\mathsf{\mathsf{H}}}_{1}^{*}(\mathbf{r})\hat{\mathsf{\mathsf{H}}}_{2}(\mathbf{r}){\rm{d}}{\mathbf{r}}\rvert^2}{1+{\mathsf{P}}_{2}\int_{\mathcal{A}}
						\lvert\hat{\mathsf{\mathsf{H}}}_{2}(\mathbf{r})\rvert^2{\rm{d}}{\mathbf{r}}}},\label{P1}
			\end{align}
			respectively, where ${\mathsf{P}}_{1}+{\mathsf{P}}_{2}\leq \sum_{k=1}^{2}{\int_{\mathcal{A}}{\lvert {\mathsf{J}}_{{\mathsf{dl}},k}({\mathbf{r}}) \rvert^2}\mathrm{d}\mathbf{r}}\le {\mathsf{P}}$.
		\end{theorem}
		\vspace{-5pt}
		\begin{IEEEproof}
			Please refer to Appendix \ref{proof_theorem_downlink-to-uplink transformation} for more details.    
		\end{IEEEproof}
		\vspace{-5pt}
		\begin{remark}\label{downlink-to-uplink transformation}
			\textbf{Theorem \ref{theorem_downlink-to-uplink transformation}} establishes the \emph{downlink-to-uplink transformation} which, given a set of \emph{downlink source currents} and a \emph{DPC encoding order}, outputs the \emph{dual uplink power allocation policy} with the same sum power as the downlink currents. This policy achieves uplink rates (using SIC decoding) equal to the downlink rates.
		\end{remark}
		\vspace{-5pt}
		\vspace{-5pt}
		\begin{remark}
			The results in \textbf{Remark \ref{uplink-to-downlink transformation}} and \textbf{Remark \ref{downlink-to-uplink transformation}} suggest that under the same sum power constraint, the downlink DPC capacity region coincides with the capacity region of its dual uplink channel. This further verifies the existence of uplink-downlink duality in CAPA communications.
		\end{remark}
		\vspace{-5pt}
		In light of the uplink-downlink duality, the sum-rate downlink capacity remains constant regardless of the DPC order (similar to its dual uplink channel), and is given as follows.
		\vspace{-5pt} 
		\begin{theorem}
			The sum-rate capacity of the downlink CAPA-based two-user channel is given by
			\begin{subequations}\label{do_sum_capacity}
				\begin{align}
					\mathsf{C}_{\mathsf{dl}}&=\log_2\Big(1+\overline{\gamma}_{\mathsf{dl},1}\overline{\gamma}_{\mathsf{dl},2}{\mathsf{g}}_{1}
					{\mathsf{g}}_{2}\overline{\rho}+\sum\nolimits_{k=1}^{2}\overline{\gamma}_{\mathsf{dl},k}{\mathsf{g}}_{k}\Big)\\
					&=\begin{cases}
						\log _2\left( 1+\widetilde{\gamma}_{\mathsf{dl},1}(\mathsf{P})\mathsf{g}_1 \right)&		\xi \ge {\mathsf{P}}\\
						\log _2\left( 1+\widetilde{\gamma}_{\mathsf{dl},2}(\mathsf{P})\mathsf{g}_2 \right)&		\xi \le -{\mathsf{P}}\\
						\log _2\left( 1+\varepsilon_1 +\varepsilon_2 +\varepsilon_1 \varepsilon_2 \overline{\rho } \right) &		\mathsf{else}\\
					\end{cases} ,    
				\end{align}
			\end{subequations}
			where $\widetilde{\gamma}_{\mathsf{dl},k}(x)\triangleq \frac{A_{{\mathsf{u}},k}k_{0}^{2}\eta ^2}{4\pi \sigma _{k}^{2}}x$ for $k\in\{1,2\}$, $\varepsilon_1=\widetilde{\gamma }_{\mathsf{dl},1}(\frac{{\mathsf{P}}-\xi}{2})\mathsf{g}_1$, and $\varepsilon_2 =\widetilde{\gamma }_{\mathsf{dl},2}(\frac{{\mathsf{P}}+\xi}{2})\mathsf{g}_2$.
		\end{theorem}
		\vspace{-5pt}
		\begin{IEEEproof}
			Eq. \eqref{do_sum_capacity} is obtained by inserting \eqref{optimal_p} into \eqref{problem_dual}.
		\end{IEEEproof}
		\vspace{-5pt}
		\begin{remark}\label{remark_downlink sum-rate capacity}
			The downlink sum-rate capacity given in \eqref{do_sum_capacity} is determined by the channel gain of each user and the channel correlation factor. This expression is applicable to an arbitrary aperture, regardless of its location, shape, and size.
		\end{remark} 
		\vspace{-5pt}
		According to the uplink-downlink duality, given a power allocation scheme $({\mathsf{P}}_1,{\mathsf{P}}_2)$ with ${\mathsf{P}}_1+{\mathsf{P}}_2={\mathsf{P}}$, we can obtain the capacity region of a dual uplink channel. The downlink capacity region is the convex hull of the union of all these dual uplink capacity regions, which is illustrated in Section~\ref{section_numerical}.
		
		{\subsubsection{Sum-Rate Achieved by Zero-Forcing Precoder}
		Although DPC is a potent capacity-achieving scheme, its
		practical implementation proves challenging. Consequently,
		non-DPC linear ZF precoding schemes hold practical
		significance due to the lower complexity. In the following analysis, we aim to evaluate the sum-rate performance achieved by the ZF precoder.
		
		Let $\mathsf{J}_{\mathsf{zf},k}(\mathbf{r})$ denote the ZF source current for user $k$ with $k\in\{1,2\}$. In order to nullify the IUI, i.e., $\int_{{\mathcal{A}}}\mathsf{J}_{\mathsf{zf},k}(\mathbf{r})\mathsf{H}_{k'}(\mathbf{r}){\rm{d}}{\mathbf{r}}=0$ for $k'\ne k$, and maximize the channel gain of user $k$, $\mathsf{J}_{\mathsf{zf},k}(\mathbf{r})$ is designed as the projection of $\mathsf{H}^*_{k}(\mathbf{r})$ onto the space orthogonal to $\mathsf{H}^*_{k'}(\mathbf{r})$, which is given by
		\begin{equation}\label{j_zf}
			\mathsf{J}_{\mathsf{zf},k}(\mathbf{r})={\mathsf{H}_{k}^{*}(\mathbf{r})}-\frac{\int_{\mathcal{A}}{\mathsf{H}_{k}^{*}}(\mathbf{r})\mathsf{H}_{k^{\prime}}(\mathbf{r})\mathrm{d}\mathbf{r}}{ \int_{\mathcal{A}}{\left| \mathsf{H}_{k^{\prime}}(\mathbf{r}) \right|^2}\mathrm{d}\mathbf{r}}\mathsf{H}_{k^{\prime}}^{*}(\mathbf{r}).
		\end{equation}
		By inserting \eqref{j_zf} into the following per-user rate expression
		\begin{align}
			\mathsf{R}_{\mathsf{dl},k}^{\mathsf{zf}}=\log _2\left( 1+\frac{| \int_{\mathcal{A}}{}\mathsf{H}_k(\mathbf{r})\mathsf{J}_{\mathsf{zf},k}(\mathbf{r})\mathrm{d}\mathbf{r} |^2}{\frac{\sigma _{k}^{2}}{A_{\mathsf{u},k}}+| \int_{\mathcal{A}}{}\mathsf{H}_k(\mathbf{r})\mathsf{J}_{\mathsf{zf},k^{\prime}}(\mathbf{r})\mathrm{d}\mathbf{r}|^2} \right)  ,
		\end{align}
		the sum-rate achieved by the ZF precoding is given as follows:
		\begin{equation}
			\mathsf{R}_{\mathsf{dl}}^{\mathsf{zf}}=\sum\nolimits_{k=1}^2{\log _2\Big( 1+\overline{\gamma }_{\mathsf{dl},k}^{\mathsf{zf}}\mathsf{g}_k\big( 1-\left| \rho \right|^2 \big) \Big)},	
		\end{equation}
		where $\overline{\gamma }_{\mathsf{dl},k}^{\mathsf{zf}} = \frac{A_{{\mathsf{u}},k}k_{0}^{2}\eta ^2}{4\pi \sigma _{k}^{2}}{\mathsf{P}}_k^{\mathsf{zf}}$ with ${\mathsf{P}}_k^{\mathsf{zf}}=\int_{\mathcal{A}}{\lvert {\mathsf{J}}_{{\mathsf{zf}},k}(\mathbf{r}) \rvert^2}\mathrm{d}\mathbf{r}$ being the optimally allocated power that can be obtained by the water-filling algorithm \cite{tse2005fundamentals,heath2018foundations}.} 
	\section{Case Studies}\label{Section: Special Cases}
	As discussed in \textbf{Remark \ref{remark_uplink sum-rate capacity}} and \textbf{Remark \ref{remark_downlink sum-rate capacity}}, the sum-rate capacity expressions derived for both uplink and downlink CAPA communications are applicable to apertures of arbitrary location, shape, and size, among other factors. In this section, we specialize ${\mathcal{A}}$ to several specific structures to unveil more system insights under the two-user scenario.
	
	\begin{figure}[!t]
		\centering
		\subfigbottomskip=0pt
		\subfigcapskip=-3pt
		\setlength{\abovecaptionskip}{3pt}
		\subfigure[Planar CAPA.]
		{
			\includegraphics[height=0.22\textwidth]{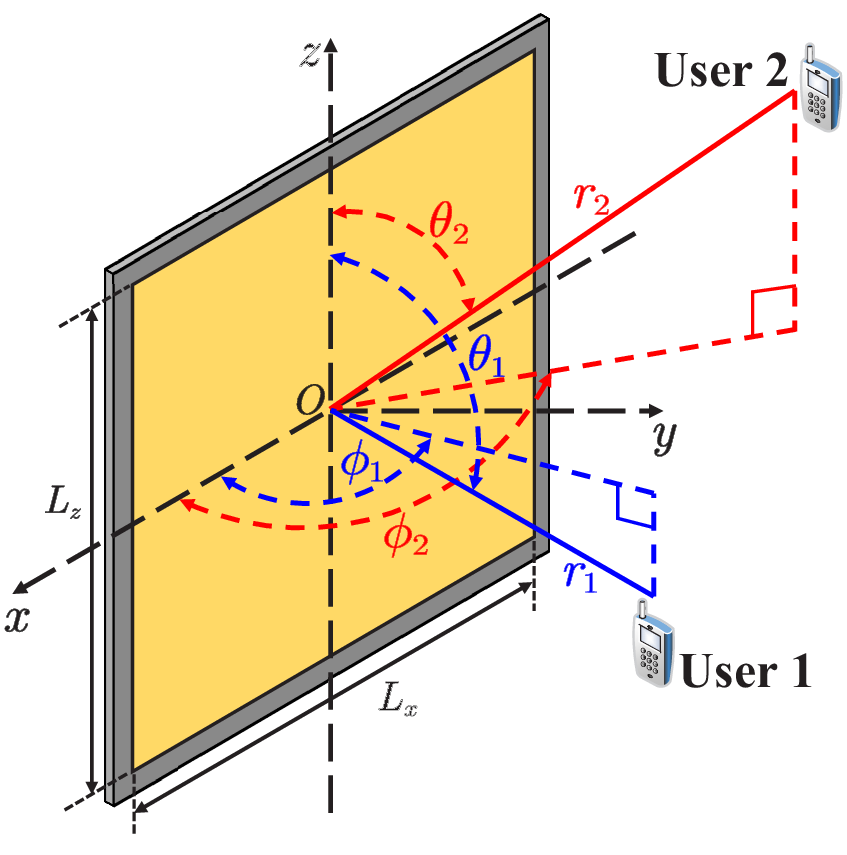}
			\label{planar_capa}	
		}
		\subfigure[Linear CAPA.]
		{
			\includegraphics[height=0.22\textwidth]{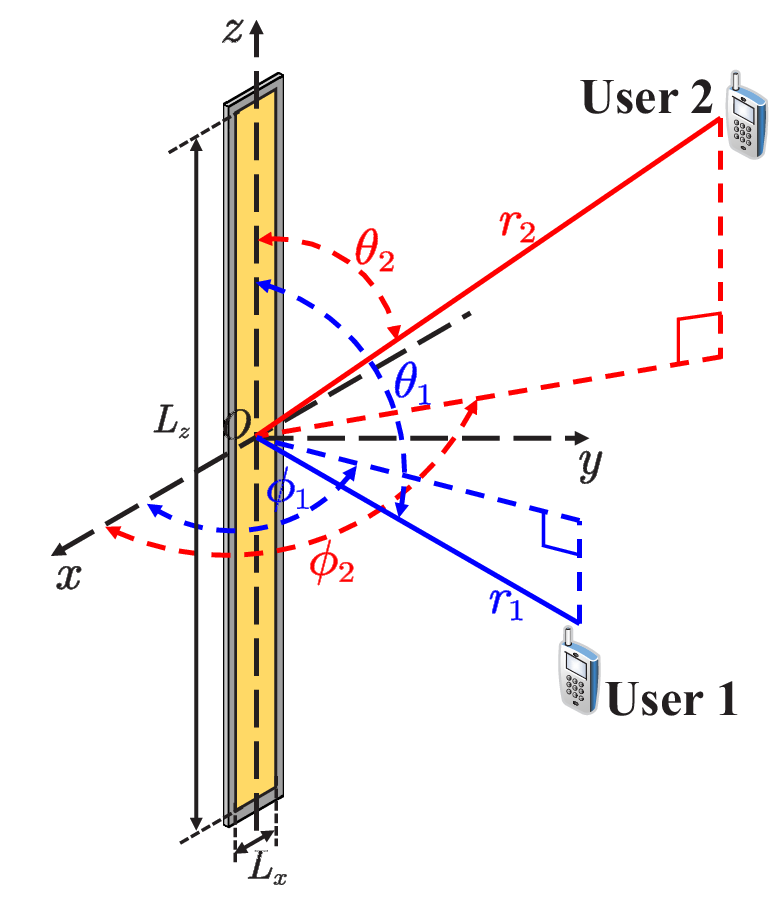}
			\label{linear_CAPA}	
		}
		\caption{Illustration of the geometries of CAPAs.}
		\label{case_study}
		\vspace{-8pt}
	\end{figure}
	
	\subsection{Planar CAPAs}
	We first consider the case where the BS array is a planar CAPA placed on the $x$-$z$ plane and centered at the origin, as shown in Fig. \ref{planar_capa}. The edges of $\mathcal{A}$ are parallel to the axes, with physical dimensions $L_x$ and $L_z$ along the $x$- and $z$-axes, respectively. In this case, we have $\mathcal{A} =\{[x,0,z]^{\mathsf{T}}|x\in [-\frac{L_x}{2},\frac{L_x}{2}],z\in [-\frac{L_z}{2},\frac{L_z}{2}]\}$. For each user $k \in \{1, 2\}$, let $r_k$ represent the distance from the center of $\mathcal{A}$ to the center of $\mathcal{A}_\mathsf{k}$, $\phi_k \in [0, \pi]$ and $\theta_k \in [0, \pi]$ denote the associated azimuth and elevation angles, respectively. Consequently, ${\mathbf{s}}_{k}=[r_k\Phi_k, r_k\Psi_k, r_k\Theta_k]^{\mathsf{T}}$, where $\Phi_k\triangleq\cos{\phi_k}\sin{\theta_k}$, $\Psi_k\triangleq\sin{\phi_k}\sin{\theta_k}$, and $\Theta_k\triangleq\cos{\theta_k}$. Based on \eqref{CAPA_LoS_Channel_Model} and \eqref{Single_User_SNR_2}, for $\mathbf{r}=[x,0,z]^{\mathsf{T}}\in\mathcal{A}$, we have
	\begin{equation}\label{case_stduy_G}
		\begin{split}
			\mathsf{G}_k(\mathbf{r})&=\frac{\sqrt{r_k\Psi _k}\mathrm{e}^{-\mathrm{j}k_0(x^2+z^2-2r_k\left( \Phi _kx+\Theta _kz \right) +r_{k}^{2})^{\frac{1}{2}}}}
			{\sqrt{4\pi}(x^2+z^2-2r_k\left( \Phi _kx+\Theta _kz \right) +r_{k}^{2})^{\frac{3}{4}}}\\
			&\triangleq\mathsf{Q}_k(x,z).
		\end{split}
	\end{equation}
	To calculate the capacity, we derive closed-form expressions of the channel gain and correlation factor as follows.
	\vspace{-5pt}
	\begin{lemma}\label{lemma_planar_capa}
		With the planar CAPA, the channel gain for each user can be calculated as follows:
		\begin{align}\label{CAPA_UPA_Channel_Gain}
			{\mathsf{g}}_k=\frac{1}{4\pi}\sum_{x\in{\mathcal{X}}_k}\sum_{z\in{\mathcal{Z}}_k}\arctan\bigg(\frac{xz/\Psi_k}{\sqrt{\Psi_k^2+x^2+z^2}}\bigg)\triangleq 
			{\mathsf{g}}_k^{\mathsf{p}},
		\end{align}
		where $k\in\{1,2\}$, ${\mathcal{X}}_k\triangleq\{\frac{L_x}{2r_k}\pm \Phi_k\}$, and ${\mathcal{Z}}_k\triangleq\{\frac{L_z}{2r_k}\pm \Theta_k\}$. The associated channel correlation factor is approximated by
		\begin{align}\label{rho_planar_capa}
			\begin{split}
				\rho =&\frac{\pi ^2A}{4n^2\sqrt{\mathsf{g}_{1}^{\mathsf{p}}\mathsf{g}_{2}^{\mathsf{p}}}}\sum_{j=1}^n\sum_{j'=1}^n{\sqrt{\left( 1-\psi _{j}^{2} \right) \big( 1-\psi _{j'}^{2} \big)}}\\
				&\times\mathsf{Q}_{1}^{*}\Big(\frac{L_x\psi _j}{2},\frac{L_z\psi _{j'}}{2}\Big)\mathsf{Q}_2\Big(\frac{L_x\psi _j}{2},\frac{L_z\psi _{j'}}{2}\Big)\triangleq \rho_{\mathsf{p}},   
			\end{split}
		\end{align}
		where $n$ is a complexity-vs-accuracy tradeoff parameter, and $\psi _j=\cos \left( \frac{\left( 2j-1 \right) \pi}{2n} \right) $ for $j\in \{1,\ldots,n\}$.
	\end{lemma}
	\vspace{-5pt}
	\begin{IEEEproof}
		Please refer to Appendix \ref{proof_lemma_planar_capa} for more details.
	\end{IEEEproof}
	Note that the expression in \eqref{rho_planar_capa} is derived using the Chebyshev–Gauss quadrature rule, as per Appendix \ref{proof_lemma_planar_capa}. 
	
	Subsequently, the closed-form expressions of sum-rate capacity can be readily obtained.
	\vspace{-5pt}
	\begin{corollary}
		The sum-rate capacity for the planar CAPA-based two-user channel can be written as follows:
		\begin{align}
			\mathsf{C}_{i}=\log _2\big(1+\overline{\gamma }_{i,1}\mathsf{g}_1^{\mathsf{p}} +
			\overline{\gamma }_{i,2}\mathsf{g}_2^{\mathsf{p}}
			+\overline{\gamma }_{i,1}\overline{\gamma }_{i,2}\mathsf{g}_1^{\mathsf{p}}\mathsf{g}_2^{\mathsf{p}}\overline{\rho}_{\mathsf{p}}\big),
		\end{align}
		where $i\in\{\mathsf{ul},\mathsf{dl}\}$, and $\overline{\rho}_{\mathsf{p}}=1-\lvert{\rho}_{\mathsf{p}}\rvert^2\in[0,1]$.
	\end{corollary}
	\vspace{-5pt}
	To shed more light on the characteristics of planar CAPAs, we investigate the asymptotic sum-rate capacity for a limiting case where $\mathcal{A}$ is assumed to be infinitely large, i.e., $L_x,L_z\rightarrow\infty$. In this case, we have $\lim_{L_x,L_z\rightarrow\infty}{\mathsf{g}}_k^{\mathsf{p}}=\frac{1}{4\pi}\frac{4\pi}{2}=\frac{1}{2}$. Regarding $\lim_{L_x,L_z\rightarrow\infty}\lvert{\rho}_{\mathsf{p}}\rvert$, while it is computationally intractable, based on the numerical results presented in \cite{boqun_jstsp,liu2024road}, we observe that $\lim_{L_x,L_z\rightarrow\infty}\lvert{\rho}_{\mathsf{p}}\rvert\ll 1$. Taken together, we can obtain the asymptotic sum-rate capacity as follows:
	\begin{align}\label{asy_planar_capa}
		\lim_{L_x,L_z\rightarrow \infty} \mathsf{C}_i\approx \log_2(1+{\overline{\gamma}_{i,1}}/{2})+\log_2(1+{\overline{\gamma}_{i,2}}/{2}),    
	\end{align}
	where $i\in\{\mathsf{ul},\mathsf{dl}\}$. The results in \eqref{asy_planar_capa} suggest that when the BS is equipped with an infinitely large planar CAPA, half of the transmission power of each user can be captured by the receiver, and the IUI vanishes.
	\vspace{-5pt}
	\begin{remark}\label{energy_conservation}
		The above observation makes intuitive sense because an infinitely large planar CAPA can receive at most half of the power transmitted by an isotropic source, while the other half of the power will never reach the surface.
	\end{remark}
	\vspace{-5pt}
	{ Additionally, since $\lim_{L_x,L_z\rightarrow\infty}\lvert{\rho}_{\mathsf{p}}\rvert\ll 1$, we have
		\begin{equation}\label{zf_large}
			\lim_{L_x,L_z\rightarrow \infty} \mathsf{R}_{i}^{\mathsf{zf}}\approx \lim_{L_x,L_z\rightarrow \infty} \mathsf{C}_{i},
		\end{equation}
		for $i\in\{\mathsf{ul},\mathsf{dl}\}$. This suggest that when the size of the CAPA is sufficiently large, the sum-rates achieved by the ZF schemes can approach the sum-rate capacities.}
	\subsection{Linear CAPAs}
	Next, we simplify the planar CAPA to a linear CAPA with $L_x\ll L_z$. In this case, the variations in currents and channel responses across the $x$-axis within $\mathcal{A}$ are negligible. The channel gain and correlation factor are calculated as follows.
	\vspace{-5pt}
	\begin{lemma}
		The channel gain for the linear CAPA is given by   
		\begin{align}
			\mathsf{g}_k=\frac{L_x\sin \phi _k\varrho_k}{4\pi r_k\sin \theta _k}, 
		\end{align}
		where $\varrho_k =\frac{L_z-2r_k\Theta _k}{\left( L_{z}^{2}-4r_k\Theta _kL_z+4r_{k}^{2} \right) ^{\frac{1}{2}}}+\frac{L_z+2r_k\Theta _k}{\left( L_{z}^{2}+4r_k\Theta _kL_z+4r_{k}^{2} \right) ^{\frac{1}{2}}}$. The channel correlation factor satisfies
		\begin{align}
			\rho =\frac{\pi L_z}{2n}\sum_{j=1}^n{\sqrt{1-\psi _{j}^{2}}}\mathsf{Q}_{1}^{*}\Big(0,\frac{L_z\psi _j}{2}\Big)\mathsf{Q}_2\Big(0,\frac{L_z\psi _j}{2}\Big).
		\end{align}
	\end{lemma}
	\vspace{-5pt}
	\begin{IEEEproof}
		When $L_x\ll L_z$, we have $\mathsf{g}_k\approx L_x\int_{-\frac{L_z}{2}}^{\frac{L_z}{2}}{\left| \mathsf{Q}_k(0,z) \right|^2\mathrm{d}z}$ and $\rho \approx L_x\int_{-\frac{L_z}{2}}^{\frac{L_z}{2}}{\mathsf{Q}_{1}^{*}(0,z)\mathsf{Q}_2(0,z)\mathrm{d}z}$. The subsequent derivation steps are similar to those shown in Appendix \ref{proof_lemma_planar_capa}, and the final results follow immediately.
	\end{IEEEproof}
	When the linear CAPA is infinitely long, i.e., $L_z\rightarrow \infty$, the asymptotic sum-rate capacity satisfies
	\begin{align}\label{Asym_SR_Linear}
		\lim_{L_z\rightarrow \infty} \mathsf{C}_i\approx \sum\nolimits_{k=1}^{2}\log_2\big(1+\frac{L_x\sin \phi _1{\overline{\gamma}_{i,k}}}{2\pi r_k\sin \theta _k}\big),
	\end{align}
	where $i\in\{\mathsf{ul},\mathsf{dl}\}$. By comparing \eqref{asy_planar_capa} with \eqref{Asym_SR_Linear}, it is found the asymptotic sum-rate capacity achieved by the linear CAPA is influenced not only by the SNR but also by the locations of the users, which is in contrast to the planar CAPA.
	
	\begin{figure}[!t]
		\centering
		\setlength{\abovecaptionskip}{2pt}
		\includegraphics[height=0.23\textwidth]{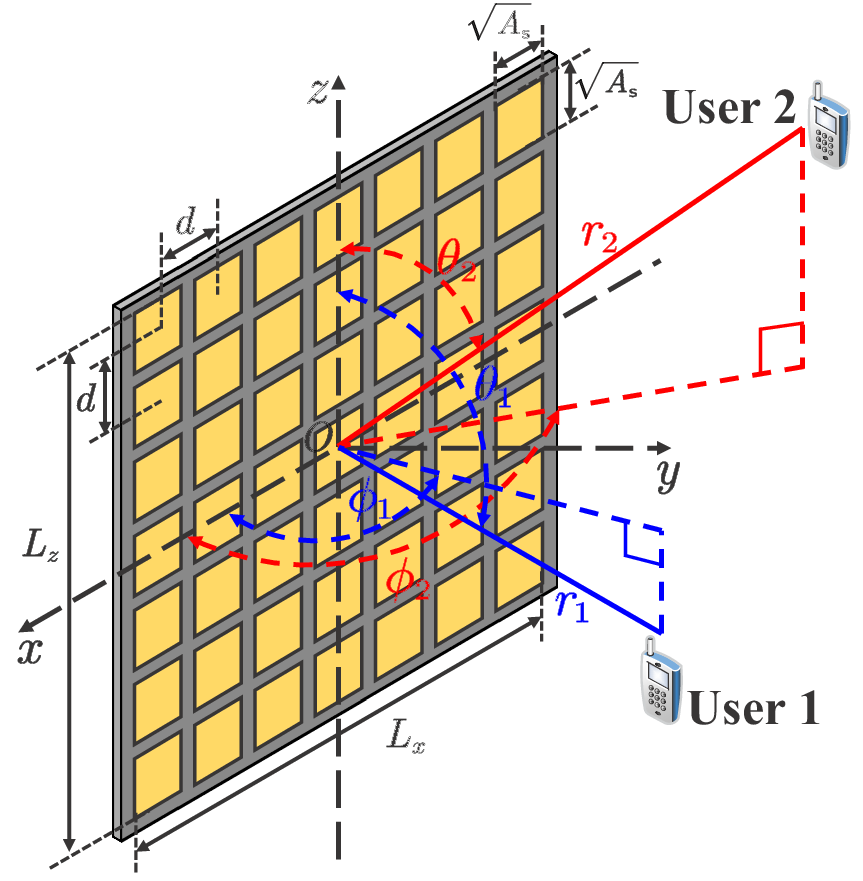}
		\caption{Illustration of a planar SPDA.}
		\vspace{-7pt}
		\label{SPD}
	\end{figure}
	
	\subsection{Planar SPDAs}
	Finally, we consider a planar SPDA with the same size as the planar CAPA, i.e., $L_x\times L_z$, comprising $M=M_zM_x$ spatially discrete elements, where $M_{x}=2\tilde{M}_x+1$ and $M_{z}=2\tilde{M}_z+1$ denote the number of antenna elements along the $x$- and $z$-axes, as depicted in Fig. \ref{SPD}. The physical dimensions of each element are indicated by $\sqrt{{A}_{\mathsf{s}}}$, and the inter-element distance is denoted as $d$, where $\sqrt{{A}_{\mathsf{s}}}\leq d\ll r_k$. In this case, the central position of each element is given by $[m_xd,0,m_zd]^{\mathsf{T}}$ for $m_x\in \mathcal{M} _x\triangleq\{0,\pm1,\ldots,\pm\tilde{M}_x\}$ and $m_z\in \mathcal{M} _z\triangleq\{0,\pm1,\ldots,\pm\tilde{M}_z\}$. Further, we have $L_x\approx M_xd$, $L_z\approx M_zd$, and $\mathcal{A}=\{[ m_xd+\ell_x ,0,m_zd+\ell_z ]^{\mathsf{T}}|\ell_x,\ell_z\in[ -\frac{\sqrt{{A}_{\mathsf{s}}}}{2},\frac{\sqrt{{A}_{\mathsf{s}}}}{2}],m_x\in\mathcal{M}_x,m_z\in\mathcal{M}_z\}$.
	\vspace{-5pt}
	\begin{lemma}\label{cor_SPD}
		The channel gain and channel correlation factor for the planar SPDA satisfy ${\mathsf{g}}_k^{\mathsf{s}}= \zeta_{\mathsf{oc}}{\mathsf{g}}_k^{\mathsf{p}}$ and $\rho_{\mathsf{s}}={A}_{\mathsf{s}}\sum_{m_x\in \mathcal{M} _x}{\sum_{m_z\in \mathcal{M} _z}{ \mathsf{Q}_{1}^{*}(m_xd,m_zd)\mathsf{Q}_2(m_xd,m_zd) }}$, respectively, where $\zeta_{\mathsf{oc}}\triangleq\frac{{A}_{\mathsf{s}}}{d^2}\in(0,1]$ is defined as the array occupation ratio. 
	\end{lemma}
	\vspace{-5pt}
	\begin{IEEEproof}
		Please refer to Appendix \ref{proof_cor_SPD} for more details.
	\end{IEEEproof}
	Therefore, the asymptotic sum-rate capacity achieved by the planar SPDA is given by
	\begin{equation}\label{Sum_Rate_Capacity_SPD}
		\lim_{M_x,M_z\rightarrow \infty} \mathsf{C}_i\approx \sum\nolimits_{k=1}^{2}\log_2(1+\zeta_{\mathsf{oc}}{\overline{\gamma}_{i,k}}/{2}),
	\end{equation}
	where $i\in\{{\mathsf{ul}},{\mathsf{dl}}\}$. Comparing \eqref{asy_planar_capa} with \eqref{Sum_Rate_Capacity_SPD} yields the following observations.
	\vspace{-5pt}
	\begin{remark}\label{Remark_CAP_SPD}
		The capacity achieved by an SPDA increases to that achieved by a CAPA when $\zeta_{\mathsf{oc}}=1$. This makes intuitive sense as an SPDA turns into a CAPA when the array occupation ratio equals one.
	\end{remark}
	\vspace{-5pt}
	{We also note that the above conclusion is derived without considering the impact of EM coupling. When MC is taken into account, the relationship between the sum-rates achieved by CAPAs and SPDAs may depend on the size of each discrete element as well as the operating frequency \cite{akrout2023super}.}
	
	\begin{figure}[!t]
		\centering
		\subfigbottomskip=3pt
		\subfigcapskip=-2pt
		\setlength{\abovecaptionskip}{3pt}   
		\subfigure[$\left| \rho \right|^2$ vs. $n$ for small aperture sizes.]
		{
			\includegraphics[height=0.175\textwidth]{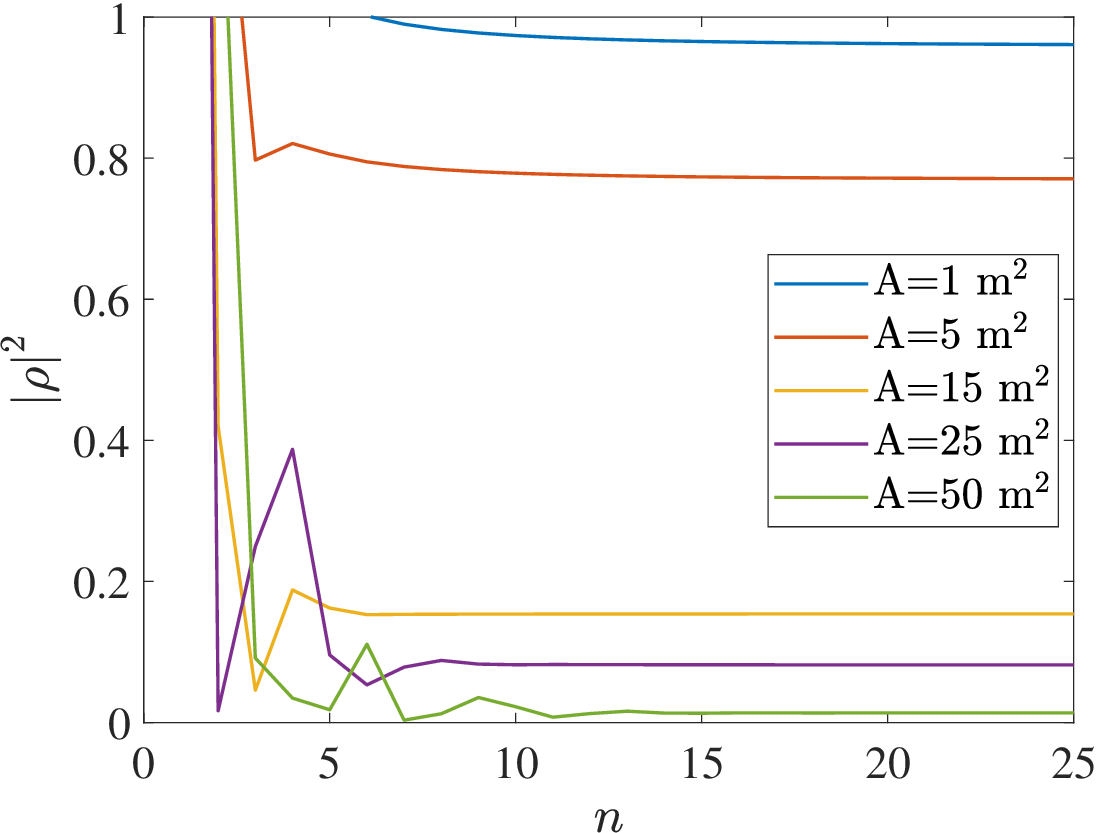}
			\label{rho1}	
		}%
		\subfigure[$\left| \rho \right|^2$ vs. $n$ with $A=10^6$ m$^2$.]
		{
			\includegraphics[height=0.175\textwidth]{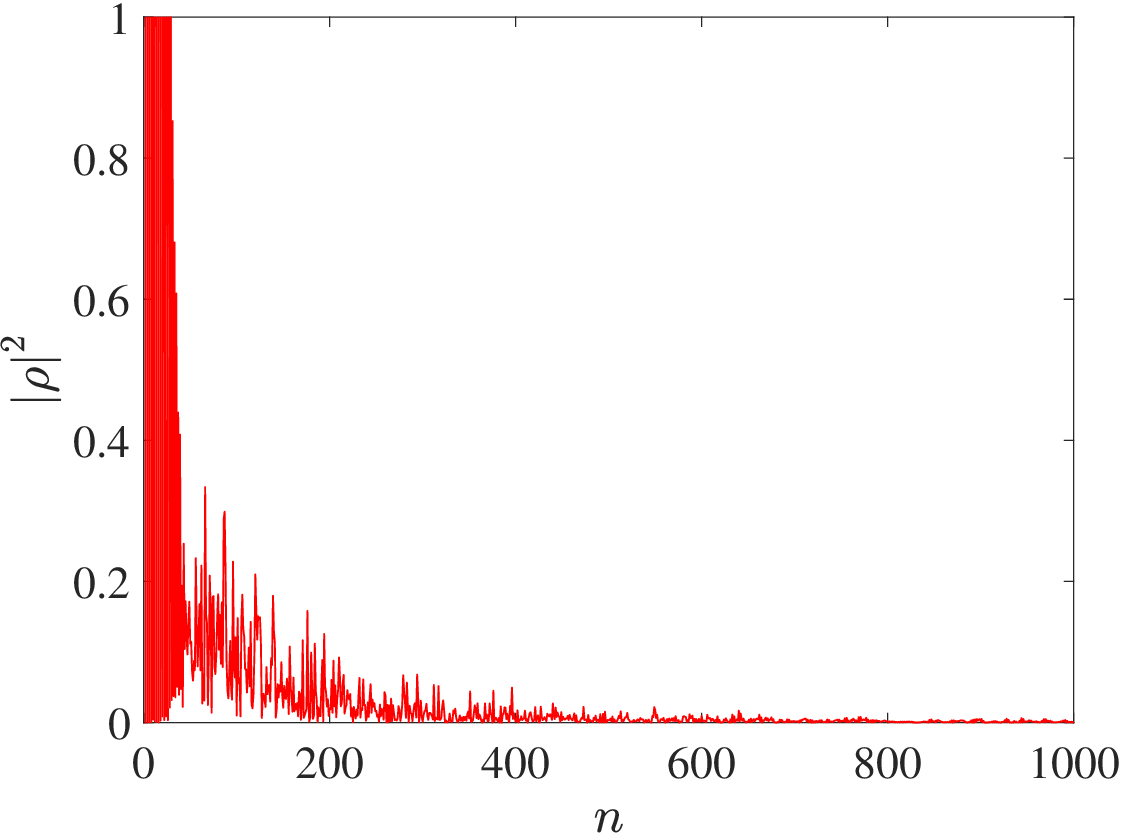}
			\label{rho2}	
		}
		\caption{{ Convergence of Chebyshev-Gauss quadrature.}}
		\label{Convergence}
		\vspace{-5pt}
	\end{figure}
	
	\begin{figure}[!t]
		\centering
		\includegraphics[height=0.28\textwidth]{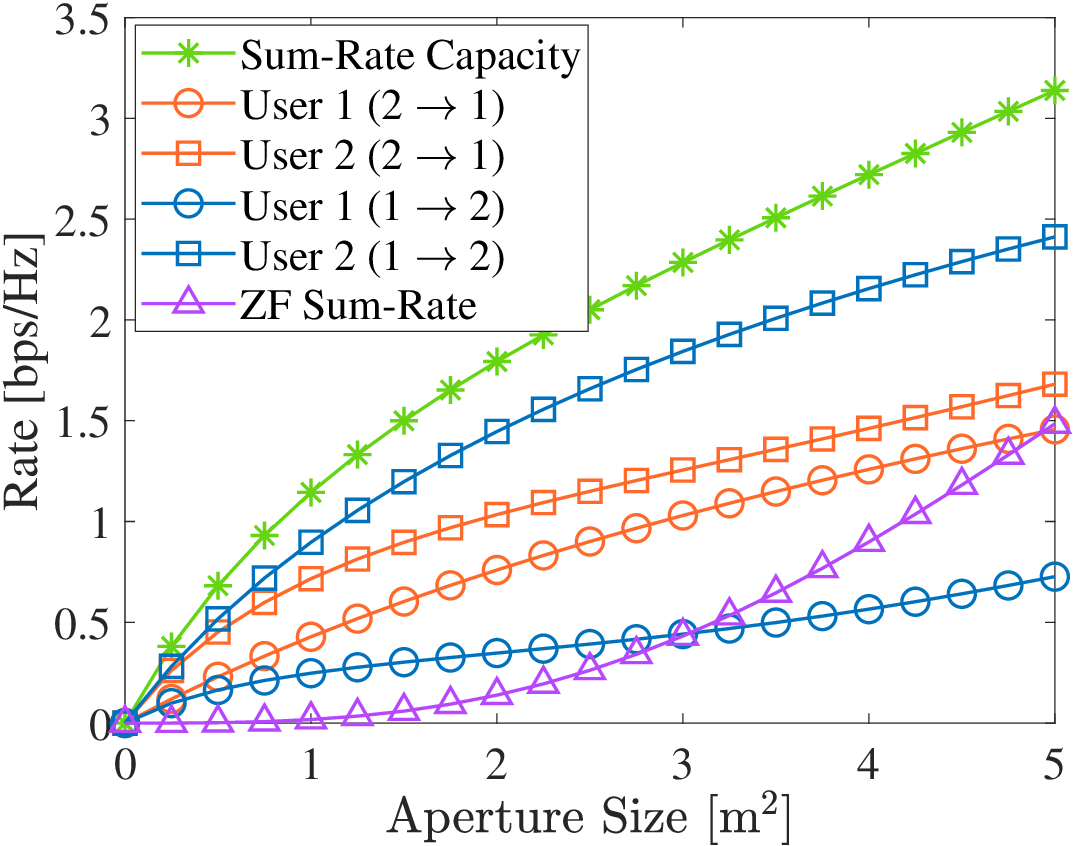}
		\caption{{ Uplink transmission rates of CAPA.}}
		\vspace{-7pt}
		\label{up_rate}
	\end{figure}
	
	\section{Numerical Results} \label{section_numerical}
	In this section, computer simulations are performed to demonstrate the performance of CAPA-based communications and verify the accuracy of the developed analytical results. For clarity, the simulations employ the planar arrays specified in Section \ref{Section: Special Cases}. Unless otherwise specified, the simulation parameters are set as follows: $\lambda = 0.125$ m, $A_{\mathsf{u},1}=A_{\mathsf{u},2}=A_{\mathsf{s}}=\frac{\lambda^2}{4\pi}$, $L_x=L_z=0.5$ m$^2$, $M_x=M_z$, $\overline{\gamma}_{\mathsf{ul},1}=30$ dB, $\overline{\gamma}_{\mathsf{ul},2}=40$ dB, $\overline{\gamma}_{\mathsf{dl},1}+\overline{\gamma}_{\mathsf{dl},2}=50$ dB with $\sigma_1^2=\sigma_2^2$, $\theta_1=\theta_2=\frac{\pi}{6}$, $\phi_1=\phi_2=\frac{\pi}{3}$, $r_1=10$ m, and $r_2=20$ m. 
	
\begin{figure}[!t]
		\centering
		\subfigbottomskip=3pt
		\subfigcapskip=-2pt
		\setlength{\abovecaptionskip}{3pt}   
		\subfigure[{ Sum-rate capacity vs. $A$.}]
		{
			\includegraphics[height=0.28\textwidth]{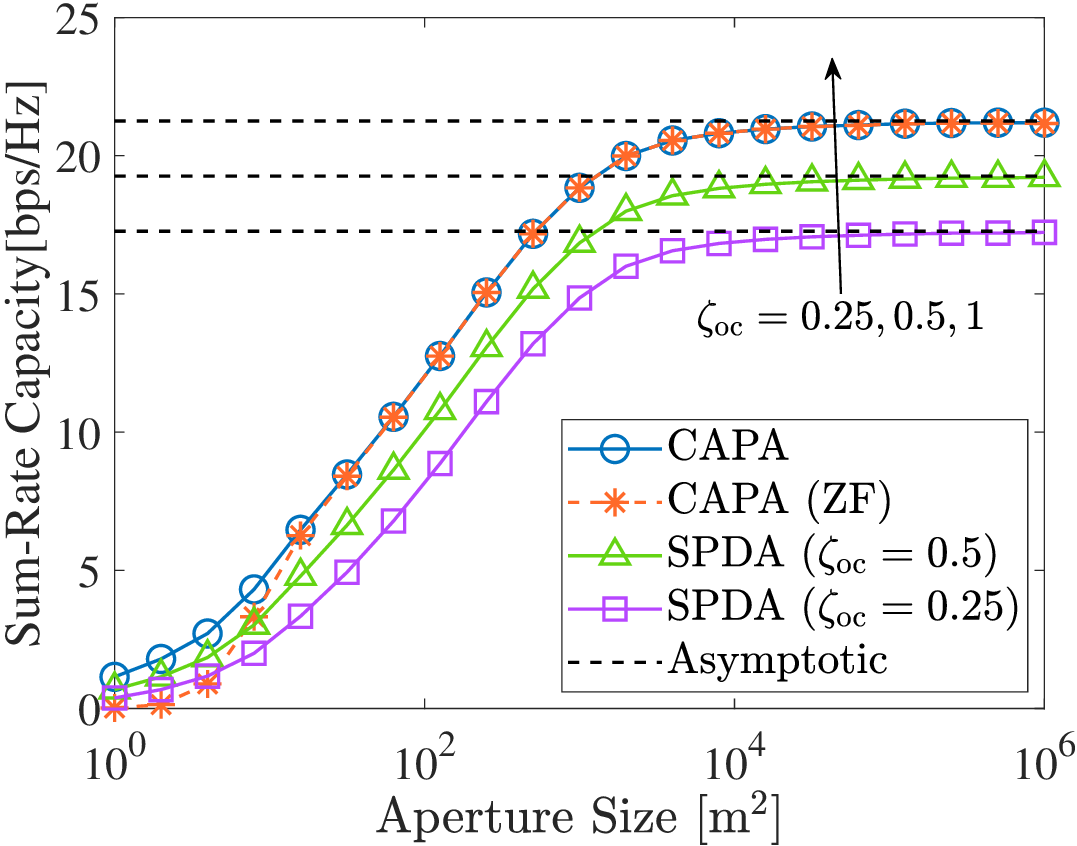}
			\label{up_size}	
		}
		\subfigure[Sum-rate capacity vs. $\zeta_{\mathsf{oc}}$.]
		{
			\includegraphics[height=0.28\textwidth]{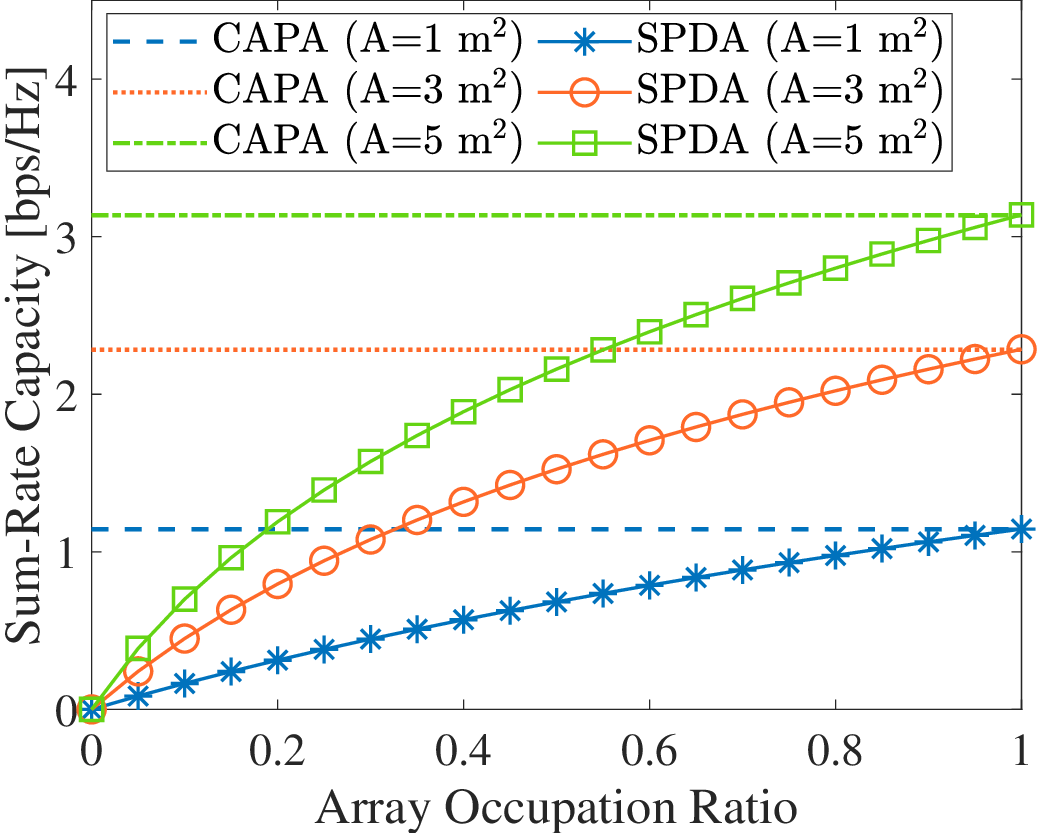}
			\label{up_aor}	
		}
		\caption{Uplink sum-rate capacity.}
		\label{up_capacity}
		\vspace{-7pt}
	\end{figure}

\begin{figure}[!t]
		\centering
		\subfigbottomskip=3pt
		\subfigcapskip=0pt
		\setlength{\abovecaptionskip}{3pt}
		\subfigure[{ Sum-rate capacity vs. $A$.}]
		{
			\includegraphics[height=0.28\textwidth]{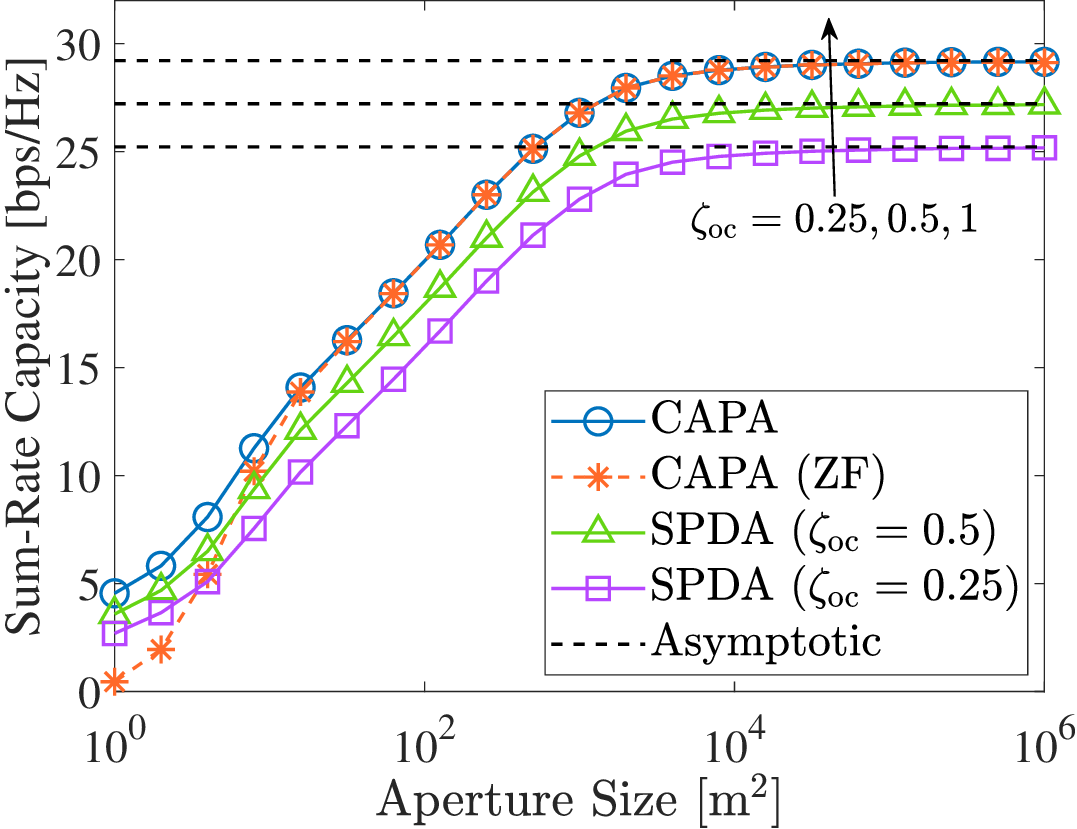}
			\label{do_size}	
		}
		\subfigure[Sum-rate capacity vs. $\zeta_{\mathsf{oc}}$.]
		{
			\includegraphics[height=0.28\textwidth]{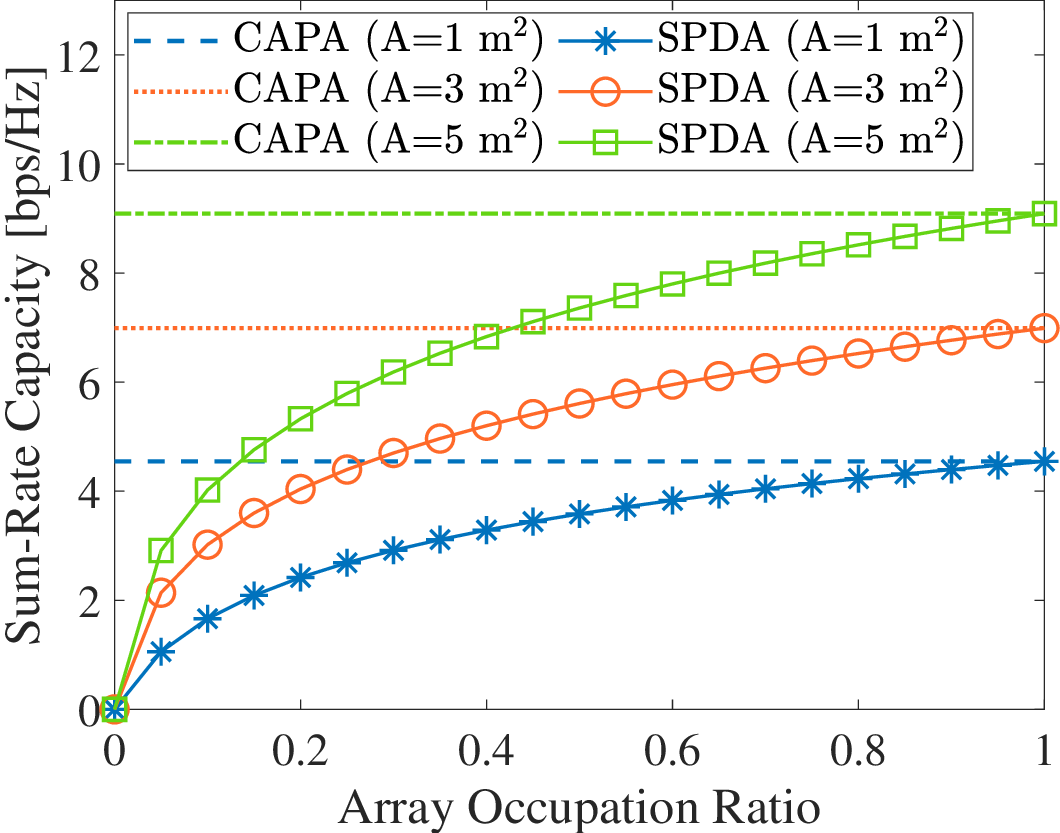}
			\label{do_aor}	
		}
		\caption{Downlink sum-rate capacity.}
		\label{do_capacity}
		\vspace{-7pt}
	\end{figure}

	{To select an appropriate value for the complexity-vs-accuracy tradeoff parameter $n$ in the Chebyshev-Gauss quadrature given in \eqref{rho_planar_capa}, we first analyze the convergence behavior of $\left| \rho \right|^2$ with respect to $n$ for different aperture sizes, as illustrated in {\figurename} {\ref{Convergence}}. The results indicate that for all aperture sizes depicted in {\figurename} {\ref{rho1}}, $\left| \rho \right|^2$ stabilizes once $n$ exceeds $20$. This means that using $20$ terms for summation in \eqref{rho_planar_capa} can precisely approximate the integral $\int_{{\mathcal{A}}}{\mathsf{G}}_1^*({\mathbf{r}}){\mathsf{G}}_2({\mathbf{r}}){\rm{d}}{\mathbf{r}}$ contained in the channel correlation factor $\left| \rho \right|^2=\frac{\left|\int_{{\mathcal{A}}}{\mathsf{G}}_1^*({\mathbf{r}}){\mathsf{G}}_2({\mathbf{r}}){\rm{d}}{\mathbf{r}}\right|^2}{{{\mathsf{g}}_1{\mathsf{g}}_2}}$. Additionally, we observe that larger aperture sizes require higher values of $n$ for the Chebyshev-Gauss quadrature to converge. To ensure convergence for extremely large array sizes, we conduct further simulations by examining an aperture size of $10^6$ m$^2$ in {\figurename} {\ref{rho2}}. The results reveal that convergence is effectively achieved when $n=1000$. Based on these observations and to maintain a balance between computational complexity and accuracy, we set $n=1000$ for {\figurename} {\ref{up_size}} and {\ref{do_size}} to illustrate asymptotic rate behavior for extremely larger arrays, while using $n=20$ for the other results.}
	
	\subsection{Uplink CAPA Communications}
	{\figurename} {\ref{up_rate}} illustrates the uplink per-user rate, sum-rate capacity, and the sum-rate achieved by the ZF detector in terms of the BS aperture size. It can be observed from this graph that as the aperture size $A$ grows, all the presented rates increase. This increase is attributed to the larger aperture size enhancing channel gain and reducing channel correlation, which in turn enhances the dedicated signal and diminishes the IUI, as discussed in \eqref{asy_planar_capa}. {Furthermore, it is important to note that the sum-rate achieved by the linear ZF detector is lower than the sum-rate capacity achieved by the proposed optimal SIC-based detector.}  
	
	To gain further insights into the capacity limit as the BS aperture size $A$ approaches infinity, we plot the sum-rate capacity against the array size for both the CAPA and SPDA in {\figurename} {\ref{up_size}}. It is evident that the CAPA outperforms the SPDA for arbitrary array size. As anticipated, increasing the aperture size causes the sum-rate capacity to approach its upper bound, as derived in \eqref{asy_planar_capa} and \eqref{Sum_Rate_Capacity_SPD}. It is worth mentioning that the asymptotic sum-rate capacity for an infinitely large aperture is a finite value, subject to the principle of energy conservation \cite{boqun_jstsp}. This verifies our discussions in \textbf{Remark \ref{energy_conservation}}. { Moreover, as the aperture size of the CAPA grows, the sum-rate achieved by the ZF detector asymptotically approaches the sum-rate capacity, which is consistent with the results in \eqref{zf_large}. This implies that the linear ZF detector, despite its lower implementation complexity, can deliver performance comparable to the optimal SIC scheme in the large-scale CAPA regime.} To compare the performance gap between CAPA and the conventional SPDA in terms of sum-rate capacity, we plot the sum-rate capacity for SPDAs as a function of the array occupation ratio $\zeta_{\mathsf{oc}}$ in {\figurename} {\ref{up_aor}}. As shown in the graph, the sum-rate capacity for an SPDA gradually converges with that of a CAPA as the array occupation ratio increases towards $1$. This observation validates the statements in \textbf{Remark \ref{Remark_CAP_SPD}}. 
	
	\begin{figure}[!t]
		\centering
		\subfigbottomskip=0pt
		\subfigcapskip=0pt
		\setlength{\abovecaptionskip}{2pt}
		\subfigure[Capacity region vs. $A$]
		{
			\includegraphics[height=0.28\textwidth]{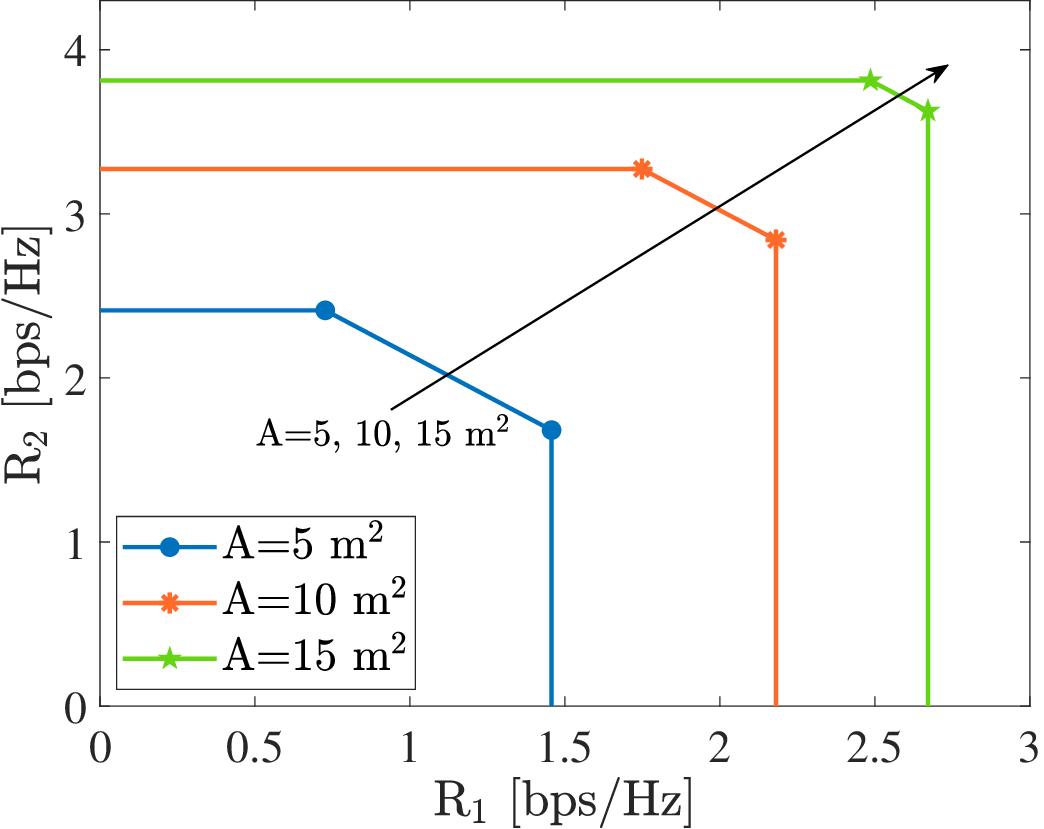}
			\label{up_region_size}	
		}
		\subfigure[{ Capacity regions with MC, $A=1$m$^2$ and $d=\frac{\lambda}{3}$.}]
		{
			\includegraphics[height=0.28\textwidth]{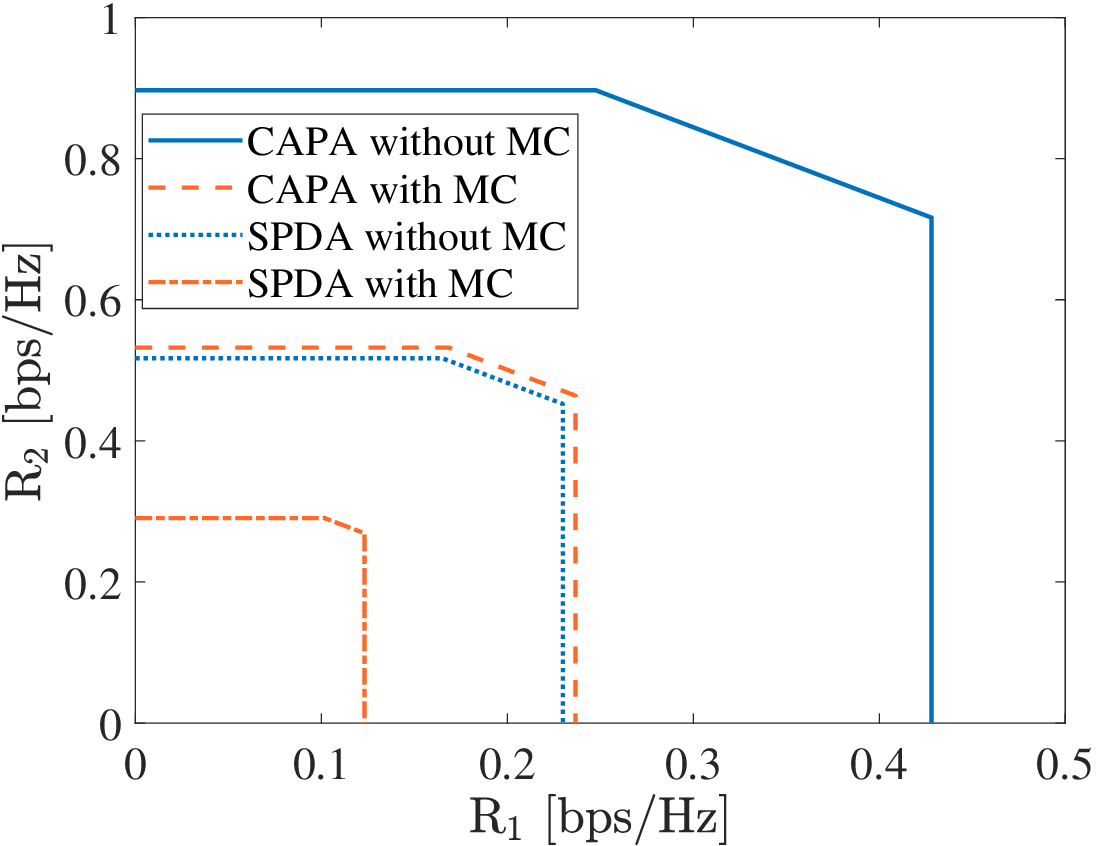}
			\label{up_region_mc}	
		}
		\caption{Uplink capacity regions.}
		\label{up_region}
		\vspace{-15pt}
	\end{figure}

	{\figurename} {\ref{up_region}} shows the uplink capacity region achieved by the CAPA, where the two corner points represent the achieved rates by \emph{SIC decoding} order $1\rightarrow2$ and $2\rightarrow1$. The rate tuple on the line segment connecting these two points is achieved by the \emph{time-sharing strategy}. In {\figurename} {\ref{up_region_size}}, the capacity regions achieved by the CAPA for various aperture sizes are illustrated. It can be observed that the capacity region expands with increasing $A$, shifting from a pentagonal to a rectangular shape. This phenomenon occurs because the IUI decreases as the aperture size increases, which is consistent with the results shown in {\figurename} {\ref{up_rate}}. 
	
	\begin{figure}[!t]
		\centering
		\includegraphics[height=0.28\textwidth]{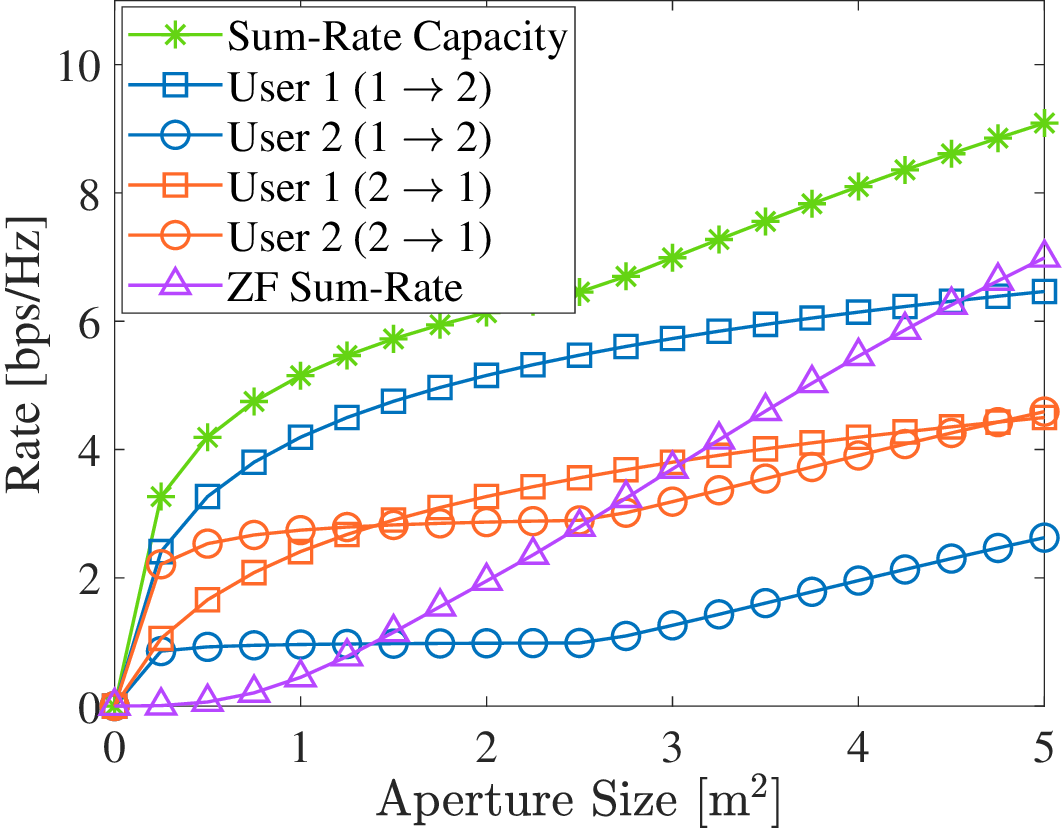}
		\caption{{ Downlink transmission rates of CAPA.}}
		\vspace{-5pt}
		\label{do_rate}
	\end{figure}
	
	{To further examine the impact of MC, {\figurename} {\ref{up_region_mc}} illustrates the rate regions with and without MC for both the CAPA and the SPDA. Since an exact MC model for continuous apertures remains an open research question, we approximate the CAPA as an SPDA with densely packed, edge-to-edge antenna elements \cite{castellanos2024electromagnetic}, where the element size is set as $\sqrt{A_{\mathsf{s}}}=d=\frac{\lambda}{3}$. In this case, the coupling matrix is modeled as $\left( z_a+z_t \right) \left( \mathbf{Z}+z_t\mathbf{I} \right) ^{-1}$, where $z_a$ represents the antenna impedance, $z_t$ denotes the termination impedance of each element, and $\mathbf{Z}$ is the mutual impedance matrix \cite{balanis2016antenna}. We set $z_a=z_t=50~\Omega$ and model the mutual impedance elements as $\left[ \mathbf{Z} \right] _{i,j}=0.1\frac{\mathrm{e}^{-\mathrm{j}k_0d_{ij}}}{d_{ij}^{2}}$, where $d_{ij}$ represents the distance between elements. From {\figurename} {\ref{up_region_mc}}, we observe that MC significantly degrades the performance of both CAPA and SPDA systems. However, despite the performance degradation caused by MC, the capacity region of CAPA remains larger than that of SPDA, highlighting its inherent advantage in terms of channel capacity.}

	\begin{figure}[!t]
		\centering
		\subfigbottomskip=0pt
		\subfigcapskip=0pt
		\subfigure[CAPA capacity region vs. $A$.]
		{
			\includegraphics[height=0.28\textwidth]{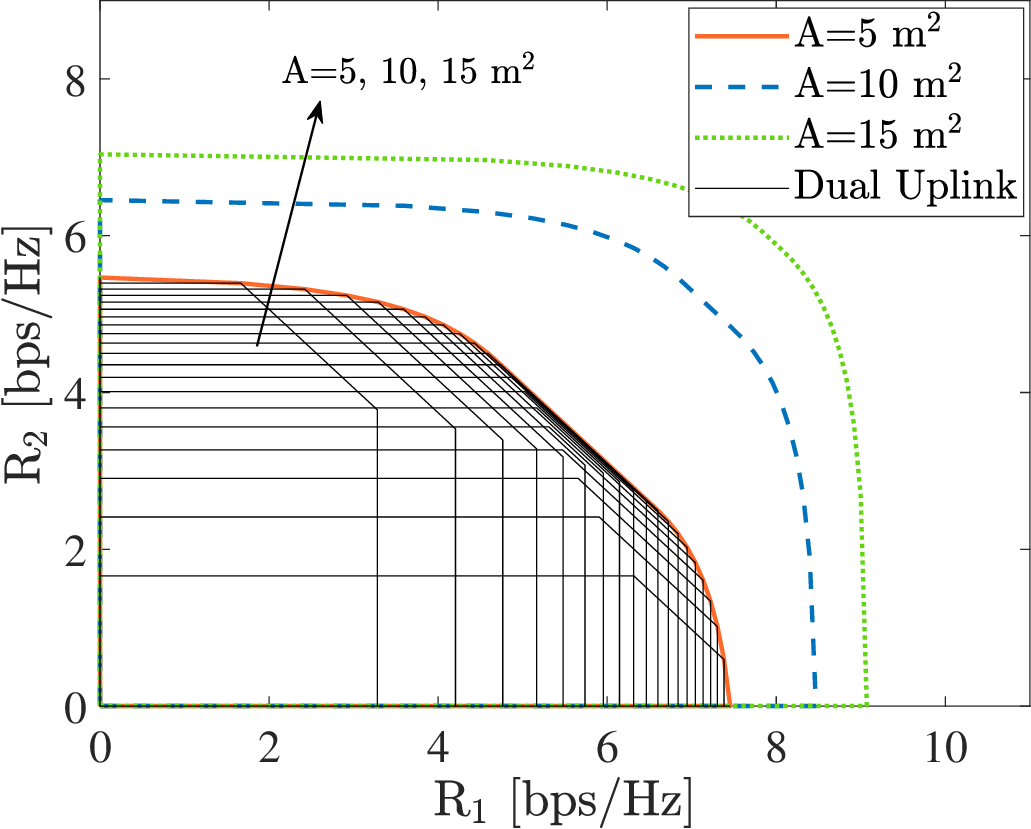}
			\label{do_region_size}	
		}
		\subfigure[{Capacity regions with MC, $A=1$m$^2$ and $d=\frac{\lambda}{3}$.}]
		{
			\includegraphics[height=0.29\textwidth]{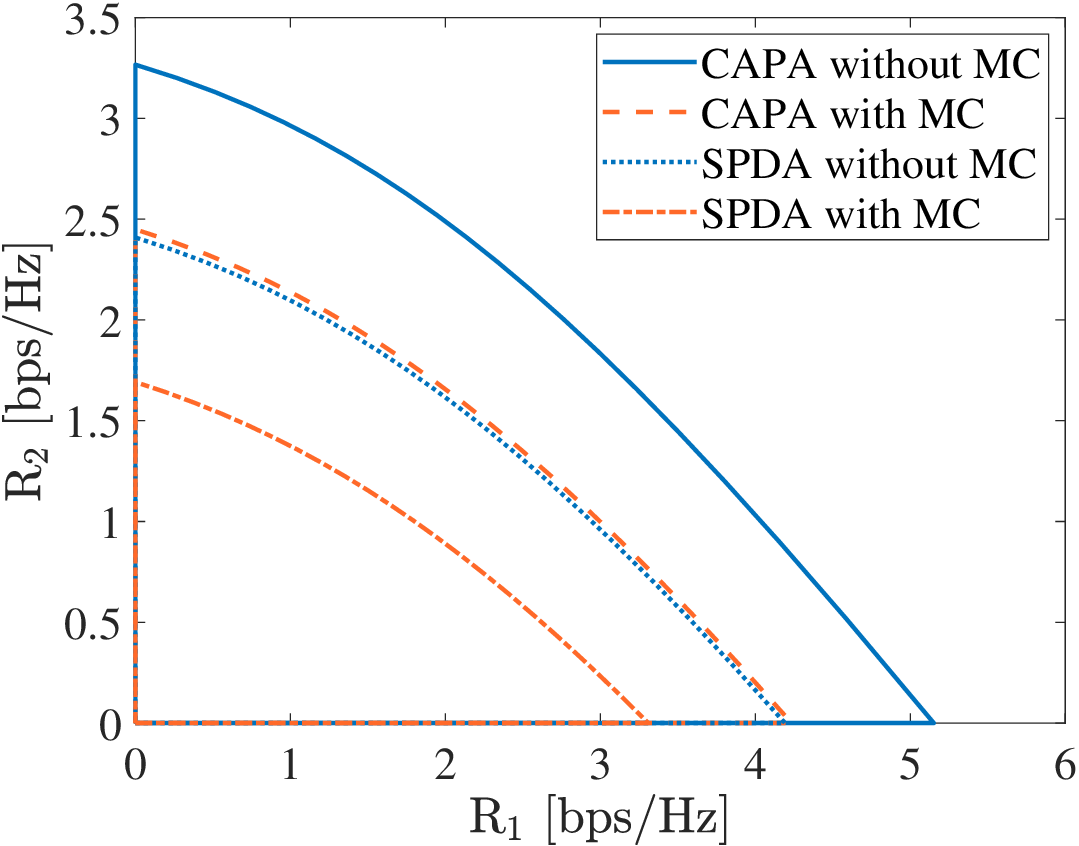}
			\label{do_region_mc}	
		}
		\caption{Downlink capacity regions.}
		\label{do_region}
		\vspace{-7pt}
	\end{figure}
	
	\subsection{Downlink CAPA Communications}
	{ {\figurename} \ref{do_rate} plots the per-user rates, the sum-rate capacity, and the sum-rate achieved by the ZF precoder as functions of the aperture size in downlink CAPA communications. A notable performance gap is observed between the sum-rate achieved by the linear ZF precoder and the sum-rate capacity. {\figurename} \ref{do_size} and {\figurename} \ref{do_aor} illustrate the downlink sum-rate capacity with respect to the aperture size and array occupation ratio, respectively. Similar to the uplink case, as the aperture size increases, the downlink sum-rate capacity converges to its upper bound, and the sum-rate achieved by the ZF scheme converges to the capacity.} Additionally, for the SPDA, the achieved sum-rate capacity is positively correlated with the array occupation ratio, further suggesting that the CAPA outperforms the SPDA in terms of channel capacity..
	
	{\figurename} \ref{do_region} illustrates the downlink capacity region achieved by the CAPA. The downlink capacity region is depicted by the convex hull of all the corresponding dual uplink capacity regions with the same sum power budget \cite{goldsmith}. This means that the boundary of the downlink capacity region is formed by the corner points of the dual uplink capacity regions, as illustrated in {\figurename} \ref{do_region_size}. It is also observed that the downlink capacity region of the CAPA expands with the aperture size $A$. { In Fig.~\ref{do_region_mc}, the MC effects are demonstrated using the same method as in the uplink case.} Additionally, in downlink scenario, the CAPA also achieves a more extensive capacity region than the conventional SPDA, highlighting its superiority.
	\section{Conclusion} \label{conclusion}
	This article has proposed an analytically tractable framework for CAPA-based multiuser communications by utilizing continuous operators to address the continuous nature of the EM field. This framework significantly differs from the conventional matrix-based transmission framework for SPDA-based communications. Based on this novel framework, we have characterized the capacity limits of uplink and downlink channels for CAPAs in both single-user and multiuser cases. For each scenario, we derived closed-form expressions for the capacity-achieving detectors or source currents along with the corresponding capacity. Additionally, we analyzed the channel capacity by specializing the derived results to several typical array structures. Through both theoretical analyses and numerical simulations, we have demonstrated that CAPA exhibits a higher sum-rate capacity and a more extensive capacity region compared to the conventional SPDA. This underscores CAPA's potential as a promising transmission paradigm for future wireless networks.
	\section*{Acknowledgement}
	The authors would like to thank the editor, Dr. Amine Mezghani, and anonymous reviewers for helpful comments that greatly improved the quality of the article.
	\begin{appendix}
		\setcounter{equation}{0}
		\renewcommand\theequation{A\arabic{equation}}
		\subsection{Proof of Lemma \ref{Lemma_Noise_Distribution}}\label{Proof_Lemma_Noise_Distribution}
		The mean of $\int_{{\mathcal{A}}}{\mathsf{V}}_{\mathsf{ul}}^{*}(\mathbf{r}){\mathsf{N}}_\mathsf{ul}(\mathbf{r}){\rm{d}}{\mathbf{r}}$ is calculated as follows:
		\begin{equation}
			{\mathbbmss{E}}\left\{\int_{{\mathcal{A}}}{\mathsf{V}}_{\mathsf{ul}}^{*}(\mathbf{r}){\mathsf{N}}_\mathsf{ul}(\mathbf{r}){\rm{d}}{\mathbf{r}}\right\}
			=\int_{{\mathcal{A}}}{\mathsf{V}}_{\mathsf{ul}}^{*}(\mathbf{r}){\mathbbmss{E}}\{{\mathsf{N}}_\mathsf{ul}(\mathbf{r})\}{\rm{d}}{\mathbf{r}},
		\end{equation}
		which, together with the fact that ${\mathbbmss{E}}\{{\mathsf{N}}_\mathsf{ul}(\mathbf{r})\}=0$, yields ${\mathbbmss{E}}\left\{\int_{{\mathcal{A}}}{\mathsf{V}}_{\mathsf{ul}}^{*}(\mathbf{r}){\mathsf{N}}_\mathsf{ul}(\mathbf{r}){\rm{d}}{\mathbf{r}}\right\}=0$. The variance is given by
		\begin{align}\label{Noise_Field_Covariance}
			&{\mathbbmss{E}}\left\{\int_{{\mathcal{A}}}{\mathsf{V}}_{\mathsf{ul}}^{*}(\mathbf{r}){\mathsf{N}}_\mathsf{ul}(\mathbf{r}){\rm{d}}{\mathbf{r}}
			\int_{{\mathcal{A}}}{\mathsf{V}}_{\mathsf{ul}}(\mathbf{r}'){\mathsf{N}}_{\mathsf{ul}}^*(\mathbf{r}'){\rm{d}}{\mathbf{r}}'\right\}\\
			&=\int_{{\mathcal{A}}}\int_{{\mathcal{A}}}{\mathsf{V}}_{\mathsf{ul}}^*(\mathbf{r})
			{\mathsf{V}}_{\mathsf{ul}}(\mathbf{r}')
			{\mathbbmss{E}}\{{\mathsf{N}}_{\mathsf{ul}}(\mathbf{r}){\mathsf{N}}_{\mathsf{ul}}^*(\mathbf{r}')\}{\rm{d}}{\mathbf{r}}{\rm{d}}{\mathbf{r}}'\\
			&=\int_{{\mathcal{A}}}{\mathsf{V}}_{\mathsf{ul}}(\mathbf{r}')\int_{{\mathcal{A}}}{\mathsf{V}}_{\mathsf{ul}}^*(\mathbf{r})
			\sigma^2\delta(\mathbf{r}-\mathbf{r}'){\rm{d}}{\mathbf{r}}{\rm{d}}{\mathbf{r}}'\\
			&\overset{\clubsuit}{=}\int_{{\mathcal{A}}}{\mathsf{V}}_{\mathsf{ul}}(\mathbf{r}')(\sigma^2{\mathsf{V}}_{\mathsf{ul}}^*(\mathbf{r}')){\rm{d}}{\mathbf{r}}'
			=\sigma^2\int_{{\mathcal{A}}}\lvert{\mathsf{V}}_{\mathsf{ul}}(\mathbf{r})\rvert^2{\rm{d}}{\mathbf{r}},
		\end{align}
		where the step $\clubsuit$ is based on the fact that $\int_{{\mathcal{A}}}\delta({\mathbf{x}}-{\mathbf{x}}_0)f(\mathbf{x}){\rm{d}}{\mathbf{x}}=f({\mathbf{x}}_0)$ with $f(\cdot)$ being an arbitrary function defined on ${\mathcal{A}}$. 
		
		\vspace{-5pt}
		\subsection{Proof of Lemma \ref{Lemma_IN_Whiten}}\label{Proof_Lemma_IN_Whiten}
		Firstly, we prove that the autocorrelation function of ${\mathsf{Z}}_{\mathsf{w}}(\mathbf{r})$ satisfies ${\mathbbmss{E}}\{{\mathsf{Z}}_{\mathsf{w}}(\mathbf{r}){\mathsf{Z}}_{\mathsf{w}}^*(\mathbf{r}')\}=
		{\sigma}^2\delta(\mathbf{r}-\mathbf{r}')$. Specifically, ${\mathbbmss{E}}\{{\mathsf{Z}}_{\mathsf{w}}(\mathbf{r}){\mathsf{Z}}_{\mathsf{w}}^*(\mathbf{r}')\}$ can be calculated as follows:
		\begin{align}
			&{\mathbbmss{E}}\{{\mathsf{Z}}_{\mathsf{w}}(\mathbf{r}){\mathsf{Z}}_{\mathsf{w}}^*(\mathbf{r}')\}\nonumber\\
			&=\int_{{\mathcal{A}}}\int_{{\mathcal{A}}}{\mathsf{W}}_{{\mathsf{Z}}}(\mathbf{r},\mathbf{x}){\mathsf{W}}_{{\mathsf{Z}}}^*(\mathbf{r}',\mathbf{x}')
			{\mathbbmss{E}}\{{\mathsf{Z}}(\mathbf{x}){\mathsf{Z}}^*(\mathbf{x}')\}{\rm{d}}\mathbf{x}{\rm{d}}\mathbf{x}'\\
			&=\int_{{\mathcal{A}}}{\mathsf{W}}_{{\mathsf{Z}}}^*(\mathbf{r}',\mathbf{x}')
			\int_{{\mathcal{A}}}{\mathsf{W}}_{{\mathsf{Z}}}(\mathbf{r},\mathbf{x})
			{\mathsf{R}}_{{\mathsf{Z}}}(\mathbf{x},{\mathbf{x}}'){\rm{d}}\mathbf{x}{\rm{d}}\mathbf{x}'.\label{Proof_Lemma_IN_Whiten_Step1}
		\end{align}
		By substituting \eqref{Autocorrelatioon_Interference_Noise} and \eqref{whiten_transform} into \eqref{Proof_Lemma_IN_Whiten_Step1} and then calculating the inner integral with respect to $\mathbf{x}$, we obtain
		\begin{equation}\label{Denominator_Calculation_Forward_1}
			\begin{split}
				&\int_{{\mathcal{A}}}{\mathsf{W}}_{{\mathsf{Z}}}(\mathbf{r},\mathbf{x})
				{\mathsf{R}}_{{\mathsf{Z}}}(\mathbf{x},{\mathbf{x}}'){\rm{d}}\mathbf{x}=\sigma^2(\delta({\mathbf{r}}-{\mathbf{x}}')\\
				&+(\overline{\gamma}_{\mathsf{ul},1}+\mu_1+\mu_1\overline{\gamma}_{\mathsf{ul},1}{\mathsf{g}}_1){\mathsf{G}}_1(\mathbf{r}){\mathsf{G}}_1^*(\mathbf{x}')).
			\end{split}
		\end{equation}
		By continuously substituting \eqref{Denominator_Calculation_Forward_1} and \eqref{whiten_transform} into \eqref{Proof_Lemma_IN_Whiten_Step1}, we next calculate the outer integral in terms of ${{\mathbf{x}}}'$, which yields
		\begin{equation}\label{Denominator_Calculation_Forward_2}
			\begin{split}
				&{\mathbbmss{E}}\{{\mathsf{Z}}_{\mathsf{w}}(\mathbf{r}){\mathsf{Z}}_{\mathsf{w}}^*(\mathbf{r}')\}=\sigma^2(\delta({\mathbf{r}}-{\mathbf{r}}')+{\mathsf{G}}_1(\mathbf{r}){\mathsf{G}}_1^*(\mathbf{r}')\\
				&\times(2\mu_1+\overline{\gamma}_{\mathsf{ul},1}+2\mu_1\overline{\gamma}_{\mathsf{ul},1}{{\mathsf{g}}_{1}}+\mu_1^2{{\mathsf{g}}_{1}}
				+\overline{\gamma}_{\mathsf{ul},1}\mu_1^2{{\mathsf{g}}_{1}^2})).
			\end{split}
		\end{equation}
		With $\mu_1=-\frac{1}{{\mathsf{g}}_{1}}\pm\frac{1}{{\mathsf{g}}_{1}
			\sqrt{1+\overline{\gamma}_{\mathsf{ul},1}{\mathsf{g}}_{1}}}$, it is readily shown that
		\begin{align}\label{mu_equation}
			2\mu_1+\overline{\gamma}_{\mathsf{ul},1}+2\mu_1\overline{\gamma}_{\mathsf{ul},1}{{\mathsf{g}}_{1}}+\mu_1^2{{\mathsf{g}}_{1}}
			+\overline{\gamma}_{\mathsf{ul},1}\mu_1^2{{\mathsf{g}}_{1}^2}=0.    
		\end{align}
		Therefore, we have ${\mathbbmss{E}}\{{\mathsf{Z}}_{\mathsf{w}}(\mathbf{r}){\mathsf{Z}}_{\mathsf{w}}^*(\mathbf{r}')\}={\sigma}^2\delta(\mathbf{r}-\mathbf{r}')$. 
		
		Next, we prove that ${\mathsf{W}}_{{\mathsf{Z}}}(\mathbf{r}',\mathbf{r})$ is an invertible linear transformation.
		We define
		\begin{align}\label{w_bar}
			\overline{\mathsf{W}}_{{\mathsf{Z}}}(\mathbf{r},\mathbf{r}')\triangleq\delta(\mathbf{r}-\mathbf{r}')-\frac{\mu_1}{1+\mu_1 {\mathsf{g}}_{1}}
			{\mathsf{G}}_1(\mathbf{r}){\mathsf{G}}_1^*(\mathbf{r}').
		\end{align}
		For an arbitrary function ${f}(\mathbf{r})$ defined in $\mathbf{r}\in{\mathcal{A}}$, we have
		\begin{equation}\label{Proof_Lemma_IN_Invertible_Step1}
			\begin{split}
				&\int_{{\mathcal{A}}}\overline{\mathsf{W}}_{{\mathsf{Z}}}(\mathbf{x},\mathbf{r}')\int_{{\mathcal{A}}}{\mathsf{W}}_{{\mathsf{Z}}}(\mathbf{r}',\mathbf{r})f(\mathbf{r})
				{\rm{d}}\mathbf{r}{\rm{d}}\mathbf{r}'\\
				&=\int_{{\mathcal{A}}}{f}(\mathbf{r})\int_{{\mathcal{A}}}
				\overline{\mathsf{W}}_{{\mathsf{Z}}}(\mathbf{x},\mathbf{r}'){\mathsf{W}}_{{\mathsf{Z}}}(\mathbf{r}',\mathbf{r})
				{\rm{d}}\mathbf{r}'{\rm{d}}\mathbf{r}.
			\end{split}
		\end{equation}
		By inserting \eqref{whiten_transform} and \eqref{w_bar} into \eqref{Proof_Lemma_IN_Invertible_Step1} and utilizing the fact that $\int_{{\mathcal{A}}}\delta({\mathbf{x}}-{\mathbf{x}}_0)f(\mathbf{x}){\rm{d}}{\mathbf{x}}=f({\mathbf{x}}_0)$, we calculate the inner integral in terms of $\mathbf{r}'$ as follows: 
		\begin{equation}\label{Proof_Lemma_IN_Invertible_Step2}
			\begin{split}
				&\int_{{\mathcal{A}}}
				\overline{\mathsf{W}}_{{\mathsf{Z}}}(\mathbf{x},\mathbf{r}'){\mathsf{W}}_{{\mathsf{Z}}}(\mathbf{r}',\mathbf{r})
				{\rm{d}}\mathbf{r}'=\delta({\mathbf{x}}-{\mathbf{r}})+{\mathsf{G}}_1(\mathbf{x})\\
				&\times{\mathsf{G}}_1^*(\mathbf{r})\bigg(\mu_1-\frac{\mu_1+\mu_1^2{\mathsf{g}}_1}{1+\mu_1{\mathsf{g}}_1}\bigg)=\delta({\mathbf{x}}-{\mathbf{r}}).
			\end{split}
		\end{equation}
		Plugging \eqref{Proof_Lemma_IN_Invertible_Step2} into \eqref{Proof_Lemma_IN_Invertible_Step1} gives
		\begin{equation}\label{Proof_Lemma_IN_Invertible_Step4}
			\begin{split}
				&\int_{{\mathcal{A}}}\overline{\mathsf{W}}_{{\mathsf{Z}}}(\mathbf{x},\mathbf{r}')\int_{{\mathcal{A}}}{\mathsf{W}}_{{\mathsf{Z}}}(\mathbf{r}',\mathbf{r})f(\mathbf{r})
				{\rm{d}}\mathbf{r}{\rm{d}}\mathbf{r}'\\
				&=\int_{{\mathcal{A}}}{f}(\mathbf{r})\delta({\mathbf{x}}-{\mathbf{r}}){\rm{d}}\mathbf{r}={f}(\mathbf{x}),
			\end{split}
		\end{equation}
		which suggests that the inversion of ${\mathsf{W}}_{{\mathsf{Z}}}(\mathbf{r}',\mathbf{r})$ is $\overline{\mathsf{W}}_{{\mathsf{Z}}}(\mathbf{r},\mathbf{r}')$. 
		
		\subsection{Proof of Theorem \ref{Theorem_MMSE_SNR_First_User}}\label{Proof_Theorem_MMSE_SNR_First_User}
		By definition, $\overline{\mathsf{H}}_2(\mathbf{r}')$ can be calculated as follows:
		\setlength\abovedisplayskip{4.5pt}
		\setlength\belowdisplayskip{4.5pt}
		\begin{align}
			\overline{\mathsf{H}}_2(\mathbf{r}')&=
			\int_{{\mathcal{A}}}(\delta(\mathbf{r}'-\mathbf{r})+\mu_1
			{\mathsf{G}}_1(\mathbf{r}'){\mathsf{G}}_1^*(\mathbf{r})){\mathsf{H}}_2(\mathbf{r}){\rm{d}}\mathbf{r}\\
			&={{\rm{j}}k_0\eta}/{\sqrt{4\pi}}({\mathsf{G}}_2(\mathbf{r}')+\mu_1\sqrt{{\mathsf{g}}_1{\mathsf{g}}_2}\rho{\mathsf{G}}_1(\mathbf{r}')).
			\label{Proof_Theorem_MMSE_SNR_First_User_Step1}
		\end{align}
		Inserting \eqref{Proof_Theorem_MMSE_SNR_First_User_Step1} into \eqref{Two_User_MMSE_SNR} yields
		\begin{align}
			\gamma_{\mathsf{ul},2}&=\overline{\gamma}_{\mathsf{ul},2}
			\int_{{\mathcal{A}}}\lvert{\mathsf{G}}_2(\mathbf{r}')+\mu_1\sqrt{{\mathsf{g}}_1{\mathsf{g}}_2}\rho{\mathsf{G}}_1(\mathbf{r}')\rvert^2{\rm{d}}{\mathbf{r}}'\\
			&=\overline{\gamma}_{\mathsf{ul},2}\int_{{\mathcal{A}}}\lvert{\mathsf{G}}_2(\mathbf{r}')\rvert^2{\rm{d}}{\mathbf{r}}'+\overline{\gamma}_{\mathsf{ul},2}{\mu_1^2}{\mathsf{g}}_1{\mathsf{g}}_2\lvert\rho
			\rvert^2\int_{{\mathcal{A}}}\lvert{\mathsf{G}}_1(\mathbf{r}')\rvert^2{\rm{d}}{\mathbf{r}}'\notag\\
			&+2\overline{\gamma}_{\mathsf{ul},2}\mu_1\sqrt{{\mathsf{g}}_1{\mathsf{g}}_2}\Re\left\{\rho\int_{{\mathcal{A}}}{\mathsf{G}}_1({\mathbf{r}}){\mathsf{G}}_2^*({\mathbf{r}}){\rm{d}}{\mathbf{r}}\right\}\\
			&=\overline{\gamma}_{\mathsf{ul},2}{\mathsf{g}}_2(1+{\mathsf{g}}_1\lvert\rho\rvert^2\mu_1({\mu_1}{\mathsf{g}}_1+2)).\label{Proof_Theorem_MMSE_SNR_First_User_Step2}
		\end{align}
		Furthermore, based on \eqref{mu_equation}, we have
		\begin{equation}
			\begin{split}
				\mu_1({\mu_1}{{\mathsf{g}}_{1}}+2)=-\overline{\gamma}_{\mathsf{ul},1}(1+\mu_1{\mathsf{g}}_{1})^2,
			\end{split}
		\end{equation}
		which, together with \eqref{Proof_Theorem_MMSE_SNR_First_User_Step2}, yields
		\begin{align}\label{Proof_Theorem_Sum_Rate_Capacity_Step1}
			\gamma_{\mathsf{ul},2}
			=\overline{\gamma}_{\mathsf{ul},2}{\mathsf{g}}_2(1-\overline{\gamma}_{\mathsf{ul},1}{\mathsf{g}}_1\lvert\rho\rvert^2(1+\mu_1{\mathsf{g}}_{1})^2).
		\end{align}
		Plugging $\mu_1=-\frac{1}{{\mathsf{g}}_{1}}\pm\frac{1}{{\mathsf{g}}_{1}
			\sqrt{1+\overline{\gamma}_{\mathsf{ul},1}{\mathsf{g}}_{1}}}$ into \eqref{Proof_Theorem_Sum_Rate_Capacity_Step1} gives
		\begin{align}\label{Proof_Theorem_Sum_Rate_Capacity_Step2}
			\gamma_{\mathsf{ul},2}=\overline{\gamma}_{\mathsf{ul},2}\mathsf{g}_2\left( 1-\frac{\overline{\gamma}_{\mathsf{ul},1}\mathsf{g}_1\left| \rho \right|^2}{1+\overline{\gamma}_{\mathsf{ul},1}\mathsf{g}_1} \right) .
		\end{align}
		The final results follow immediately.
		\subsection{Proof of Theorem \ref{theorem_optimal_current}}\label{proof_theorem_optimal_current}
		Our objective is to verify that the presented source current distributions $\{{\mathsf{J}}_{{\mathsf{dl}},k}({\mathbf{r}})\}_{k=1}^{2}$ are able to: \romannumeral1) achieve the downlink sum-rate capacity $\log_2(1+\overline{\gamma }_{\mathsf{dl},1}\mathsf{g}_1+\overline{\gamma }_{\mathsf{dl},2}\mathsf{g}_2+\overline{\gamma}_{\mathsf{dl},1}\overline{\gamma }_{\mathsf{dl},2}\mathsf{g}_1\mathsf{g}_2\overline{\rho})$, and \romannumeral2) satisfy the sum power constraint $\sum_{k=1}^{2}{\int_{\mathcal{A}}{\lvert {\mathsf{J}}_{{\mathsf{dl}},k}({\mathbf{r}}) \rvert^2}\mathrm{d}\mathbf{r}}={\mathsf{P}}_{1}+{\mathsf{P}}_{2}= {\mathsf{P}}$.
		
		Substituting \eqref{J1} into $\lvert \int_{{\mathcal{A}}}\hat{\mathsf{H}}_{1}({\mathbf{r}}){\mathsf{J}}_{{\mathsf{dl}},1}({\mathbf{r}}){\rm{d}}{\mathbf{r}} \rvert^2$ gives
		\begin{equation}
			\begin{split}
				&\left\lvert \int_{{\mathcal{A}}}\hat{\mathsf{H}}_{1}({\mathbf{r}}){\mathsf{J}}_{{\mathsf{dl}},1}({\mathbf{r}}){\rm{d}}{\mathbf{r}}\right\rvert^2=\frac{{{\mathsf{P}}_{1}}{\mathsf{P}}_{2}\lvert\int_{\mathcal{A}}
					\hat{\mathsf{H}}_{1}^{*}(\mathbf{r})\hat{\mathsf{H}}_{2}(\mathbf{r}){\rm{d}}{\mathbf{r}}\rvert^2}{-1-{\mathsf{P}}_{2}\int_{\mathcal{A}}
					\lvert\hat{\mathsf{H}}_{2}(\mathbf{r})\rvert^2{\rm{d}}{\mathbf{r}}}\\
				&+{{\mathsf{P}}_{1}}\int_{\mathcal{A}}
				\lvert\hat{\mathsf{H}}_{1}(\mathbf{r})\rvert^2{\rm{d}}{\mathbf{r}}=\overline{\gamma }_{\mathsf{dl},1}\mathsf{g}_1\left( 1-\frac{\overline{\gamma }_{\mathsf{dl},2}\mathsf{g}_2|\rho |^2}{1+\overline{\gamma }_{\mathsf{dl},2}\mathsf{g}_2} \right),
			\end{split}
		\end{equation}
		which yields 
		\begin{align}
			\mathsf{R}_{\mathsf{dl},1}^{2\rightarrow 1}&=\log_2\left(1+\left\lvert \int_{{\mathcal{A}}}\hat{\mathsf{H}}_{1}({\mathbf{r}}){\mathsf{J}}_{{\mathsf{dl}},1}({\mathbf{r}}){\rm{d}}{\mathbf{r}} \right\rvert^2\right)\\
			&=\log_2\left( 1+\overline{\gamma }_{\mathsf{dl},1}\mathsf{g}_1\left( 1-\frac{\overline{\gamma }_{\mathsf{dl},2}\mathsf{g}_2|\rho |^2}{1+\overline{\gamma }_{\mathsf{dl},2}\mathsf{g}_2} \right) \right).\label{D1}
		\end{align}
		Moreover, inserting \eqref{J2} into \eqref{do_R2} gives
		\begin{align}
			&\mathsf{R}_{\mathsf{dl},2}^{2\rightarrow 1}=\log _2\left( 1+\frac{\lvert \int_{{\mathcal{A}}}\hat{\mathsf{H}}_{2}({\mathbf{r}}){\mathsf{J}}_{{\mathsf{dl}},2}({\mathbf{r}}){\rm{d}}{\mathbf{r}} \rvert^2}{1+\lvert \int_{{\mathcal{A}}}\hat{\mathsf{H}}_{2}({\mathbf{r}}){\mathsf{J}}_{{\mathsf{dl}},1}({\mathbf{r}}){\rm{d}}{\mathbf{r}} \rvert^2} \right)\\
			&=\log _2\left( 1+{\mathsf{P}}_2\frac{\lvert \int_{\mathcal{A}}
				\lvert\hat{\mathsf{H}}_{2}(\mathbf{r})\rvert^2{\rm{d}}{\mathbf{r}} \rvert^2}{\int_{\mathcal{A}}
				\lvert\hat{\mathsf{H}}_{2}(\mathbf{r})\rvert^2{\rm{d}}{\mathbf{r}}} \right)\\
			&=\log _2\left( 1+\overline{\gamma }_{\mathsf{dl},2}\mathsf{g}_2 \right).\label{D2}
		\end{align}
		Inserting \eqref{D1} and \eqref{D2} into $\mathsf{R}_{\mathsf{dl},1}^{2\rightarrow 1}+\mathsf{R}_{\mathsf{dl},2}^{2\rightarrow 1}$ gives the sum-rate capacity for a given $({\mathsf{P}}_1,{\mathsf{P}}_2)$, i.e., $\log_2(1+\overline{\gamma }_{\mathsf{dl},1}\mathsf{g}_1+\overline{\gamma }_{\mathsf{dl},2}\mathsf{g}_2+\overline{\gamma}_{\mathsf{dl},1}\overline{\gamma }_{\mathsf{dl},2}\mathsf{g}_1\mathsf{g}_2\overline{\rho})$, which concludes the proof of the first part.
		
		We next turn to the sum power constraint. Note that
		\begin{align}
			\int_{\mathcal{A}}{\lvert {\mathsf{J}}_{{\mathsf{dl}},1}(\mathbf{r}) \rvert^2}\mathrm{d}\mathbf{r}=\int_{\mathcal{A}}{\int_{\mathcal{A}}{\mathsf{J}}_{{\mathsf{dl}},1}(\mathbf{r})\delta \left( \mathbf{r}-\mathbf{r}' \right) {\mathsf{J}}_{{\mathsf{dl}},1}^{*}(\mathbf{r}')}\mathrm{d}\mathbf{r}\mathrm{d}\mathbf{r}'. \label{J1_power}
		\end{align}
		On the other hand, based on \eqref{J2}, $\int_{\mathcal{A}}{\lvert {\mathsf{J}}_{{\mathsf{dl}},2}(\mathbf{r}) \rvert^2}\mathrm{d}\mathbf{r}$ can be calculated as follows:
		\begin{align}
			&\int_{\mathcal{A}}{\lvert {\mathsf{J}}_{{\mathsf{dl}},2}(\mathbf{r}) \rvert^2}\mathrm{d}\mathbf{r}
			={{\mathsf{P}}_{2}}+{{\mathsf{P}}_{2}}\left\lvert\int_{\mathcal{A}}
			\hat{\mathsf{\mathsf{H}}}_{2}(\mathbf{r}){\mathsf{J}}_{{\mathsf{dl}},1}({\mathbf{r}}){\rm{d}}{\mathbf{r}}\right\rvert^2\\
			&={{\mathsf{P}}_{2}}+{{\mathsf{P}}_{2}}\int_{\mathcal{A}}\int_{\mathcal{A}}{\mathsf{J}}_{{\mathsf{dl}},1}({\mathbf{r}})
			\hat{\mathsf{\mathsf{H}}}_{2}(\mathbf{r})\hat{\mathsf{\mathsf{H}}}_{2}^{*}(\mathbf{r}'){\mathsf{J}}_{{\mathsf{dl}},1}^{*}({\mathbf{r}}){\rm{d}}{\mathbf{r}}{\rm{d}}{\mathbf{r}}.\label{J2power}
		\end{align}
		The sum of \eqref{J1_power} and \eqref{J2power} satisfies
		\begin{equation}\label{sum_power_1} 
			\begin{split}
				\sum_{k=1}^{2}{\int_{\mathcal{A}}{\lvert {\mathsf{J}}_{{\mathsf{dl}},k}({\mathbf{r}}) \rvert^2}\mathrm{d}\mathbf{r}}&={{\mathsf{P}}_{2}}+
				\int_{\mathcal{A}}\int_{\mathcal{A}}{\mathsf{J}_{{\mathsf{dl}},1}(\mathbf{r})}(\delta ( \mathbf{r}-\mathbf{r}' )\\
				&+{{\mathsf{P}}_{2}}{\hat{\mathsf{H}}}_2(\mathbf{r})\hat{\mathsf{H}}_{2}^{*}(\mathbf{r}^{\prime}))
				\mathrm{d}\mathbf{r}{\mathsf{J}}_{{\mathsf{dl}},1}^{*}(\mathbf{r}')\mathrm{d}\mathbf{r}'.
			\end{split}
		\end{equation}
		Substituting \eqref{J1} into \eqref{sum_power_1} yields
		\begin{align}
			&\int_{\mathcal{A}}\int_{\mathcal{A}}{\mathsf{J}_{{\mathsf{dl}},1}(\mathbf{r})}(\delta ( \mathbf{r}-\mathbf{r}' )+{{\mathsf{P}}_{2}}{\hat{\mathsf{H}}}_2(\mathbf{r})
			\hat{\mathsf{H}}_{2}^{*}(\mathbf{r}'))\mathrm{d}\mathbf{r}{\mathsf{J}}_{{\mathsf{dl}},1}^{*}(\mathbf{r}')\mathrm{d}\mathbf{r}'\nonumber\\
			&={\mathsf{P}}_1\frac{\int_{\mathcal{A}}\int_{\mathcal{A}}f_{\mathsf{n}}(\mathbf{r})(\delta ( \mathbf{r}-\mathbf{r}' )+{{\mathsf{P}}_{2}}{\hat{\mathsf{H}}}_2(\mathbf{r})
				\hat{\mathsf{H}}_{2}^{*}(\mathbf{r}'))\mathrm{d}\mathbf{r}f_{\mathsf{n}}^{*}(\mathbf{r}')\mathrm{d}\mathbf{r}'}{\int_{\mathcal{A}}
				\lvert\hat{\mathsf{H}}_{1}(\mathbf{r})\rvert^2{\rm{d}}{\mathbf{r}}-\frac{{\mathsf{P}}_{2}\lvert\int_{\mathcal{A}}
					\hat{\mathsf{H}}_{1}^{*}(\mathbf{r})\hat{\mathsf{H}}_{2}(\mathbf{r}){\rm{d}}{\mathbf{r}}\rvert^2}{1+{\mathsf{P}}_{2}\int_{\mathcal{A}}
					\lvert\hat{\mathsf{H}}_{2}(\mathbf{r})\rvert^2{\rm{d}}{\mathbf{r}}}},\label{Sum_Power_Constraint_Num_Ini1}
		\end{align}
		where $f_{\mathsf{n}}(\mathbf{r})\triangleq\hat{\mathsf{H}}_{1}^{*}(\mathbf{r})-\frac{{\mathsf{P}}_{2}\int_{\mathcal{A}}
			\hat{\mathsf{H}}_{1}^{*}(\mathbf{r})\hat{\mathsf{H}}_{2}(\mathbf{r}){\rm{d}}{\mathbf{r}}}{1+{\mathsf{P}}_{2}\int_{\mathcal{A}}
			\lvert\hat{\mathsf{H}}_{2}(\mathbf{r})\rvert^2{\rm{d}}{\mathbf{r}}}\hat{\mathsf{H}}_{2}^{*}(\mathbf{r})$.
		The inner integral in the numerator of \eqref{Sum_Power_Constraint_Num_Ini1} can be calculated as follows:
		\begin{align}
			&\int_{\mathcal{A}}f_{\mathsf{n}}(\mathbf{r})(\delta ( \mathbf{r}-\mathbf{r}' )+{{\mathsf{P}}_{2}}{\hat{\mathsf{H}}}_2(\mathbf{r})
			\hat{\mathsf{H}}_{2}^{*}(\mathbf{r}'))\mathrm{d}\mathbf{r}\\
			&=\hat{\mathsf{H}}_{1}^{*}(\mathbf{r}')+\bigg(1-\frac{1+{\mathsf{P}}_{2}\int_{\mathcal{A}}\lvert\hat{\mathsf{H}}_{2}(\mathbf{r})\rvert^2{\rm{d}}{\mathbf{r}}}{1+{\mathsf{P}}_{2}\int_{\mathcal{A}}
				\lvert\hat{\mathsf{H}}_{2}(\mathbf{r})\rvert^2{\rm{d}}{\mathbf{r}}}\bigg)\nonumber\\
			&\times{\mathsf{P}}_{2}\bigg(\int_{\mathcal{A}}
			\hat{\mathsf{H}}_{1}^{*}(\mathbf{r})\hat{\mathsf{H}}_{2}(\mathbf{r}){\rm{d}}{\mathbf{r}}\bigg)\hat{\mathsf{H}}_{2}^{*}(\mathbf{r}')
			=\hat{\mathsf{H}}_{1}^{*}(\mathbf{r}'),
		\end{align}
		based on which the outer integral in the numerator of \eqref{Sum_Power_Constraint_Num_Ini1} is calculated as follows:
		\begin{equation}
			\begin{split}
				\int_{\mathcal{A}}\!\hat{\mathsf{H}}_{1}^{*}(\mathbf{r}')f_{\mathsf{n}}^{*}(\mathbf{r}')\mathrm{d}\mathbf{r}'\!\!=\!-\frac{{\mathsf{P}}_{2}\lvert\int_{\mathcal{A}}\!
					\hat{\mathsf{H}}_{1}^{*}(\mathbf{r})\hat{\mathsf{H}}_{2}(\mathbf{r}){\rm{d}}{\mathbf{r}}\rvert^2}{1+{\mathsf{P}}_{2}\int_{\mathcal{A}}
					\lvert\hat{\mathsf{H}}_{2}(\mathbf{r})\rvert^2{\rm{d}}{\mathbf{r}}}\!+\!\!\int_{\mathcal{A}}\!\lvert\hat{\mathsf{H}}_{1}(\mathbf{r})\rvert^2{\rm{d}}{\mathbf{r}}.
			\end{split}
		\end{equation}
		As a result, we obtain
		\begin{align}
			\eqref{Sum_Power_Constraint_Num_Ini1}=
			{\mathsf{P}}_1\frac{\int_{\mathcal{A}}
				\lvert\hat{\mathsf{H}}_{1}(\mathbf{r})\rvert^2{\rm{d}}{\mathbf{r}}-\frac{{\mathsf{P}}_{2}\lvert\int_{\mathcal{A}}
					\hat{\mathsf{H}}_{1}^{*}(\mathbf{r})\hat{\mathsf{H}}_{2}(\mathbf{r}){\rm{d}}{\mathbf{r}}\rvert^2}{1+{\mathsf{P}}_{2}\int_{\mathcal{A}}
					\lvert\hat{\mathsf{H}}_{2}(\mathbf{r})\rvert^2{\rm{d}}{\mathbf{r}}}}{\int_{\mathcal{A}}
				\lvert\hat{\mathsf{H}}_{1}(\mathbf{r})\rvert^2{\rm{d}}{\mathbf{r}}-\frac{{\mathsf{P}}_{2}\lvert\int_{\mathcal{A}}
					\hat{\mathsf{H}}_{1}^{*}(\mathbf{r})\hat{\mathsf{H}}_{2}(\mathbf{r}){\rm{d}}{\mathbf{r}}\rvert^2}{1+{\mathsf{P}}_{2}\int_{\mathcal{A}}
					\lvert\hat{\mathsf{H}}_{2}(\mathbf{r})\rvert^2{\rm{d}}{\mathbf{r}}}}={\mathsf{P}}_1,
		\end{align}
		which, together with \eqref{sum_power_1}, suggests that $\sum_{k=1}^{2}{\int_{\mathcal{A}}{\lvert {\mathsf{J}}_{{\mathsf{dl}},k}({\mathbf{r}}) \rvert^2}\mathrm{d}\mathbf{r}}={\mathsf{P}}_{1}+{\mathsf{P}}_{2}$. As the downlink sum-rate equals that of its dual uplink channel under the same sum power constraint, the downlink capacity is achieved when the CAPA transmits at maximum power, i.e., ${\mathsf{P}}_{1}+{\mathsf{P}}_{2}={\mathsf{P}}$, which concludes the proof of the second part.

		\subsection{Proof of Theorem \ref{theorem_downlink-to-uplink transformation}}\label{proof_theorem_downlink-to-uplink transformation}
		Our objective is to verify that the presented power allocation policy $({\mathsf{P}}_1,{\mathsf{P}}_2)$ is able to: \romannumeral1) ensure the uplink rates equal the downlink rates shown in \eqref{do_R1} and \eqref{do_R2}, and \romannumeral2) satisfy the sum power constraint ${\mathsf{P}}_{1}+{\mathsf{P}}_{2}\leq\sum_{k=1}^{2}{\int_{\mathcal{A}}{\lvert {\mathsf{J}}_{{\mathsf{dl}},k}({\mathbf{r}}) \rvert^2}\mathrm{d}\mathbf{r}}\leq {\mathsf{P}}$. The first part can be directly proved by substituting \eqref{P2} and \eqref{P1} into \eqref{Dual_Uplink_R1} and \eqref{Dual_Uplink_R2}. We next turn to the sum power constraint.
		
		Following the derivation steps in Appendix \ref{Proof_Lemma_IN_Whiten}, it is proved that the inversion of ${\mathsf{U}}(\mathbf{r}',\mathbf{r})$ is $\overline{\mathsf{U}}(\mathbf{r},\mathbf{r}')$, where ${\mathsf{U}}(\mathbf{r}',\mathbf{r})=\delta(\mathbf{r}'-\mathbf{r})+\mu_2{\mathsf{G}}_2(\mathbf{r}'){\mathsf{G}}_2^*(\mathbf{r})$ with $\mu_2=-\frac{1}{{\mathsf{g}}_{2}}\pm\frac{1}{{\mathsf{g}}_{2}\sqrt{1+\overline{\gamma}_{\mathsf{ul},2}{\mathsf{g}}_{2}}}$ and $\overline{\mathsf{U}}(\mathbf{r},\mathbf{r}')=\delta(\mathbf{r}-\mathbf{r}')-\frac{\mu_2}{1+\mu_2 {\mathsf{g}}_{2}}{\mathsf{G}}_2(\mathbf{r}){\mathsf{G}}_2^*(\mathbf{r}')$, i.e., $\int_{\mathcal{A}}\overline{\mathsf{U}}(\mathbf{r},{\mathbf{x}}){\mathsf{U}}(\mathbf{x},\mathbf{r}'){\rm{d}}{\mathbf{x}}=
		\int_{\mathcal{A}}{\mathsf{U}}(\mathbf{r},{\mathbf{x}})\overline{\mathsf{U}}(\mathbf{x},\mathbf{r}'){\rm{d}}{\mathbf{x}}=\delta({\mathbf{r}}-{\mathbf{r}}')$. Furthermore, it is worth noting that $\int_{\mathcal{A}}
		\hat{\mathsf{\mathsf{H}}}_{1}(\mathbf{r}){\mathsf{J}}_{{\mathsf{dl}},1}({\mathbf{r}}){\rm{d}}{\mathbf{r}}=
		\int_{\mathcal{A}}\int_{\mathcal{A}}
		\hat{\mathsf{\mathsf{H}}}_{1}(\mathbf{r}')\delta(\mathbf{r}'-\mathbf{r}){\mathsf{J}}_{{\mathsf{dl}},1}({\mathbf{r}}){\rm{d}}{\mathbf{r}}'{\rm{d}}{\mathbf{r}}\triangleq {\diamond}$. It follows that
		\begin{align}
			{\diamond}&=\int_{\mathcal{A}}\int_{\mathcal{A}}\int_{\mathcal{A}}
			\hat{\mathsf{\mathsf{H}}}_{1}(\mathbf{r}'){\mathsf{U}}(\mathbf{r}',{\mathbf{x}})\overline{\mathsf{U}}(\mathbf{x},\mathbf{r}){\mathsf{J}}_{{\mathsf{dl}},1}({\mathbf{r}}){\rm{d}}{\mathbf{r}}
			{\rm{d}}{\mathbf{x}}{\rm{d}}{\mathbf{r}}'\nonumber\\
			&=\int_{\mathcal{A}}\int_{\mathcal{A}}
			\hat{\mathsf{\mathsf{H}}}_{1}(\mathbf{r}'){\mathsf{U}}(\mathbf{r}',{\mathbf{x}}){\rm{d}}{\mathbf{r}}'
			\int_{\mathcal{A}}\overline{\mathsf{U}}(\mathbf{x},\mathbf{r}){\mathsf{J}}_{{\mathsf{dl}},1}({\mathbf{r}}){\rm{d}}{\mathbf{r}}
			{\rm{d}}{\mathbf{x}},
		\end{align}
		which, together with the Cauchy's inequality, yields
		\begin{align}\label{Cauchy's inequality}
			\lvert{\diamond}\rvert^2\leq\int_{\mathcal{A}}\lvert{\diamond}_1(\mathbf{x})\rvert^2{\rm{d}}{\mathbf{x}}\int_{\mathcal{A}}\lvert{\diamond}_2(\mathbf{x})\rvert^2{\rm{d}}{\mathbf{x}},
		\end{align}
		where ${\diamond}_1(\mathbf{x})\triangleq\int_{\mathcal{A}}\overline{\mathsf{U}}(\mathbf{x},\mathbf{r}){\mathsf{J}}_{{\mathsf{dl}},1}({\mathbf{r}}){\rm{d}}{\mathbf{r}}
		={\mathsf{J}}_{{\mathsf{dl}},1}({\mathbf{x}})-\frac{\mu_2{\mathsf{G}}_2^*(\mathbf{x})}{1+\mu_2 {\mathsf{g}}_{2}}\int_{\mathcal{A}}{\mathsf{G}}_2(\mathbf{r}){\mathsf{J}}_{{\mathsf{dl}},1}({\mathbf{r}}){\rm{d}}{\mathbf{r}}$ and ${\diamond}_2(\mathbf{x})\triangleq \int_{\mathcal{A}}
		\hat{{\mathsf{H}}}_{1}(\mathbf{r}'){\mathsf{U}}(\mathbf{r}',{\mathbf{x}}){\rm{d}}{\mathbf{r}}'=\mu_2{\mathsf{G}}_2(\mathbf{x})
		\int_{\mathcal{A}}{\mathsf{G}}_2^*(\mathbf{r}')\hat{\mathsf{\mathsf{H}}}_{1}(\mathbf{r}'){\rm{d}}{\mathbf{r}}'+\hat{\mathsf{\mathsf{H}}}_{1}(\mathbf{x})$. Consequently, we have
		\begin{align}
			&\int_{\mathcal{A}}\lvert{\diamond}_2(\mathbf{x})\rvert^2{\rm{d}}{\mathbf{x}}=\int_{\mathcal{A}}\lvert\hat{{\mathsf{H}}}_{1}(\mathbf{x})\rvert^2{\rm{d}}{\mathbf{x}}+
			\Big\lvert\int_{\mathcal{A}}{\mathsf{G}}_2^*(\mathbf{r}')\hat{\mathsf{\mathsf{H}}}_{1}(\mathbf{r}'){\rm{d}}{\mathbf{r}}'\Big\rvert^2\nonumber\\
			&\times(2\mu_2+\mu_2^2{\mathsf{g}}_2)=
			\int_{\mathcal{A}}\lvert\hat{{\mathsf{H}}}_{1}(\mathbf{x})\rvert^2{\rm{d}}{\mathbf{x}}-
			\frac{\lvert\int_{\mathcal{A}}{\mathsf{G}}_2(\mathbf{r}')\hat{\mathsf{\mathsf{H}}}_{1}^*(\mathbf{r}'){\rm{d}}{\mathbf{r}}'\rvert^2}{(1+\overline{\gamma}_{\mathsf{dl},2}{\mathsf{g}}_2)
				/\overline{\gamma}_{\mathsf{dl},2}}\nonumber\\
			&=\int_{\mathcal{A}}\lvert\hat{\mathsf{\mathsf{H}}}_{1}(\mathbf{r})\rvert^2{\rm{d}}{\mathbf{r}}-\frac{{\mathsf{P}}_{2}\lvert\int_{\mathcal{A}}
				\hat{\mathsf{\mathsf{H}}}_{1}^{*}(\mathbf{r})\hat{\mathsf{\mathsf{H}}}_{2}(\mathbf{r}){\rm{d}}{\mathbf{r}}\rvert^2}{1+{\mathsf{P}}_{2}\int_{\mathcal{A}}
				\lvert\hat{\mathsf{\mathsf{H}}}_{2}(\mathbf{r})\rvert^2{\rm{d}}{\mathbf{r}}}, 
		\end{align}
		which, together with \eqref{P1} and \eqref{Cauchy's inequality}, yields ${\mathsf{P}}_1\leq \int_{\mathcal{A}}\lvert{\diamond}_1(\mathbf{x})\rvert^2{\rm{d}}{\mathbf{x}}$. Furthermore, we obtain
		\begin{align}
			&\int_{\mathcal{A}}\lvert{\diamond}_1(\mathbf{x})\rvert^2{\rm{d}}{\mathbf{x}}=\int_{\mathcal{A}}\lvert{\mathsf{J}}_{{\mathsf{dl}},1}({\mathbf{r}})\rvert^2{\rm{d}}{\mathbf{r}}-
			\Big\lvert\int_{\mathcal{A}}{\mathsf{G}}_2(\mathbf{r}){\mathsf{J}}_{{\mathsf{dl}},1}({\mathbf{r}}){\rm{d}}{\mathbf{r}}\Big\rvert^2\nonumber\\
			&\times\Big(\frac{2\mu_2}{1+\mu_2 {\mathsf{g}}_{2}}-\Big(\frac{\mu_2}{1+\mu_2 {\mathsf{g}}_{2}}\Big)^2{\mathsf{g}}_{2}\Big).
		\end{align}
		Recalling that $\mu_2=-\frac{1}{{\mathsf{g}}_{2}}\pm\frac{1}{{\mathsf{g}}_{2}\sqrt{1+\overline{\gamma}_{\mathsf{dl},2}{\mathsf{g}}_{2}}}$ yields $\frac{2\mu_2}{1+\mu_2 {\mathsf{g}}_{2}}-(\frac{\mu_2}{1+\mu_2 {\mathsf{g}}_{2}})^2{\mathsf{g}}_{2}=-\overline{\gamma}_{\mathsf{dl},2}$, which yields
		\begin{align}
			\int_{\mathcal{A}}\lvert{\diamond}_1(\mathbf{x})\rvert^2{\rm{d}}{\mathbf{x}}=\int_{\mathcal{A}}\lvert{\mathsf{J}}_{{\mathsf{dl}},1}({\mathbf{r}})\rvert^2{\rm{d}}{\mathbf{r}}+{\mathsf{P}}_2
			\Big\lvert\int_{\mathcal{A}}\hat{\mathsf{H}}_2(\mathbf{r}){\mathsf{J}}_{{\mathsf{dl}},1}({\mathbf{r}}){\rm{d}}{\mathbf{r}}\Big\rvert^2.
		\end{align}
		As proved above, we have ${\mathsf{P}}_1\leq \int_{\mathcal{A}}\lvert{\diamond}_1(\mathbf{x})\rvert^2{\rm{d}}{\mathbf{x}}$, and thus the powers satisfy
		\begin{align}
			{\mathsf{P}}_1+{\mathsf{P}}_2\leq \int_{\mathcal{A}}\lvert{\mathsf{J}}_{{\mathsf{dl}},1}({\mathbf{r}})\rvert^2{\rm{d}}{\mathbf{r}}+{\mathsf{P}}_2
			\Big(1+\Big\lvert\int_{\mathcal{A}}\hat{\mathsf{H}}_2(\mathbf{r}){\mathsf{J}}_{{\mathsf{dl}},1}({\mathbf{r}}){\rm{d}}{\mathbf{r}}\Big\rvert^2\Big),
		\end{align}
		which, together with \eqref{P2}, yields
		\begin{align}
			{\mathsf{P}}_1+{\mathsf{P}}_2&\leq \int_{\mathcal{A}}\lvert{\mathsf{J}}_{{\mathsf{dl}},1}({\mathbf{r}})\rvert^2{\rm{d}}{\mathbf{r}}+
			\frac{\lvert\int_{\mathcal{A}}
				\hat{\mathsf{\mathsf{H}}}_{2}(\mathbf{r}){\mathsf{J}}_{{\mathsf{dl}},2}({\mathbf{r}}){\rm{d}}{\mathbf{r}}\rvert^2}{\int_{\mathcal{A}}\lvert\hat{\mathsf{\mathsf{H}}}_{2}(\mathbf{r})\rvert^2
				{\rm{d}}{\mathbf{r}}},\nonumber\\
			&\overset{\heartsuit}{\leq} \int_{\mathcal{A}}\lvert{\mathsf{J}}_{{\mathsf{dl}},1}({\mathbf{r}})\rvert^2{\rm{d}}{\mathbf{r}} + \int_{\mathcal{A}}\lvert{\mathsf{J}}_{{\mathsf{dl}},2}({\mathbf{r}})\rvert^2{\rm{d}}{\mathbf{r}}\leq {\mathsf{P}},\label{Downlink_Power_Constraint_Proof}
		\end{align}
		where the step $\heartsuit$ in \eqref{Downlink_Power_Constraint_Proof} is due to the Cauchy's inequality. This completes the proof of \textbf{Theorem \ref{theorem_downlink-to-uplink transformation}}.
		
		\subsection{Proof of Lemma \ref{lemma_planar_capa}}\label{proof_lemma_planar_capa}
		Based on \eqref{case_stduy_G}, the channel gain for the planar CAPA can be calculated as follows:
		\begin{align}
			&\mathsf{g}_k=\int_{\frac{L_z}{-2}}^{\frac{L_z}{2}}{\int_{\frac{L_x}{-2}}^{\frac{L_x}{2}}{\mathsf{Q}_{k}^{*}(x,z)\mathsf{Q}_k(x,z)\mathrm{d}x}\mathrm{d}z}
			=\frac{r_k\Psi _k}{4\pi}\int_{-\frac{L_z}{2}}^{\frac{L_z}{2}}\nonumber\\
			&\times{\int_{-\frac{L_x}{2}}^{\frac{L_x}{2}}}{{(x^2+z^2-2r_k\left( \Phi _kx+\Theta _kz \right) +r_{k}^{2})^{-\frac{3}{2}}\mathrm{d}x}\mathrm{d}z}.
		\end{align}
		Subsequently, we can calculate the inner integral with the aid of \cite[Eq. (2.264.5)]{integral} and the outer integral with the aid of \cite[Eq. (2.284)]{integral}, which yields the results in \eqref{CAPA_UPA_Channel_Gain}. The channel correlation factor can be written as follows:
		\begin{align}
			\rho =\frac{1}{\sqrt{\mathsf{g}_1\mathsf{g}_2}}\int_{-\frac{L_z}{2}}^{\frac{L_z}{2}}{\int_{-\frac{L_x}{2}}^{\frac{L_x}{2}}{\mathsf{Q}_{1}^{*}(x,z)\mathsf{Q}_2(x,z)\mathrm{d}x}\mathrm{d}z}.
		\end{align}
		The above integrals can be calculated by applying the Chebyshev-Gauss quadrature rule, i.e., $\int_{-1}^1{\frac{f\left( x \right)}{\sqrt{1-x^2}}\mathrm{d}}x\approx \frac{\pi}{n}\sum_{j=1}^n{f\left( x_j \right)}$ with $x_j=\cos \left( \frac{\left( 2j-1 \right) \pi}{2n} \right) $, leading to the results in \eqref{rho_planar_capa}. This completes the proof of \textbf{Lemma \ref{lemma_planar_capa}}.
		\subsection{Proof of Lemma \ref{cor_SPD}}\label{proof_cor_SPD}
		In the case of an planar SPDA, due to the small element size compared to the distance between the BS and the user, i.e., $\sqrt{{A}_{\mathsf{s}}}\ll r_k$, the variation of the channel across an element is negligible. Thus, the channel gain can be written as follows:
		\begin{align}\label{e1}
			\mathsf{g}_k=\sum\nolimits_{m_x\in \mathcal{M} _x}{\sum\nolimits_{m_z\in \mathcal{M} _z}{{{A}_{\mathsf{s}}}\left| \mathsf{Q}_k(m_xd,m_zd) \right|}}^2.
		\end{align}
		Defining $\epsilon_k=\frac{d}{r_k} \ll 1$, we can rewrite \eqref{e1} as follows:
		\begin{align}\label{e2}
			\begin{split}
				\mathsf{g}_k=\frac{A\Psi _k}{4\pi r_{k}^{2}}
				\sum_{m_x\in \mathcal{M} _x}\sum_{m_z\in \mathcal{M} _z}f_{\mathsf{sp}}^{(k)}\left( m_x\epsilon_k,m_z\epsilon_k \right) ,
			\end{split}
		\end{align}
		where $f_{\mathsf{sp}}^{(k)}\left( x,z \right) \triangleq (x^2+z^2-2\Phi_kx-2\Omega_kz+1)^{-\frac{3}{2}}$ is defined within the square area ${\mathcal{C}}_k\triangleq\{ (x,z) \mid -\frac{M_x\epsilon _k}{2}\leq x\leq \frac{M_x\epsilon _k}{2},-\frac{M_z\epsilon _k}{2}\leq z\leq \frac{M_z\epsilon _k}{2} \}  $, which is then divided into $M_xM_z$ sub-squares, each with an area $\epsilon _k^2$. As $\epsilon _k\ll 1$, we have $f_{\mathsf{sp}}^{(k)}\left( x,z \right) \approx f_{\mathsf{sp}}^{(k)}\left( m_x\epsilon _k,m_z\epsilon _k \right) $ for $\forall \left( x,z \right) \in \left\{ \left( x,z \right) \mid \left( m_x-\frac{1}{2} \right) \epsilon _k\leq x\leq \left( m_x+\frac{1}{2} \right) \epsilon _k,\left( m_z-\frac{1}{2} \right) \epsilon _k\leq z \right.$ $\left.\leq \left( m_z+\frac{1}{2} \right) \epsilon _k \right\}$. It follows from the concept of double integral that
		\begin{align}
			\sum_{m_x\in \mathcal{M} _x}\sum_{m_z\in \mathcal{M} _z}\!\!\!{f_{\mathsf{sp}}^{(k)}\left( m_x\epsilon _k,m_z\epsilon _k \right) \epsilon _k^2}\approx \!\!\iint_{{\mathcal{C}}_k}{f_{\mathsf{sp}}^{(k)}\left( x,z \right) \mathrm{d}x\mathrm{d}z}.  
		\end{align}
		Consequently, \eqref{e2} can be rewritten as follows:
		\begin{equation}
			\begin{split}
				\mathsf{g}_k\approx\frac{\zeta_{\mathsf{oc}} \Psi _k}{4\pi}\int_{-\frac{M_z\epsilon _k}{2}}^{\frac{M_z\epsilon _k}{2}}{\int_{-\frac{M_x\epsilon _k}{2}}^{\frac{M_x\epsilon _k}{2}}}f_{\mathsf{sp}}^{(k)}\left( x,z \right)\mathrm{d}x\mathrm{d}z, 
			\end{split}
		\end{equation}
		which can be calculated by utilizing \cite[Eqs. (2.264.5) \& (2.284)]{integral}. Furthermore, the expression of $\rho_{\mathsf{s}}$ follows from its definition, which completes the entire proof. 
	\end{appendix}
	\bibliographystyle{IEEEtran}
	\bibliography{mybib}
\end{document}